\documentstyle[12pt]{article}
\newcommand{\p}[1]{(\ref{#1})}
\newcommand{\be}{\begin{equation}}
\newcommand{\bea}{\begin{eqnarray}}
\newcommand{\ee}{\end{equation}}
\newcommand{\eea}{\end{eqnarray}}

\def\theequation{\arabic{section}.\arabic{equation}}

\def\dsll{\not {\! \partial}}
\def\psisl{\not {\! \! \psi}}

\def\xisl{\not {\! \xi}}
\def\esl{\not {\! \epsilon}}
\def\ssl{\not {\! \cal S}}

\def\ssl{\not {\! \cal S}}
\def\theequation{\arabic{section}.\arabic{equation}}
\begin{document}
\topmargin -1cm \oddsidemargin=0.25cm\evensidemargin=0.25cm
\setcounter{page}0
\renewcommand{\thefootnote}{\fnsymbol{footnote}}
\begin{titlepage}
\begin{flushright}
DFTT-10/2008\\
TUW-08-06
\end{flushright}
\vskip .7in
\begin{center}
{\Large \bf Gauge Invariant Lagrangians for Free and Interacting
Higher Spin Fields. A Review of the BRST formulation.} \vskip .7in
{\large Angelos Fotopoulos$^a$}\footnote{e-mail: {\tt
foto@to.infn.it}} and {\large Mirian Tsulaia$^b$}\footnote{e-mail:
{\tt tsulaia@hep.itp.tuwien.ac.at}} \vskip .2in {$^a$\it
Dipartimento di Fisica Teorica dell'Universit\`a di Torino and
INFN\\Sezione di Torino,
via P. Giuria 1, I-10125 Torino, Italy} \\
\vskip .2in {$^b$ \it Institute for Theoretical Physics, Vienna University of Technology,
Wiedner Hauptstrasse 8-10, 1040 Vienna, Austria}\\

\end{center}
\vskip .7in
\begin{abstract}

We give a detailed review of the construction of  gauge invariant Lagrangians for free and interacting
  higher spin fields using the BRST approach developed over
 the past few years. \\

\center{{\it 'Extended version of the contribution to the volume
dedicated to the 60th-birthday of Prof I.L. Buchbinder`}}

\end{abstract}

\vfill

\end{titlepage}
\tableofcontents
\setcounter{equation}0
\section{Introduction}
~Higher ~spin ~gauge ~theories ~have ~attracted  ~considerable ~interest
~during ~the ~last ~decade.
 ~Other ~than ~being ~a ~fascinating topic by
itself, Higher Spin field theory has attracted a significant
amount of attention due to its close relation with string - and M
- theory.

The study of higher spin  gauge theories is notoriously difficult
and demanding. Even for free higher spin gauge fields it is
highly nontrivial to construct Lagrangians that yield higher spin
field equations with enough gauge invariance to remove nonphysical
polarizations - ghosts - from the spectrum. Moreover, the
requirement of gauge invariance severely  restricts   the possible
gravitational backgrounds where free fields with spin greater than
two can consistently propagate.  To date, only constant
curvature backgrounds - Minkowski, de Sitter (dS) and Anti-de
Sitter (AdS) spaces - are known to support consistent propagation
of higher spin gauge fields.

Interacting higher spin gauge fields are much harder to deal with.
An important landmark was reached with the understanding of
\cite{Fradkin:1986qy} -- \cite{Vasiliev:1990en} that the AdS
background can accommodate consistent self - interactions of
massless higher spin fields. An important property of this
construction is that the coupling constants of massless higher
spin  interactions are proportional to positive powers of the AdS
radius and therefore the naive flat space--time limit cannot be
defined. This picture has two crucial features: the presence of an
infinite tower of massless higher spin fields and nonlocality.

The results of \cite{Fradkin:1986qy} -- \cite{Vasiliev:1990en}
have been obtained in the ``frame--like'' formulation of
higher spin fields, which is a highly nontrivial generalization of
the MacDowell--Mansouri--Stelle--West
\cite{MacDowell:1977jt}--\cite{Stelle:1979aj} formulation of
Anti de Sitter gravity and  the higher spin fields are encoded in
generalized vierbeins and spin connections. It seems therefore to
be of  extreme importance to understand the results of
\cite{Fradkin:1986qy} -- \cite{Vasiliev:1990en} in terms of the so
called ``metric--like'' formulation of the higher spin fields,
where the basic objects are ``customary'' tensor fields of
arbitrary rank and symmetry.

In the present review we concentrate mainly on the gauge invariant
``metric--like'' formulation of the free and interacting higher
spin fields \cite{Pashnev:1997rm}--\cite{Barnich:2005bn} (see
\cite{Fronsdal:1978rb}--\cite{Pashnev:1989gm}  for earlier work).
Particular emphasis will be made on the method of  BRST
constructions \cite{Pashnev:1997rm}--\cite{Buchbinder:2006ge},
which is analogous to the BRST approach in Open String field
Theory \cite{Neveu:1986mv}-- \cite{Gross:1986ia}. The BRST method
is based upon the principle of gauge invariance. Namely, the free
Lagrangians must possess enough gauge invariance to remove
nonphysical states-ghosts- from the spectrum, while interactions
are constructed via consistent deformations of the ``free''
abelian gauge transformations.

There is however a crucial difference from string theory. In
string theory the infinite dimensional conformal symmetry of the
two dimensional world--sheet is translated into space--time gauge
invariance, by building the corresponding BRST charge in terms of
Virasoro generators and constructing the corresponding field
theory Lagrangian. In the BRST approach for higher spin gauge
fields the only constraint we impose is that of gauge invariance,
since the corresponding world sheet-description is not known yet.
One can try, however, to find a connection with the ``high energy
limit'' \cite{Gross:1987ar}--\cite{Bonelli:2003kh}
 of string theory, at least as a formal tool
for a better understanding of interacting higher spin fields. The
studies in this direction are far from  complete, and the
best we can do here is to present a nontrivial model of
interacting higher spin fields, derived from Open String Field
Theory (OSFT).

 We will start our discussion from the
construction of various free
 Lagrangians for
massless and massive bosonic higher spin fields, which belong to
reducible and irreducible representations of an arbitrary
dimension Poincare and Anti de--Sitter groups. In the second part
of the review we turn to the interactions and discuss the general
method as well as particular examples of the interactions between
higher spin fields. In the last part of the paper we shall briefly
describe the case of half-integer higher spin fields.

Until  now several excellent reviews on the subject are available
\cite{Vasiliev:2004qz} .
 We believe that the present one along with \cite{Bekaert:2003uc} will
provide a useful completion to the already existing ones. As we have
already mentioned  this is not a comprehensive review and therefore
many important and interesting topics have not been included here
such as (apart form the ``frame--like'' formulation of higher spin
fields mentioned above \cite{Fradkin:1986qy} --
\cite{Vasiliev:1990en}, see also \cite{Sezgin:2005pv})
\begin{itemize}

\item The nonlocal formulation
\cite{Francia:2002aa}-\cite{Francia:2002pt},

\item The method of tensorial (super)spaces \cite{Bandos:1998vz}--\cite{Fedoruk:2006jm}

\item The method for construction of cubic interaction vertices in the
light --cone gauge \cite{Metsaev:2007rn}

\item Topics of AdS/CFT correspondence
\cite{Sundborg:2000wp}--\cite{Prokushkin:1999xq}

\item Fields belonging to irreducible representations of Poincare
and AdS groups
 with mixed symmetry
\cite{Aulakh:1986cb}--\cite{Alkalaev:2006rw}

\item Partial masslessness of higher spin fields
\cite{Deser:2001us}

\item Connections to matrix models, brane world scenario and other
implications \cite{Saitou:2006ca}--\cite{Benincasa:2007xk}

\end{itemize}

\section{Free Massless Fields }

\setcounter{equation}0\subsection{Reducible representations of the
Poincare group} Let us start with the simplest example of massless
reducible representations of the Poincare group. The field we
would like to describe is a tensor field of arbitrary rank $s$,
completely symmetric in its indexes or in  other words, a field
with spin $s$. Since the field is massless it has to satisfy the
Klein--Gordon equation
\begin{equation}\label{111} \Box \varphi_{\mu_1 \mu_2...
\mu_s}(x)=0.
\end{equation}
Further, in order not to have propagation of states with negative
norm-- ghosts , the higher spin field should satisfy the
transversality condition
\begin{equation}\label{222}
\partial^{\mu_1} \varphi_{\mu_1 \mu_2... \mu_s}(x)=0.
\end{equation}
Our aim is therefore to construct a free Lagrangian which gives
the mass-shell and transversality conditions as a result of its
equations of motion.

The simplest example of this construction is the Maxwell field
in arbitrary ${\cal D}$ dimensional Minkowski space--time  described by the Lagrangian
\begin{equation}
{\cal L}= A_\mu \Box A^\mu - A_{\mu}\partial^\mu \partial^\nu A_\nu
\end{equation}
which is invariant under the gauge transformations
\begin{equation}
\delta A_\mu(x)=\partial_\mu \lambda.
\end{equation}
One can further impose the Lorentz gauge condition
\begin{equation}\label{trM}
\partial^\mu A_\mu(x)=0
\end{equation}
which is consistent with on-shell gauge invariance, provided the
parameter of gauge transformations is constrained as $\Box
\Lambda(x)=0$. After imposing the Lorentz gauge condition the
Maxwell field satisfies the usual Klein--Gordon equation
\begin{equation}
\Box A_\mu(x)=0.
\end{equation}
The residual gauge transformation with the restricted parameter
$\Lambda$ and the transversality condition remove nonphysical
polarization and one is left with only physical polarizations
which satisfy massless Klein --Gordon equations.

Our aim is to generalize this construction to the case of
arbitrary spin. In order to achieve this let us introduce an
auxiliary Fock space spanned by oscillators $\alpha_\mu$ and
$\alpha^+_\mu$, which satisfy commutation relations
\begin{equation}\label{osc}
[\alpha_\mu, \alpha^+_\nu] = g_{\mu \nu}, \quad g^{\mu \nu}= diag (-1,1,...,1)
\end{equation}
and consider a state in this space
\begin{equation} \label{state}
|\Phi \rangle = \frac{1}{s!}\Phi_{\mu_1 \mu_2..
\mu_s}(x)\alpha^{+,\mu_1} \alpha^{+,\mu_2}... \alpha^{+,\mu_s}|0
\rangle_\alpha
\end{equation}
with
\begin{equation}
\alpha_\mu |0\rangle_\alpha =0.
\end{equation}
In this auxiliary Fock space differentiation and trace operations are realized via the operators
\begin{equation}\label{op1}
l_0 = p^\mu p_\mu, \quad l^+_1=  \alpha^{+}_\mu p^\mu, \quad l_1 =  \alpha_{\mu} p^\mu,
\quad p_\mu =-i\partial_\mu,
\end{equation}
\begin{equation}\label{op2}
M^+= \frac{1}{2}\alpha^{+,\mu} \alpha_{\mu}^+, \quad M= \frac{1}{2}\alpha^\mu \alpha_{\mu}.
\end{equation}
\begin{equation}
{(l_1)}^\dagger = l_1^+, \quad {(M)}^\dagger=M^+.
\end{equation}
It is straightforward to check that the action of operators
(\ref{op1})-- (\ref{op2}) on the state (\ref{state}) is translated
to the action on a symmetric tensor field $\Phi_{\mu_1, \mu_2,..
\mu_s}(x)$ as follows
\begin{equation}\label{dic1}
l_0 |\Phi \rangle \rightarrow - \Box \Phi_{\mu_1 \mu_2.. \mu_s},
\end{equation}
\begin{equation}\label{dic2}
l_1 |\Phi \rangle \rightarrow - i\partial^{\mu_1} \Phi_{\mu_1
\mu_2.. \mu_s}, \quad l_{1}^+ |\Phi \rangle \rightarrow -
i\partial_{(\mu_{s+1}} \Phi_{\mu_1 \mu_2.. \mu_s)}
\end{equation}
\begin{equation}\label{dic3}
M|\Phi \rangle \rightarrow \frac{1}{2}\Phi^{\mu_1}{}_{\mu_1
\mu_3.. \mu_s}, \quad M^+|\Phi \rangle \rightarrow g_{(\mu_1
\mu_2}\Phi_{\mu_3.. \mu_{s+2})}.
\end{equation}
In order to describe reducible massless higher spin modes it is enough to consider the
operators $l_0, l_1$, $l_{1}^+$. The reason behind this is that  was mentioned
above at the end
 the physical field should satisfy
massless Klein--Gordon and transversality conditions. These conditions
are described by the equations
\begin{equation}
l_0  |\Phi \rangle= l_1 |\Phi \rangle=0,
\end{equation}
while the operator $l_{1}^+$ should be included because it is
hermitian conjugate to $l_{1}$ and we would like to have a
hermitian Lagrangian. We can compute the algebra between $l_0,
l_1$ and $l_{1}^+$ which is
\begin{equation}\label{al1}
[l_1, l_{1}^+]= l_0, \quad [l^{\pm }_1 , l_0 ]  =0.
\end{equation}
The commutation relations with the operators $M^\pm$ are
\begin{equation}\label{al2}
[M, l_{1}^+]= l_1, \quad [l_1, M^+]= l_{1}^+
\end{equation}
The next step is to construct a nilpotent BRST charge for this
system of operators. The reason is that having obtained a nilpotent BRST charge
$Q^2=0$
it is
straightforward to construct the gauge invariant hermitian Lagrangian of the
form $\langle \Phi|Q|\Phi\rangle$ which is invariant under the
gauge transformations $\delta |\Phi \rangle =Q|\Lambda \rangle$.
Following the standard method\footnote{If Grassman even operators
$G_i$ form a Lie algebra $[G_i, G_j]= U_{ij}^k G_k$,
with $U_{ij}^k$ being structure constants
 then for each
operator $G_i$ one introduces a pair of Grassman -- odd variables
$c_i$ and $b_i$ with anti-commutation relations $\{ c_i, b_j
\}=\delta_{ij}$ and then constructs a nilpotent BRST charge $Q=
c_i G_i + \frac{1}{2}U^k_{ij}c_j c_i b_k$.} we introduce Grassman
- odd ghost variables $c_0, c_{1}^+, c_{1}$ with ghost number one
and corresponding momenta $b_0, b_{1}, b_{1}^+$ with ghost number
$-1$ with the only nonzero anti commutation relations
\begin{equation}
\{c_0, b_0 \}= \{c_1, b_{1}^+ \}= \{c_{1}^+, b_1 \}=1.
\end{equation}
For the system at hand the BRST charge has the simple form
\begin{equation} \label{BRST}
Q= c_0 l_0 + c_1 l_{1}^+ +c_{1}^+ l_1 -c_{1}^+c_1 b_0,
\end{equation}
which obviously satisfies the nilpotency property $Q^2=0$. Finally,
before obtaining an explicit form of the Lagrangian one needs to
define a ghost vacuum, which is taken conventionally as
\begin{equation}
b_0|0 \rangle _{gh.}=c_1|0 \rangle _{gh.}=b_1|0 \rangle _{gh.}=0
\end{equation}
while the other ghost variables act as creators. This choice is
clear from the form of terms linear in ghosts in the BRST charge
(\ref{BRST}). Indeed we would like the operators $l_0$ and $l_1$
to annihilate the physical states, when considering $Q|\Phi
\rangle =0$ as the equations of motion and therefore the choice of
$c_0$ and $c_{1}^+$ as creators is natural. Therefore $b_0$ and
$b_1$ are annihilators. We choose $b_{1}^+$ as creator and
therefore $c_{1}$ as annihilator because one needs one creation
operator with ghost number $-1$ to assign proper ghost numbers to
the basic field and gauge transformation parameters as we shall
see in a moment. We define a Fock vacuum in our enlarged space as
\begin{equation}
|0\rangle = |0\rangle_\alpha \otimes |0\rangle_{gh.}
\end{equation}
Then one can write a gauge invariant Lagrangian
\begin{equation}\label{L}
L= \int d c_0 \langle \Phi | Q |\Phi \rangle
\end{equation}
which leads to the equations of motion
\begin{equation}\label{EOM}
Q|\Phi \rangle=0,
\end{equation}
 and  is invariant under gauge transformations
\begin{equation}\label {GT}
\delta |\Phi \rangle= Q |\Lambda \rangle.
\end{equation}
From the previous equations it is simple to obtain the general
form of $|\Phi \rangle$ and $|\Lambda \rangle$. Indeed, the
Lagrangian must have ghost number zero. The Grassmanian
integration (i.e., differentiation) over the variable $c_0$ has
ghost number $-1$, while the BRST charge has ghost number $+1$ as
 can be seen from (\ref{BRST}). Therefore, the field $|\Phi
\rangle$ must have ghost number zero. Since $|\Phi \rangle$ has
ghost number zero and $Q$ has ghost number $+1$, then $|\Lambda
\rangle$ must have ghost number $-1$. Finally, one has
\begin{equation} \label{trexp}
|\Phi \rangle=|\varphi \rangle + c_{1}^+b_{1}^+|D\rangle + c_0
b_{1}^+|C\rangle
\end{equation}
\begin{equation}\label{lambdat}
|\Lambda \rangle = b_{1}^+|\lambda \rangle
\end{equation}
where the fields $|\varphi \rangle,|D\rangle, |C\rangle  $ so
named triplet in \cite{Francia:2002pt}, \cite{Sagnotti:2003qa},
(see also \cite{Ouvry:1986dv}, \cite{Burdik:2001hj} for earlier
work ) depend only on the oscillators $\alpha^+_\mu$ and have
ghost number zero. Putting the expansion (\ref{trexp}) into the
Lagrangian (\ref{L}), integrating over the bosonic ghost zero mode
$c_0$, according to the rules
\begin{equation}
\int dc_0 \langle 0| c_0|0 \rangle =1, \quad \int dc_0 \langle 0| |0
\rangle =0
\end{equation}
 and
performing normal ordering for the rest of the ghost variables, thus
effectively integrating them out, (for example $\langle A|b_1
c_{1}^+|B\rangle = \langle A||B\rangle$,
$\langle A|c_1b_{1}^+|B\rangle = \langle A||B\rangle$,
$\langle A|b_1 c_1 c_1^+b_{1}^+|B\rangle = -\langle A||B\rangle$
) one obtains the Lagrangian
\begin{eqnarray}\label{Ltriplet1}
{\cal L}& =&  \langle \varphi | l_0   | \varphi \rangle - \langle D
| l_0  |
D \rangle + \langle C|| C \rangle   \nonumber \\
&& - \langle \varphi | l_{1}^+ | C \rangle + \langle D | l_1  |C
\rangle - \langle C | l_1 | \varphi \rangle + \langle C| l_{1}^+ | D
\rangle
\end{eqnarray}
and equations of motion
\begin{equation}\label{trf1}
l_0|\varphi \rangle = l_{1}^+|C \rangle
\end{equation}
\begin{equation}\label{trf2}
l_0 |D \rangle = l_1|C \rangle
\end{equation}
\begin{equation}\label{trf3}
|C \rangle = l_{1}^+ |D \rangle - l_{1}|\varphi \rangle
\end{equation}
while the gauge transformation rule (\ref{GT}) gives
\begin{equation}\label{GTt1}
\delta |\varphi \rangle = l_{1}^+|\lambda \rangle, \quad \delta |D
\rangle = l_{1}|\lambda \rangle, \quad \delta |C \rangle =
l_{0}|\lambda \rangle.
\end{equation}
These equations are simple to derive. For example the gauge
transformation rules (\ref{GTt1}) can be easily obtained from the
explicit form of the BRST charge (\ref{BRST}) acting on
(\ref{lambdat}) namely
\begin{equation}
Q|\Lambda \rangle = (l_{1}^+ + c_0 b_{1}^+ l_0 + c_{1}^+b_{1}^+
l_{1})|\lambda \rangle.
\end{equation}
Comparing this with (\ref{trexp}) one gets (\ref{GTt1}).

From the equations of motion (\ref{trf1})--(\ref{trf3}) and the
gauge transformation rules (\ref{GTt1}) one can easily obtain the
$\alpha^+_{\mu}$ oscillator content of $|\varphi \rangle$,
$|C\rangle$, $|D \rangle$ and $|\lambda \rangle$. Since the
operators $l_{1}^+, l_0, l_{1}$ have $+1,0,$ and $-1$ oscillator
numbers respectively one can conclude that
\begin{equation}\label{components1}
|\varphi\rangle =\frac{1}{s!} \varphi_{\mu_1 \ldots \mu_s}(x)
\alpha^{+,\mu_1} \ldots \alpha^{+,\mu_s}\; |0\rangle
\end{equation}
\begin{equation}\label{components2}
 |D \rangle= \frac{1}{(s-2)!}\, D_{\mu_1 \ldots
\mu_{s-2}}(x) \alpha^{+,\mu_1} \ldots \alpha^{+, \mu_{s-2}} \,
c_{1}^+ \, b_{1}^+ \; |0\rangle \
\end{equation}
\begin{equation}\label{components3}
| C\rangle=  \frac{-i}{(s-1)!}\, C_{\mu_1 \ldots \mu_{s-1}}(x)
\alpha^{+,\mu_1}  \ldots \alpha^{+, \mu_{s-1}} \, b_{1}^+ \;
|0\rangle ,
\end{equation}
while the corresponding gauge transformation parameter $|\Lambda
\rangle$ is,
\begin{equation}
|\Lambda\rangle \ =\ \frac{i}{(s-1)!}\,
\Lambda_{\mu_1\mu_2...\mu_{s-1}}(x) \, \alpha^{+,\mu_1} \ldots
\alpha^{+,\mu_{s-1}} b_{1}^+ \, |0\rangle \ .
\end{equation}
One can easily rewrite the gauge transformation rules, the
Lagrangian and the equations of motion in tensorial notation. For
example to obtain the Lagrangian one has to put expressions
(\ref{components1})--(\ref{components3}) into the Lagrangian
(\ref{Ltriplet1}) and perform the normal ordering with respect to
the oscillators $\alpha^+_\mu$ and $\alpha_\mu$. The result
is
\begin{eqnarray}
{\cal L} &=& - \, \frac{1}{2}\ (\partial_\mu
\varphi_{\mu_1...\mu_s}) (\partial^\mu \varphi^{\mu_1...\mu_s})
 \ + \ s\,
(\partial^{\mu_{s}} \varphi_{\mu_1...\mu_{s-1} \mu_s}) \, C^{\mu_1...\mu_{s-1}} \nonumber \\
& + & s(s-1)\, (\partial^{\mu_{s-1}} C_{\mu_1...\mu_{s-2} \mu_{s-1}} )\, D^{\mu_1...\mu_{s-2}} \nonumber \\
&+&  \ \frac{s(s-1)}{2} \ (\partial_\mu D_{\mu_1...\mu_{s-2}})
(\partial^\mu D^{\mu_1... \mu_{s-2}}) \nonumber \\
 &-& \frac{s}{2} \,
 C^{\mu_1 \mu_2...\mu_{s-1}} C_{\mu_1 \mu_2...\mu_{s-1}}. \label{LtripletB}
\end{eqnarray}
In a similar  manner\footnote{In our notations the symmetrization
is without a factorial in the denominator, for example
 $\partial_{(\mu}A_{\nu)}= \partial_\mu A_\nu + \partial_{\nu}A_\mu.$}
\begin{eqnarray}
&&\Box \; \varphi_{\mu_1 \mu_2...\mu_s} \ = \ \partial_{(\mu_1 } C_{\mu_2 ...\mu_s)} \ , \nonumber \\
&& \partial^{\mu_s} \varphi_{\mu_1 \mu_2...\mu_{s-1} \mu_s}  -
\partial_{(\mu_{s-1}} \, D_{\mu_1 \mu_2...\mu_{s-2})} \ = \ C_{\mu_1 \mu_2...\mu_{s-1}} \ , \nonumber \\
&& \Box \; D_{\mu_1 \mu_2...\mu_{s-2}} \ = \ \partial^{\mu_{s-1}}
C_{\mu_1 \mu_2...\mu_{s-1}} \ , \label{flattriplet}
\end{eqnarray}
and
\begin{eqnarray}
&& \delta \varphi_{\mu_1 \mu_2...\mu_s} \ = \ \partial_{(\mu_1} \, \Lambda_{\mu_2...\mu_s)} \ , \nonumber \\
&& \delta  C_{\mu_1 \mu_2...\mu_{s-1}}  \ = \ \Box \; \Lambda_{\mu_1 \mu_2...\mu_{s-1}} \ , \nonumber \\
&& \delta D_{\mu_1 \mu_2...\mu_{s-2}} \ = \ \partial^{\mu_{s-1}}
\Lambda_{\mu_1 \mu_2...\mu_{s-1}} \ . \label{flattripletgauge}
\end{eqnarray}

Finally let us discuss the spectrum of the system. Obviously the
Lagrangian contains auxiliary fields alongside with  physical
ones. The field $C$ is obviously auxiliary since it can be
integrated out via its equations of motion. The field $D$ has the
wrong sign kinetic term. However it does not propagate since it is
''pure gauge". As a result one can show that the only physical
polarizations are transverse components of spin $s, s-2,...1/0$,
 depending on whether the rank of the tensor $\varphi$ is even or
odd, described by the traceless parts of $\varphi$, $\varphi'-2D
\dots$ respectively.

It is instructive to demonstrate the gauge fixing procedure with a
simple example for a field $\varphi$ of rank $2$. The case of an
arbitrary spin can be treated in a completely analogous way. For a
spin--$2$ triplet there should exist a field $\varphi_{\mu \nu}$
of rank $2$, a field $C_\mu(x)$ of rank $1$ and a field $D(x)$ of
rank zero. The triplet equations take a rather simple form in this
case \cite{Francia:2002pt}:
\begin{equation}\label{graveqn}
\Box \varphi_{\mu \nu} = \partial_\mu C_\nu + \partial_\nu C_\mu
\end{equation}
\begin{equation}\label{tr2}
 C_{\mu} = \partial_\nu \varphi_{\mu}{}^{\nu} - \partial _\mu D
\end{equation}
\begin{equation}
\Box D = \partial_\mu C^\mu.
\end{equation}
The system is invariant under the gauge transformations
\begin{equation}
\delta \varphi_{\mu \nu} = \partial_\mu \Lambda_\nu +
\partial_\nu \Lambda_\mu,  \quad \delta C_\mu = \Box \Lambda_\mu,
\quad \delta D = \partial_\mu \Lambda^\mu.
\end{equation}
Let us also introduce  a traceless field $\tilde \varphi_{\mu \nu}$
\begin{equation}
\tilde \varphi_{\mu \nu} = \varphi_{\mu \nu} - \frac{1}{{\cal D}}
g_{\mu \nu}\varphi^{'}, \quad \varphi^{'} = g^{\mu \nu}
\varphi_{\mu \nu}.
\end{equation}

In order to see the physical polarizations described by these
equations one can use the light-cone gauge fixing procedure i.e.,
eliminate $\tilde \varphi_{++},\tilde \varphi_{+i}$ and $\tilde
\varphi_{+-}$ of the field $\tilde \varphi_{\mu \nu}$ using the
gauge transformation parameter $\Lambda_{\mu}$. The other
nonphysical polarizations $\tilde \varphi_{--},\tilde
\varphi_{-i}$ as well as the field $C_\mu$ are eliminated by the
field equations. Therefore, one is left with the physical degrees
of freedom $\tilde \varphi_{ij}$ $(i,j, = 1,..., {\cal D} -2)$
which correspond to the spin $2$ field and a gauge invariant
scalar
\begin{equation}\label{scalar}
 \tilde D = \varphi^{'} -2D.
\end{equation}
The ``orthogonal''  scalar $\overline D = D- \frac{1}{{\cal D}}
\varphi^\prime$ is ``pure gauge'' and does not propagate.
Therefore we have obtained a gauge invariant Lagrangian
description of two free fields with spins $2$ and $0$.

It is an even simpler exercise to show that for a triplet with
spin $1$ one obtains the usual Maxwell Lagrangian, after
elimination of the only auxiliary field $C$. Indeed, the Lagrangian
(\ref{LtripletB}) takes the form
\begin{equation}
{\cal L} = - \, \frac{1}{2}\ (\partial_\mu
\varphi_{\nu}) (\partial^\mu \varphi^{\nu})
 \ + \ \,
(\partial^{\mu} \varphi_{\mu}) \, C
 - \frac{1}{2} \,
 C^2.
\end{equation}
and after elimination of the field $C$ via its equation of motion
$C=\partial_\mu \varphi^\mu$ (compare with (\ref{trM})) one
obtains the usual Maxwell Lagrangian.

{\bf String theory derivation}

A formal
way
\cite{Bengtsson:1986ys}
 to construct the nilpotent BRST charge
of the previous subsection
is to start with the BRST charge for the open bosonic string
\begin{equation}\label{B1}
{\cal Q} = \sum_{k, l=-\infty}^{+\infty}(C_{-k} L_k - \frac{1}{2}
(k-l):C_{-k}C_{-l}B_{k+l}:)-C_0,
\end{equation}
 perform the rescaling of  oscillator variables
\begin{equation}\label{B2}
c_k = \sqrt{2 \alpha^\prime}C_k, \qquad b_k = \frac{1}{\sqrt{2
\alpha^\prime}}B_k, \qquad c_0 =  \alpha^\prime C_0, \qquad b_0 =
\frac{1}{ \alpha^\prime}B_0,
\end{equation}
$$
\alpha^\mu_k \rightarrow  \sqrt{k} \alpha^\mu_k
$$
and then take  $\alpha^\prime \rightarrow \infty$. In this way one
obtains a BRST charge
\begin{equation}\label{hebrst}
 Q = c_0 l_0 + \tilde{Q} - b_0 {\cal M}
\end{equation}
\begin{equation}\label{B3}
\tilde{Q} = \sum_{k=1}^\infty ( c_k l^{+}_k + c_k^+ l_{k} ), \quad
{\cal M}=\sum_{k=1}^\infty c^{+}_k c_k, \quad l_0 = p^\mu p_\mu,
\qquad l_k^{+}= p^\mu \alpha_{k \mu}^+
\end{equation}
which is nilpotent in any space-time dimension. The oscillator
variables obey the usual (anti)commutation relations
\begin{equation}\label{B4}
[\alpha_\mu^k, \alpha_\nu^{l,+} ] = \delta^{kl} \eta_{\mu \nu},
\quad \{ c^{k,+}, b^l \} = \{ c^k, b^{l,+} \} = \{ c_0^k , b_0^l \}
= \delta^{kl}\,,
\end{equation}
and the vacuum in the Fock space is defined as
\begin{equation}\label{B5}
\alpha^\mu_k |0\rangle =  0, \quad
 c_k|0\rangle =  0  \quad k>0 , \qquad
b_k|0\rangle\ = \ 0 \qquad k \geq 0.
\end{equation}
Let us note that one can take the value of $k$ to be any fixed
number without affecting the nilpotency of the BRST charge
(\ref{hebrst}). Fixing the value of $k$ to be $k=1$ one obtains the description
of totally symmetric massless higher spin fields, with spins $s,
s-2,..1/0$ of the previous subsection,
  whereas for an arbitrary value of $k$ one has the so called
''generalized triplet"
\begin{equation}
\label{gentri} |\Phi  \rangle =
 \frac{c^+_{k_1}\dots c^+_{k_p} b^+_{l_1}\dots b^+_{l_p}}{{(p!)}^2}
|D^{l_1, \dots l_p}_{k_1, \dots l_p}\rangle + \frac{c_0
c^+_{k_1}\dots c^+_{k_{p-1}} b^+_{l_1}\dots b^+_{l_p}}{(p-1)! p!}
|C^{l_1, \dots l_p}_{k_1, \dots k_{p-1}}\rangle, \nonumber
\end{equation}
where the vectors $|D^{k_1, \dots k_p}_{l_1, \dots l_p}\rangle $ and
$|C^{k_1, \dots k_p}_{l_1, \dots l_p}\rangle $ are expanded only in
terms of oscillators $\alpha^{\mu +}_k$, and the first term in the
ghost expansion of (\ref{gentri}) with $p=0$ corresponds to the
state $|\varphi \rangle$ in (\ref{trexp}). One can show that
 the whole spectrum of the open bosonic string decomposes
into an infinite number of generalized triplets, each of them
 describing a finite number of fields with mixed symmetries
\cite{Sagnotti:2003qa}. In other words we can actually justify the
way the BRST charge for generalized triplets was obtained from the
BRST charge of the open bosonic string since its cohomology
classes correctly describe the degrees of freedom of massless
bosonic fields belonging to mixed symmetry representations of the
Poincare group (see e.g. \cite{Sagnotti:2003qa}). So taking the
point of view that, in the high energy limit the whole spectrum of
the bosonic string collapses to zero mass, which becomes
infinitely degenerate, one can take the BRST charge (\ref{hebrst})
as the one which correctly describes this spectrum.

\subsection{Reducible representations of the AdS group}

The description of the reducible massless representations of an
arbitrary ${\cal D} $ dimensional Anti de Sitter group follows the
same lines as for the case of reducible massless representations
of the Poincare group \cite{Bengtsson:1990un},
\cite{Sagnotti:2003qa}, \cite{Fotopoulos:2006ci}.

 Here we give some basic definitions concerning ${\cal D}$
dimensional Anti de Sitter space. More detailed treatment can be
found in \cite{Brink:2000ag} or in reviews \cite{Aharony:1999ti}-- \cite{deWit:2002vz}.

AdS space is a vacuum solution of Einstein equations with a negative
cosmological constant. Its Riemann tensor has the form
\begin{equation} R_{\mu\nu\rho\sigma} \ = \
\frac{1}{L^2} \, \left( g_{\mu\rho} \, g_{\nu\sigma} \ - \
g_{\nu\rho} \, g_{\mu\sigma} \right) \  \label{RADS},
\end{equation} where $L$ is the AdS radius, $L \rightarrow \infty$
corresponds to  the flat space--time limit and $g_{\mu \nu}$ is the
metric of $AdS_{\cal D}$. It is convenient to
represent   ${\cal D}$ dimensional AdS space with coordinates
$x^\mu$ ($\mu = 0.,.. ,{\cal D} -1$) and signature $(1,{\cal D}-1)$
as a hyperboloid in ${\cal D}+1$ dimensional flat space with
signature $(2,{\cal D}-1)$, parameterized  by coordinates $y^A$
($A = 0.,.. ,{\cal D} $). The coordinates in this ambient space
obey the condition

\begin{equation}
\eta_{AB}y^A y^B = - L^2, \quad \eta_{AB} \eta^{AB} = {\cal D}+1,
\quad \eta_{AB} =(-,+,+ ... +,-).
\end{equation}
Therefore, the isometry group is a pseudo-orthogonal group of
rotations $SO({\cal D}-1,2)$ and the AdS space itself is
isomorphic to the coset $SO({\cal D}-1,2)/SO({\cal D}-1,1)$.
 In order to simplify the
equations we set the radius of the AdS space to unity and restore it when writing down the field equations.

The AdS isometry group is noncompact and therefore its  unitary
representations are infinite dimensional. In order to build them it
is convenient to rewrite the $SO({\cal D}-1,2)$ algebra
\begin{equation}
[J_{AB}, J_{CD}]=\eta_{BC}J_{AD} -\eta_{AC}J_{BD} - \eta_{BD}J_{AC}+
\eta_{AD}J_{BC},
\end{equation}
$$
 \quad J_{AB} = - J_{BA}, \quad
{(J_{AB})}^+ = -J_{AB},
$$
in a different form. Namely, after taking the following linear
combinations
 \begin{equation} J_a^{\pm} =
(-iJ_{0a} \pm  J_{{\cal D}a}), \quad a= 1,...,{\cal D}-1
\end{equation}
 \begin{equation} H= iJ_{0{\cal D}},
\end{equation}  one obtains the commutation relations
 \bea
\left[H, J_a^{\pm}\right] = \pm J_a^{\pm} \nonumber \\
\left[J_a^-, J_b^{+}\right] = 2(H\delta_{ab} + J_{ab}) \nonumber \\
\left[J_{ab}, J_{\pm c}\right] = \delta_{bc}J_{\pm a} -
\delta_{ac}J_{\pm b}. \eea as well as
\begin{equation}
[J_{ab}, J_{cd}]=\eta_{bc}J_{ad} -\eta_{ac}J_{bd} - \eta_{bd}J_{ac}+
\eta_{ad}J_{bc}.
\end{equation}
From these commutation relations one can conclude that the AdS
isometry group has  a maximal compact subgroup
  $SO(2) \otimes
SO({\cal D}-1)$,
 spanned by generators $H$ and $J_{ab}$ respectively.
These  operators correspond to one dimensional and ${\cal D}-1$
dimensional  rotations. The operator  $H$ is the energy operator
on AdS while time on AdS is defined as the variable conjugate to
$H$. Therefore,  the time variable is  compact and energy
eigenvalues are quantized having integer values in order for a
wave function to be single valued\footnote{However, for physical
applications  one usually considers the
 covering space
of AdS, where time is uncompactified \cite{Aharony:1999ti}}. The
quadratic Casimir operator in this basis has the form
\begin{equation} \label{Casimir}
{\cal C}_2 = - \frac{1}{2}J^{AB}J_{AB}= H(H - {\cal D} +1) -
\frac{1}{2}J^{ab}J_{ab} - J_a^+J_a^-
\end{equation}
Infinite dimensional unitary representations of the AdS group are
obtained  from the ``lowest weight states" $|E_0, s\rangle $,
which is a representation of $SO(2) \otimes SO({\cal D}-1)$. The
latter therefore is
 characterized by its energy eigenvalue and a Young tableaux with
labels $s=(s_1,s_2,..,s_k)$, $k = [\frac{{\cal D} -1 }{2}])$. A
lowest weight state is annihilated by all operators $J_a^-$ \bea
J_a^-|E_0, s\rangle =0. \eea Then the other states of each
representation are obtained by successively applying operators
$J_a^+$ on the lowest weight state
\begin{equation}
J_{a_1}^+ J_{a_2}^+ ... J_{a_k}^+|E_0, s\rangle
\end{equation}
The crucial point is that representations obtained in this way do
not always have a positive norm. Therefore, when building new states
with the help of operators $J_a^+$ one has to check their norm. For
some special values of $E_0$ and $s$ the norm is equal to zero.
There is a unitarity bound on the energy $E_0$ below which the
states get negative norms and should be excluded form the physical
spectrum. The unitarity bound is saturated (norm of states becomes
zero) for states with $E_0$ and $s$ related via
\begin{equation}
E_0 = s_1 + {\cal D} - t_1 -2
\end{equation}
where $t_1$ is the number of rows of maximal length $s_1$ in the
corresponding Young tableaux.\footnote{There might be some extra
states which saturate the unitarity bound. For example in the case
of ${\cal D} =4$ there are two states for scalar massless fields
with $E_0=1$ and $E_0=2$. These states have the same quadratic
Casimir operator but correspond to different asymptotic behaviour
on the AdS boundary \cite{Fronsdal:1974ew}}. The states which
saturate the unitarity bound are identified with massless fields
on AdS space -- time. These states decouple from the original
multiplet along with their descendants since their scalar product
with the other states is zero.
 This effect is known as a `multiplet shortening``
 and it is interpreted as an enhancement of gauge symmetry.

Fields whose energy is  above the unitarity bound are massive
representations of AdS space. Both massive and massless fields on
AdS have flat space -- time counterparts i.e., one can take the
usual flat space limit to obtain massless and massive fields
propagating through Minkowski space-time. However there is one
more type of field on AdS, which have no flat space -time
analogue. These are called singletons. For example the unitarity
bound for a spinless singleton is $E_0 = \frac{1}{2}({\cal D} -3).$
Singletons do not admit a proper field theoretical description in
AdS bulk, rather they are described as boundary degrees on
freedom.

Let us turn to a field theoretical description of massless fields on
$AdS_{\cal D}$. In order to obtain  wave equations describing massless fields
with an arbitrary integer value of spin on an AdS background one has to
find a relation between the quadratic Casimir operator and the
D'Alembertian. The result for totally symmetric representations of an
AdS group i.e., when $s=(s,0,..0)$, is \cite{Metsaev:1994ys}
   \be \label{uba} (\nabla^2 - \frac{(s-2)(s+{\cal D}-3)}{L^2})
F_{A_1 A_2...A_s}(y)=0. \ee where
\begin{equation}\nabla^A =
\theta^{AB}\frac{\partial}{\partial y^B}, \quad \theta^{AB} =
\eta^{AB}+\frac{y^A y^B}{L^2}, \quad \nabla^2= \nabla^A \nabla_A.
\end{equation} A possible way to see where
the condition  (\ref{uba}) comes from, is to introduce a auxiliary
Fock space spanned by a set of oscillators \be\label{osci} [
\alpha^A , \alpha^{B+} ] \ = \ \eta^{AB} \ ,   \ee and consider a
state in this Fock space
\begin{equation}
|\Phi \rangle\ = \frac{1}{s!}F_{A_1A_2...A_s} \alpha^{A_1 +}
\alpha^{A_2+}... \alpha^{A_s + }|0 \rangle.
\end{equation}
The generators of  $SO(2,{\cal D}-1)$ can be represented as
\begin{equation} \label{JAB}
J^{AB} = L^{AB}+M^{AB},
\end{equation}
where the orbital part $L^{AB}$ and spin part $M^{AB}$ have the
form
\begin{equation} \label{LAB}
L^{AB} = y^A \nabla^B - y^B \nabla^A, \quad M^{AB} = \alpha^{A+}
\alpha^B - \alpha^{B+} \alpha^A.
\end{equation}
A field $|\Phi \rangle$ in this Fock space is required to satisfy
the mass--shell condition
\begin{equation} \label{msh}
(\nabla^2 - m^2)|\Phi \rangle\ =0,
\end{equation}
where $m^2$ is  a ``mass -- like`` parameter  to be determined,
divergencelessness condition
\begin{equation}
\alpha^A \nabla_A|\Phi \rangle\ =0,
\end{equation}
 and transversality condition
\begin{equation}
y^A \alpha_A|\Phi \rangle\ =0.
\end{equation}
The requirement of invariance of these equations under gauge
transformations
\begin{equation} \delta |\Phi \rangle\ =\alpha^{A+}
\nabla_A|\Lambda_1 \rangle\ +y^A \alpha^+_A|\Lambda_2 \rangle\
\end{equation}
leads to the mass--shell  equation (\ref{uba}). If one computes
the explicit form of the quadratic Casimir operator in terms of
realization (\ref{JAB})--(\ref{LAB}), one finds its eigenvalues  $
 <{\cal C}_2>$. Comparing equation
\begin{equation}
 ({\cal C}_2 -<{\cal C}_2>)|\Phi \rangle =0
 \end{equation}
 with (\ref{uba}) one obtains the expression for the unitarity
 bound in an alternative way.

Using the formulas given in the Appendix and the relations
\begin{equation}
\nabla^\mu \tilde \Phi_{\mu \mu_1 ... \mu_s} =\frac{\partial
y^{A_1}}{\partial x^{\mu_1}}... \frac{\partial y^{A_s}}{\partial
x^{\mu_s}}(\nabla^A + ({\cal D} + s)y^A) \tilde \Phi_{A A_1...
A_s}, \label{tran1}
\end{equation}
\begin{equation}
\nabla_{(\mu_1} \tilde \Phi_{ \mu_2 ... \mu_s)} =\frac{\partial
y^{A_1}}{\partial x^{\mu_1}}... \frac{\partial y^{A_s}}{\partial
x^{\mu_s}}(\partial_{(A_1}\Phi_{A_2 ... A_s)}
 + (s-1)y^A \eta_{(A_1 A_2} \tilde \Phi_{A A_3 ... A_s)}),
\label{tran2}
\end{equation}
  \begin{eqnarray} \nonumber
\Box \tilde \Phi_{\mu_1... \mu_s} &=& \frac{\partial
y^{A_1}}{\partial x^{\mu_1}}... \frac{\partial y^{A_s}}{\partial
x^{\mu_s}}( \nabla^2 \Phi_{A_1...A_s}   - s\Phi_{A_1... A_s} +2s
\partial_{(A_1} y^A\Phi_{A A_2...A_s)}\\
&& + s(s-1)y^A y^B\eta_{(A_1 A_2}\Phi_{A B A_3...A_s)} ),
\end{eqnarray}
\be \Box \tilde \Phi_{A_1 ... A_s} = \nabla^2 \tilde \Phi_{A_1
...A_s} \ee one can relate equations in the $x$ and $y$ spaces.
For example, the massless Klein --Gordon equation in the ``$y$-
space'' (\ref{uba}) when written in the ``$x$- space'' is
 \be
\label{ubax} (\Box - \frac{(s-2)(s+{\cal D}-3)-s}{L^2}) F_{\mu_1
\mu_2...\mu_s}(x)=0. \ee where $\Box$ is the D'Alembertian of the
AdS space--time. Below we shall work in the ``$x$- space''
\cite{Sagnotti:2003qa}, the corresponding equations in the ``$y$--
space'' can be found in \cite{Bengtsson:1990un},
\cite{Fotopoulos:2006ci}.

In order to describe a triplet on $\cal D$ dimensional Anti
de--Sitter space it is convenient to introduce the set of
oscillators $(\alpha^{\mu +},\alpha^\mu)$, which can be obtained
from the ones in (\ref{osci}) and
 the AdS vielbein
 \begin{equation} [ \alpha_\mu , \alpha^{ +}_\nu ] \ = \ g_{\mu\nu},  \quad \alpha^\mu =e^\mu_a
 \alpha^a
\ , \end{equation}
 where $g_{\mu \nu}$ denotes the AdS metric.
The ordinary partial derivative is replaced by the operator
\begin{equation} p_\mu \ = \  -\; i \, \left(
\partial_\mu + \omega_{\mu}^{ab} \, \alpha_{\; a}^+\,
 \alpha_{ \; b} \right) \ . \label{covcurve}
\end{equation}
Acting with $p_\mu$ on a state in Fock space
\begin{equation}\label{PHI} |\Phi\rangle \ =\
\frac{1}{(s)!}\, \varphi_{\mu_1\mu_2...\mu_{s}}(x) \, \alpha^{\mu_1
+} \ldots \alpha^{\mu_s +}  \, |0\rangle \ ,
\end{equation}
produces the proper covariant derivative
\begin{equation}\label{defp}
p_\mu|\Phi\rangle \ =-\frac{i}{(s)!}\alpha^{\mu_1 +} \ldots
\alpha^{\mu_s +} \nabla_\mu \, \varphi_{\mu_1\mu_2...\mu_{s}}(x)
|0\rangle,
\end{equation}
\begin{equation}
\langle \Phi |p_\mu =\langle 0|\alpha^{\mu_1 } \ldots \alpha^{\mu_s
} \nabla_\mu \, \varphi_{\mu_1\mu_2...\mu_{s}}(x)\frac{i}{(s)!},
\end{equation}
where in (\ref{covcurve})   $\omega_\mu^{ab}$ denotes the spin --
connection on AdS and $\nabla_\mu$ is the AdS covariant derivative.
These operators satisfy commutation relations
  \begin{equation} [p_\mu,p_\nu] \
=  \frac{1}{L^2}(\alpha_{\; \mu}^+ \, \alpha_{\; \nu} \, -\,
\alpha_{\; \nu}^+ \, \alpha_{\; \mu} \ ) , \end{equation} due to the
expression (\ref{RADS}) for the Riemann  tensor.

Further, let us introduce the following operators

 D'Alembertian
operator
\begin{equation} \label{lapl} l_0 \ = \
g^{\mu \nu} \, ( p_\mu p_\nu \, + \, i\; \Gamma^\lambda_{\mu \nu}\;
p_\lambda )\ = \ p^a \; p_a \, - \ i \, \omega_a{}^{ab} \, p_b
\end{equation} which  acts on Fock-space states as the proper
D'Alembertian operator
\begin{equation} l_0|\Phi\rangle \
=-\frac{1}{(s)!} \alpha^{\mu_1 +} \ldots \alpha^{\mu_s +}
 \Box \, \varphi_{\mu_1\mu_2...\mu_{s}}(x) \,
  \, |0\rangle \
\end{equation}

Divergence operator
\begin{equation}
l_1 =  \alpha^\mu p_\mu \label{divergence}
\end{equation}
which acts on a state in the Fock space as divergence
\begin{equation} l_1|\Phi\rangle \
=-\frac{i}{(s-1)!}\alpha^{\mu_2 +} \ldots \alpha^{\mu_s +}
 \nabla_{\mu_1} \,
\varphi^{\mu_1}{}_{\mu_2\mu_3...\mu_{s}}(x) \,  \, |0\rangle \
\end{equation}

 Symmetrized exterior derivative operator,
\begin{equation}
l_1^+ =  \alpha^{\mu +} p_\mu \label{exteriord}
\end{equation}

\begin{equation} l_1^+|\Phi\rangle \
=-\frac{i}{(s+1)!} \alpha^{\mu +}\alpha^{\mu_1 +} \ldots
\alpha^{\mu_s +} \nabla_{\mu} \,
\varphi_{\mu_1\mu_2\mu_3...\mu_{s}}(x) \,   \, |0\rangle \
\end{equation}
which is the hermitian conjugate to the operator $l_1$ with respect to the
scalar product
\begin{equation}
\int d^{\cal D} x \sqrt{-g} \langle \Phi_1||\Phi_2 \rangle .
\end{equation}
 It is straightforward to obtain the commutation relations of the
  algebra generated by these operators. The commutator between
  $l_{1}^+$ and
  $l_1$ becomes
\be [ l_1 , l_{1}^+ ] \ = \ \tilde{l_0} \ , \ee where the modified
D'Alembertian is \be \tilde l_0 \ = \ l_0 \ - \ \frac{1}{L^2}\,
\left( -{\cal D} \, + \, \frac{{\cal D}^2}{4} \, + \, 4 \,
M^\dagger \; M \, - \, N^2 \, + \, 2\, N \right). \  \ee Here \be N
\ = \ \alpha^{+,\mu} \alpha_{\mu} \ + \ \frac{\cal D}{2} \
\label{Nop} \ee counts the number of indices of the Fock-space
fields, up to the space-time dimension ${\cal D}$, while \be M \ =
\ \frac{1}{2}\; \alpha^\mu \alpha_{\mu} \ \label{Mop} \ee
takes traces of the Fock-space fields.

The emergence of these new operators enlarges the algebra, that now
includes the additional commutators
\begin{eqnarray}
&&
[ M^\dagger \; , \; l_1] \ = \ -\, l_{1}^+ \ , \nonumber \\
&&[\tilde l_0 \; , \; l_1] \ = \ \frac{2}{L^2}\, l_1 \, - \,
\frac{4}{L^2}N \, l_1 \, + \,
      \frac{8}{L^2}\, l_{1}^+ \; M \ , \nonumber\\
&&[ N \; , \; l_1 ] \ = \ -\; l_1  \ , \label{tripletalg1}
\end{eqnarray}
and their hermitian conjugates, together with
\begin{equation}
[N \; , \; M] \ = \ - \; 2\; M \ , \quad
 [M^\dagger \; , \; N] \ =
\ - \, 2\; M^\dagger \ , \quad [M^\dagger \; , \; M] \ = \ -\, N \ ,
\label{tripletalg2}
\end{equation}
that define an $SO(1,2)$ subalgebra.

Note that (\ref{tripletalg1}) and (\ref{tripletalg2}) actually
define a \emph{non-linear algebra}, and therefore the associated
BRST charge should be naively constructed with the recipe of
\cite{ThierryMieg:1987um} (see also  \cite{Buchbinder:2001bs},
\cite{Buchbinder:2007au}--\cite{Isaev:2008vi}). As in
\cite{Buchbinder:2001bs}, however, this would introduce a larger
set of ghosts and corresponding fields, going beyond the triplet
structure. The latter case which will lead to the description of
irreducible higher spin modes will be discussed in the next
subsections. Now
  in the spirit of the flat limit for the triplet, \emph{let
us retain only the $(l^{\pm}_{1},l_0)$ constraints, treating
(\ref{tripletalg1}) as an ordinary algebra where $M$, $M^\dagger$
and $N$ play the role of ``structure constants''}. In other words,
the first step is to construct the BRST charge using the standard
formula adopted for the case of the constraints forming a Lie
algebra i.e. ignore the fact that we have structure functions
rather than structure constants. The second step is to compute the
square of the BRST operator but we now take into account the fact
that we have structure functions rather than constants. And the
third step is to add compensating terms to restore the nilpotency
of the BRST charge.
 Remarkably, this
is possible and guarantees the Lagrangian nature of corresponding
field equations.
With this proviso, one can write the
\emph{identically nilpotent} BRST charge
\begin{eqnarray} \label{brst} \nonumber
Q& =&c_0\; \left(\tilde{l}_0 \, - \, \frac{4}{L^2}  N \, + \,
\frac{6}{L^2} \right)
  \, + \, c_1\; l_{1}^+\  \, + \, c_{1}^+ \; l_1
    \, - \, c_{1}^+ \; c_1 \; b_0 \\ \nonumber
   &-& \, \frac{6}{L^2} \; c_0\; c_{1}^+ \; b_1
      \, - \, \frac{6}{L^2} \; c_0 \; b_{1}^+ \; c_1
        \, + \, \frac{4}{L^2} \; c_0 \; c_{1}^+ \; b_1 \; N
      \, + \, \frac{4}{L^2} \;c _0 \; b_{1}^+ \;  c_1 \; N \\
&-& \,\frac{8}{L^2} \; c_0 \; c_{1}^+ \; b_{1}^+ \; M \, + \,
\frac{8}{L^2} \; c_0 \; c_1 \; b_1 \; M^\dagger \, + \,
\frac{12}{L^2} \; c_0 \; c_{1}^+ \; b_{1}^+ \; c_1 \; b_1 \ .
\end{eqnarray}
The nilpotency of $Q$ ensures the consistency of the construction,
and as usual determines a BRST invariant Lagrangian of the form
(\ref{L}), and thus a Lagrangian set of equations as in (\ref{EOM}).
In component notation
\begin{eqnarray} \nonumber
{\cal L} &=& - \, \frac{1}{2}\ (\nabla_\mu \varphi)^2 \ + \ s\,
\nabla \cdot \varphi \, C \ + \ s(s-1)\, \nabla \cdot C \, D \
 + \ \frac{s(s-1)}{2} \, (\nabla_\mu D)^2 \ - \ \frac{s}{2} \,
C^2 \\ \nonumber &+& \ \frac{s(s-1)}{2L^2}\, {(\varphi^{'})}^2 \ - \
\frac{s(s-1)(s-2)(s-3)}{2L^2} \, {(D^{'})}^2
  \ - \ \frac{4s(s-1)}{L^2} \, D \, \varphi^{'} \\
&-& \ \frac{1}{2L^2} \, \left[ (s-2)({\cal D}+s-3) \, - \, s
\right] {\varphi}^2 \ + \ \frac{s(s-1)}{2L^2} \, \left[ s({\cal
D}+s-2)+6 \right]\,  D^2  . \label{LtripletBADS}
\end{eqnarray}
Here we introduced short -- hand notation.
 The
symbol $\nabla \cdot$ means divergence, while $\nabla$  is
symmetrized action of $\nabla_\mu$ on a tensor. The symbol
${}^\prime$ means that we take the trace of a field.
Multiplication of a tensor by the metric $g$ implies the
symmetrized multiplication, i.e. if $A$ is a vector $A_\mu$ we
have $g A = g_{\mu \nu} A_\rho +g_{\mu \rho} A_\nu+g_{\nu \rho}
A_\mu $. The Lagrangian leads to the corresponding field equations
\begin{eqnarray}
&& \Box \; \varphi \ = \ \nabla C + \frac{1}{L^2} \, \left\{ 8 \, g
\, D \ - \ 2 \, g \, \varphi^{'} \ + \ \left[ (2-s)(3-{\cal D}-s) -s \right]
\,
\varphi \right\} \nonumber \ , \\
&& C = \nabla \cdot \varphi - \nabla D \nonumber \ , \\
&& \Box \; D \ = \ \nabla \cdot C \ + \frac{1}{L^2} \left\{
[s({\cal D}+s-2) +6] D - 4 \varphi^{'} - 2 g D^{'} \right\} \
\label{AdStriplet}
\end{eqnarray}
which are invariant under the gauge transformations
\begin{eqnarray}
&& \delta \varphi \ = \ \nabla \, \Lambda \ , \nonumber \\
&& \delta  C  \ = \ \Box \;
\Lambda + \frac{(s-1)(3-s-{\cal
D})}{L^2}\
  \Lambda
+ \frac{2}{L^2} \,  g \; \Lambda^\prime
 \nonumber \\
&& \delta D \ = \ \nabla \cdot
\Lambda \ . \label{adstripletgauge}
\end{eqnarray}
Let us note that the invariance of the Lagrangian
(\ref{LtripletBADS}) and of the equations of motion
(\ref{AdStriplet}) under the gauge transformations
(\ref{adstripletgauge}) can be checked directly using the action
on a vector $V_\mu$ of the AdS covariant derivatives commutator
\begin{equation}\label{comut}
[ \nabla_\mu , \nabla_\nu ] \, V_\rho \ = \ \frac{1}{L^2} \left(
g_{\nu\rho} \, V_\mu \ - \
 g_{\mu\rho} \, V_\nu \right) \ .
\end{equation}
The gauge fixing procedure can be carried out as in the case of
 flat space--time. Therefore, one obtains massless spin $s$
field which is contained in the traceless part of $\varphi$,
$\varphi^\prime - 2D$, $\varphi^{\prime \prime }- 4 D^\prime
\dots$ e.t.c. describing massless fields of spin $s-2,
s-4,...,1/0$, as  was for the case of flat space--time. This
completes our discussion of the irreducible higher spin modes on
AdS.

\subsection{Irreducible representations of the Poincare group}
As we saw in the previous subsections, the inclusion of the
operators $l_0$, $l_1$ and $l_{1}^+$ leads to the description of
reducible massless higher spin fields which are subject to
mass-shell and transversality conditions (\ref{111})--(\ref{222}).
In order to describe irreducible higher spin mode one has to add
the trace operator $M= \frac{1}{2} \alpha^\mu \alpha_{\mu}$ (and its
hermitian conjugate $M^+ = \frac{1}{2}\alpha^{\mu + }\alpha_{\mu}^+$
which is needed for the hermiticity of the corresponding BRST
charge) to the initial set of operators $l_0$, $l_1$ and $l_{-1}$.
The resulting algebra now has the form (\ref{al1}), (\ref{al2}),
(\ref{tripletalg2}). The crucial point here is the appearance of
an extra operator $N=\alpha^{\mu +} \alpha_{\mu}+ \frac{{\cal
D}}{2}$ in the right hand side of the commutator $[M, M^+]$. This
operator is strictly positive because its eigenvalues are
$s+\frac{{\cal D}}{2}$ and therefore it can not annihilate any
state in the  Fock space whether it is physical or not. On the other hand,
an inclusion of $N$ into the total set of operators seems to be
unavoidable as long as we include operators $M^\pm$ into the total
set of constraints\footnote{One can check that the conversion
procedure of \cite{Faddeev:1986pc}, or use of the Dirac brackets
does not simplify the problem at hand. } since otherwise one will
deal with operators whose algebra does not close. This in turn
will cause problems with the construction of the nilpotent BRST
charge.

A way out is that although the operator $N$ can not annihilate a
state in a Fock space, we can modify the operator with some new
operator $h$ whose eigenvalues can cancel those of $N$ i.e.,
consider $\tilde N= N+h$ in such a way that the operator $\tilde
N$ will be able to annihilate a state in the Fock space. However
if one tries naively to modify the operator $N$ by simply adding
 an operator $h$ to it the algebra (\ref{tripletalg2}) will not
close.

To solve the problem
\cite{Pashnev:1998ti}, \cite{Sagnotti:2003qa}
 we therefore use the method of BRST reduction which in fact originated from the dimensional
reduction procedure described in the next section. To understand
it better let us rephrase the problem. If we declare the operator
$N$ to be a constraint then we will have to introduce a real
(since $N$ is hermitian) ghost and therefore we shall have a term
of the form $c_N N$ in the BRST charge. Our task is to eliminate
the $c_N$ rather than the operator $N$, since eliminating the
ghost $c_N$ effectively prevents $N$ from imposing a condition on the
Fock space. Combining this with the previous discussion we have to
choose the parameter $h$ to eliminate the dependence on $c_N$ of
the BRST charge, but the BRST charge should still be nilpotent.

As a first step to implement the procedure outlined above, we
introduce extra oscillators with the commutation relations
 \be [d,d^\dagger]=-1 \
\label{dosc}. \ee The states in the enlarged Fock space are
expanded, as usual, in (anti)ghost modes, and each of the
resulting terms have the form,
\begin{equation}\label{irredvarphi}
|\varphi^s\rangle \ = \ \sum_k \; |\varphi^s_{k}\rangle \ \equiv \
\sum_k \; \varphi^{k}_{i;\mu_1 \mu_2...\mu_{s-2k}}\;
\alpha^{\mu_1,+} \; \alpha^{\mu_2,+} \; ...\; \alpha
^{\mu_{s-2k},+} \; (d^{\; \dagger})^{k} \; |0\rangle \ ,
\end{equation}
where $s$ is the rank of the $k=0$ component tensor in the
expansion. With these variables one can build  new ``auxiliary''
representations of the algebra $SO(2,1)$ formed by operators $N,
M^{\pm}$, which have the form
\begin{equation} \label{auxiliary}
 M_{(aux)} \ = \, \sqrt{h +  d^\dagger d } \ d  \ ,
\quad  M^\dagger_{(aux)} \ = \ d^\dagger\, \sqrt{h + d^\dagger d } \
, \quad  N_{(aux)} \ = \ - \, 2 \, d^\dagger d \ - \ h.
\end{equation}
 Further, define
new operators, \begin{equation} \tilde{M}_\pm \ = \ M^\pm \ + \
M^\pm_{(aux)} \ , \qquad \tilde N \ = \ N \ + \ N_{(aux)} \ ,
\end{equation}
that realize again the SO(2,1) algebra  (\ref{tripletalg2}). The
nilpotent BRST charge for the resulting system is then formally
constructed, treating all operators under consideration using the standard recipe,
\begin{eqnarray}
\tilde Q&=&c_0 l_0 \, + \, c_1 \; l_{1}^+ \, + \, c_M \; \tilde
M^\dagger \, + \, c_{1}^+ \; l_1 \, + \, c_{M}^+ \; \tilde M +\, c_N
\tilde N
 \nonumber \\
&-&  \, c_{1}^+\; c_1 \; b_0 \, + \, c_{1}^+ \; b_{1}^+\; c_M \, -\,
  c_{M}^+ \; c_1 \; b_1 \nonumber \\
&+& c_{N} (2c_{M}^+b_{M} +  2b_{M}^+c_{M} + c_{1}^+b_{1} +
b_{1}^+c_{1} -3) - c_{M}^+c_{M}b_{N} \ .
\end{eqnarray}
The final step is the elimination of the term proportional to
$c_{N}$ while maintaining the nilpotency of the BRST charge. This
can be done performing the unitary transformation on the BRST charge
\begin{equation}
Q_1 \ = \ e^{-i \; \pi \; x_h} \, \tilde Q \, e^{i \; \pi \; x_h}
\ , \label{unitary}
\end{equation}
where $x_h$ is the phase-space coordinate conjugate to $h$, so
that \be [x_h, h] \ = \ i \ , \ee and
\begin{equation}
\pi \ = \ N \ - \ 2 \, d^\dagger d \ + \ 2\, c_{M}^+\, b_{M} \ +\
2\, b_{M}^+\, c_{M} \ + \ c_{1}^+\, b_{1} \ + \ b_{1}^+\, c_{1} \ -\
3 \
\end{equation}
is essentially a number operator. Note that this transformation
removes all terms depending on $c_{N}$ from the BRST charge, while
obviously preserving its nilpotency. Finally, the term containing
$b_{N}$ can be also dropped without any effect on the nilpotency.
This can be checked by direct computations but one can see it easily by the following
argument.
 Let us write $ Q_1$ in the form
\begin{equation}
 Q_1 = Q - c_{M}^+ c_{M}b_N
\end{equation}
and take its square.
By construction $Q_1$ is nilpotent, i.e., $Q_1^2=0$. On the right hand  side
 the term $c_{M}^+ c_{M}b_N$ is nilpotent too, so we are left with terms
$Q^2$ and $\{Q,c_{M}^+ c_{M}b_N \} $.
Since $Q$  does not contain the $c_N$ ghost
the anticomutator will be proportional to $b_N$.
The operator $Q^2$ does not contain neither $c_N$ and $b_N$.
Therefore each of these terms are separately nilpotent.

Finally, the BRST charge
for this system takes the form
\begin{equation}
Q \ = \ Q_1 \ + \ Q_2 \ ,
\end{equation}
with
\begin{equation}
\{ \; Q_1 \; , \; Q_2 \; \} \ = \ 0 \ , \qquad  \quad Q_1^2 \ = \
- \; Q_2^2 \ ,
\end{equation}
where
\begin{eqnarray} \nonumber
Q_1&=&c_0 l_0 \, + \, c_1 \; l_{1}^+ \, + \, c_M \; M^\dagger \, +
\, c_{1}^+ \; l_1 \, + \, c_{M}^+ \; M
 \nonumber \\
&-&  \, c_{1}^+\; c_1 \; b_0 \, + \, c_{1}^+ \; b_{1}^+\; c_M \, -\,
  c_{M}^+ \; c_1 \; b_1 \nonumber ,
\end{eqnarray}
and
\begin{eqnarray} \nonumber
\!\!\!Q_2&=&c_{M}^+ \; \sqrt{\; -\; 1\; +\; N\; -\; d^\dagger \, d
\; + \; 2\; b_{M}^+\; c_M \; + \; 2\; c_{M}^+\; b_{M} \; + \;
b_{1}^+\; c_1 \; + \; c_{1}^+\; b_1 } \ \, d
\nonumber \\
&&\!\!\!\!\!\!\!\!\!\!\!\!\!\!\! +\, d^\dagger \sqrt{\; -\; 1 \; +
\; N\; - \; d^\dagger \; d \; + \; 2\; b_{M}^+\; c_M \; + \; 2\;
c_{M}^+\; b_{M} \; + \; b_{1}^+\; c_1 \; + \; c_{1}^+\; b_1 }\ \, c_M
\, \nonumber.
\end{eqnarray}
Again, this determines a BRST invariant Lagrangian of the type
(\ref{L}) and now the most general expansions of the state vector
$|\Phi\rangle$ and of the gauge parameter $|\Lambda \rangle$ in
ghost variables are
\begin{eqnarray}
|\Phi\rangle&=&|\varphi_1\rangle \, + \, c_{1}^+ \; b_{1}^+ \;
|\varphi_2\rangle \, + \,  c_{M}^+ \; b_{M}^+ \; |\varphi_3\rangle
\, + \, c_{1}^+ \; b_{M}^+ \; |\varphi_4\rangle \nonumber
 \\
&+& \, c_{M}^+ \; b_{1}^+ \; |\varphi_5\rangle \, + \, c_{1}^+ \;
c_{M}^+\; b_{1}^+ \; b_{M}^+ \; |\varphi_6\rangle \, +\,
c_0 \; b_{1}^+ \; |C_1\rangle \nonumber\\
&+& c_0 \; b_{M}^+ \; |C_2\rangle \, + \, c_0 \; c_{1}^+ \; b_{1}^+
\; b_{M}^+ \; |C_3\rangle \, + \, c_0 \; c_{M}^+ \; b_{1}^+ \;
b_{M}^+\; |C_4\rangle \ ,
\end{eqnarray}
and
\begin{eqnarray}
|\Lambda\rangle&=&b_{1}^+ \, |\Lambda_1\rangle \, + \, b_{M}^+ \;
|\Lambda_2 \rangle \, + \,  c_{1}^+ \; b_{1}^+ \; b_{M}^+ \,
|\Lambda_3\rangle \, + \, c_{M}^+ \; b_{1}^+ \; b_{M}^+ \,
|\Lambda_4\rangle \nonumber
 \\
&+& \, c_{0} \; b_{1}^+ \; b_{M}^+ \, |\Lambda_5\rangle  \ ,
\end{eqnarray} where $|\varphi_i\rangle$ and
$|C_i\rangle$ have ghost number zero and depend \emph{only} on the
bosonic creation operators $\alpha^{\mu +}$ and $d^\dagger$ as in
(\ref{irredvarphi}). Let us also note that both the Lagrangian and
the gauge transformations are not affected by redefinitions of the
gauge parameters of the type \be \delta \; |\Lambda\rangle \ =\ Q
\; |\omega\rangle \ , \ee and in particular with \be
|\omega\rangle\ = \ b_{1}^+ \; b_{M}^+ \; |\omega_1\rangle \ . \ee
As a result, one of the gauge parameters, $|\Lambda_5\rangle$, is
inessential and can be ignored.

With this proviso, the resulting Lagrangian in the bosonic
Fock-space notation is
\begin{eqnarray} \label{ireduciblel}
{\cal L} &=& \, - \, \langle C_1|C_1\rangle \,- \, \langle
C_2|\varphi_2\rangle \, + \, \langle C_3|\varphi_3\rangle\, + \,
\langle C_4|C_4\rangle\, - \,
  \langle \varphi_2|C_2\rangle \, + \, \langle \varphi_3|C_3\rangle  \nonumber \\
&-& \langle C_1| M^\dagger|\varphi_4\rangle \, - \, \langle C_1|
l_{1}^+|\varphi_2\rangle \, + \,
  \langle C_1| l_1|\varphi_1\rangle \, - \, \langle C_2| M^\dagger|\varphi_3\rangle \,
  - \, \langle C_2| l_{1}^+ |\varphi_5\rangle  \nonumber \\
\nonumber &+&\langle C_2| M|\varphi_1\rangle \, - \, \langle C_3|
M^\dagger|\varphi_6\rangle \, + \, \langle C_3|
l_1|\varphi_5\rangle \, - \, \langle C_3| M|\varphi_2\rangle \,
+\, \langle C_4| l_{1}^+|\varphi_6\rangle \\
\nonumber &+&\langle C_4| l_1|\varphi_3\rangle \, - \, \langle
C_4| M|\varphi_4\rangle \, + \, \langle \varphi_1|
M^\dagger|C_2\rangle \, + \, \langle \varphi_1| l_{1}^+|C_1\rangle
\, -\,
\langle \varphi_1| l_0|\varphi_1\rangle \\
\nonumber &-&\langle \varphi_2| M^\dagger|C_3\rangle \, + \,
\langle \varphi_2| l_0|\varphi_2\rangle \, - \, \langle \varphi_2|
l_1|C_1\rangle \, + \, \langle \varphi_3| l_{1}^+|C_4\rangle \, +
\,
\langle \varphi_3| l_0|\varphi_3\rangle   \\
\nonumber &-&\langle \varphi_3| M|C_2\rangle \, - \, \langle
\varphi_4| M^\dagger|C_4\rangle \, + \, \langle \varphi_4|
l_0|\varphi_5\rangle \, - \, \langle \varphi_4| M|C_1\rangle \,
+\,
\langle \varphi_5| l_{1}^+|C_3\rangle  \\
\nonumber &+&\langle \varphi_5| l_0|\varphi_4\rangle \, - \,
\langle \varphi_5| l_1|C_2\rangle \,- \, \langle \varphi_6|
l_0|\varphi_6\rangle \, + \, \langle \varphi_6| l_1|C_4\rangle \,
- \,
\langle \varphi_6| M|C_3\rangle  \\
\nonumber &-&\langle C_1| d^\dagger X_{1}|\varphi_4\rangle \, - \,
\langle C_2| d^\dagger X_{2}|\varphi_3\rangle \, + \, \langle C_2|
X_{0}\; d|\varphi_1 \rangle
\, - \, \langle C_3| d^\dagger X_{4}|\varphi_6\rangle \\
\nonumber &-&  \langle C_3| X_{2}\; d|\varphi_2\rangle \, -\,
\langle C_4| X_{3}\; d|\varphi_4\rangle \, + \, \langle \varphi_1|
d^\dagger X_{0}|C_2\rangle \, - \, \langle \varphi_2| d^\dagger X_{2}|C_3\rangle \\
\label{CL} &-&\langle \varphi_3| X_{2}\; d|C_2\rangle \, - \,
\langle \varphi_4| d^\dagger X_{3}|C_4\rangle \, - \, \langle
\varphi_4| X_{1}\; d|C_1\rangle\, - \, \langle \varphi_6| X_{4}\;
d|C_3\rangle \ ,
\end{eqnarray}
where
\begin{equation}
X_{r}\ = \ \sqrt{-1 + N  - d^\dagger d + r} \ .
\end{equation}
From this Lagrangian one can derive equations of motion
\begin{eqnarray}                     \label{equation1}
&&   -M^+ |C_2\rangle - l_{1}^+ |C_1\rangle + l_0 |\varphi_1\rangle
- b^+ X_{0} |C_2\rangle =0,\\
&&   |C_2\rangle + M^+ |C_3\rangle - l_0 |\varphi_2\rangle
+ l_1 |C_1\rangle + d^+ X_{2} |C_3\rangle =0,\nonumber\\
&&   - |C_3\rangle - l_{1}^+ |C_4\rangle - l_0 |\varphi_3\rangle
+ M |c_2\rangle + X_{2} d |C_2\rangle =0,\nonumber\\
&&    M^+ |C_4\rangle - l_0 |\varphi_5\rangle + M |C_1\rangle
+ d^+ X_{3} |C_4\rangle +  X_{1} d |C_1\rangle =0,\nonumber\\
&&   -l_{1}^+ |C_3\rangle - l_0 |\varphi_4\rangle + l_1 |C_2\rangle =0,
\nonumber\\
&&    l_0 |\varphi_6\rangle - l_1 |C_4\rangle + M |C_3\rangle
+ X_{4} d |C_3\rangle =0,\nonumber\\
&&    |C_1\rangle + M^+ |\varphi_4\rangle + l_{1}^+ |\varphi_2\rangle
- l_1 |\varphi_1\rangle + d^+ X_{1} |\varphi_4\rangle =0,\nonumber\\
&&    M^+ |\varphi_3\rangle + l_{1}^+ |\varphi_5\rangle - M |\varphi_1\rangle
+ d^+ X_{2} |\varphi_3\rangle - X_{0} d |\varphi_1\rangle + |\varphi_2\rangle =0,
\nonumber\\
&&    M^+ |\varphi_6\rangle - l_1 |\varphi_5\rangle + M |\varphi_2\rangle
+ d^+ X_{4} |\varphi_6\rangle + X_{2} d |\varphi_2\rangle - |\varphi_3\rangle =0,
\nonumber\\
\label{equation10}
&& -|C_4\rangle - l_{1}^+ |\varphi_6\rangle - l_1 |\varphi_3\rangle
+ M |\varphi_4\rangle + X_{3} d |\varphi_4\rangle =0.\nonumber
\end{eqnarray}
Both the Lagrangian (\ref{ireduciblel}) and gauge transformations
(\ref{equation10}) are invariant under gauge transformations
\begin{eqnarray}\label{gauge1}
\delta |\varphi_1\rangle& =&
l_{1}^+ |\Lambda_1\rangle +
M^+ |\Lambda_2\rangle +
d^+ X_0 |\Lambda_2\rangle ,\\
\delta |\varphi_2\rangle &=&
 |\Lambda_2\rangle +
l_1 |\Lambda_1\rangle +
M^+ |\Lambda_3\rangle +
d^+ X_{2} |\Lambda_3\rangle ,\nonumber\\
\delta |\varphi_3\rangle &=&
- |\Lambda_3\rangle +
M |\Lambda_2\rangle -
l_{1}^+ |\Lambda_4\rangle +
X_{2} d |\Lambda_2\rangle  ,\nonumber\\
\delta |\varphi_4\rangle &=&
  l_1 |\Lambda_2\rangle
- l_{1}^+ |\Lambda_3\rangle,\nonumber\\
\delta |\varphi_5\rangle &=&
 M^+ |\Lambda_4\rangle
+ M |\Lambda_1\rangle
+ d^+ X_{3} |\Lambda_4\rangle
+ X_{1} d |\Lambda_1\rangle  ,\nonumber\\
\delta |\varphi_6\rangle &=&
- M |\Lambda_3\rangle
- X_{4} d |\Lambda_3\rangle
+ l_1 |\Lambda_4\rangle ,\nonumber\\
\delta |C_1\rangle &=&
  l_0 |\Lambda_1\rangle ,\nonumber\\
\delta |C_2\rangle &=&
 l_0 |\Lambda_2\rangle,\nonumber\\
\delta |C_3\rangle &=&
 l_0 |\Lambda_3\rangle,\nonumber\\
\label{gauge10}
\delta |C_4\rangle &=&
 l_0 |\Lambda_4\rangle \nonumber
\end{eqnarray}
From the field equations and the gauge transformations one can
unambiguously read the oscillator content of the vectors
$|\varphi_i \rangle$, $|C_i \rangle$ and $| \Lambda_i \rangle$. In
order to describe a spin-$s$ field, let us fix the number of
oscillators $\alpha^{\mu +}$ in the zeroth-order term of the
expansion of $| \varphi_1 \rangle$ in the oscillator $d^\dagger$,
that we shall denote by $\varphi_1^0$, to be equal to $s$. This is
actually the field $\varphi$ of the previous subsections, while
all other terms describe auxiliary  fields. The zeroth-order
components in the $d^\dagger$ oscillators for the other fields
have thus the following $\alpha^{\mu +}$ content, here summarised
in terms of the resulting total spin, displayed within brackets:
$\varphi_2^0 \; [s-2]$ , $\varphi_3^0\; [s-4]$ , $\varphi_4^0\;
[s-3]$ ,
 $\varphi_5^0\; [s-3]$ ,  $\varphi_6^0\; [s-6]$ ,  $C_1^0\; [s-1]$ ,
 $C_2^0\; [s-2]$ ,
 $C_3^0\; [s-4]$ ,  $C_4^0\; [s-5]$ ,  $\Lambda_1^0\; [s-1]$ ,
 $\Lambda_2^0\; [s-2]$ ,
 $\Lambda_3^0\; [s-4]$ ,  $\Lambda_4^0\; [s-5]$.
Moreover, the field equations and the gauge transformations show
that each power of the $d^\dagger$ oscillator reduces the number
of $\alpha^{\mu +}$ oscillators by two units, so that, for
instance, the $\varphi_1^k$ component field has $s-2k$ oscillators
of this type, and thus spin $(s-2k)$. Therefore, as anticipated,
this off-shell formulation of a spin-$s$ field requires finite
number of auxiliary fields and gauge transformation parameters,
although their total number grows linearly with $s$.

Combining the gauge transformations with the field equations, it
is possible to choose a gauge where all fields aside from
$\varphi_1^0, \varphi_2^0, \varphi_5^0$ and $C_1^0$ are
eliminated, so that one is left with a reduced set of equations
invariant under an unconstrained gauge symmetry with parameter
$\Lambda_1^0$. To this end, one first gauges away all fields
$C_i^k$ but $C_1^0$, and the residual gauge transformations are
restricted by the conditions \be l_0 \, \Lambda_1^k \ = \ 0 \quad
(k \neq 0)\quad  {\rm and}\quad  l_0 \Lambda_i^k \ = \ 0 \quad
(i=2,3,4) \quad {\rm and} \quad k \geq 0 \ . \ee The parameters
$\Lambda_1^k$ $(k \neq 0)$ and $\Lambda_4^k$ gauge away
$\varphi_5^k$ $(k \neq 0)$, while the parameters $\Lambda_3^k$
gauge away $\varphi_6^k$. The conclusion is that one is finally
left with gauge transformation parameters restricted by the
additional condition \be( M \ + \ X_{4} \, d ) \, |\Lambda_3
\rangle \ = \ 0 \ , \ee and with the help of parameters
$|\Lambda_2 \rangle$ and $|\Lambda_3 \rangle$ one can also gauge
away
 $\varphi_1^k$, $\varphi_2^k$ $(k \neq 0)$ and $\varphi_3^k$, while
$\varphi_4^k$ vanishes as a result of the field equations.

If one further eliminates the field $\varphi_5^0$ with the help of
the gauge transformation parameter $\Lambda_1^0$, the system of
equations of motion \p{equation1}, apart from the dynamical
equations for fields $|\varphi^0_{1}\rangle, \;$ and $
|\varphi^0_{2}\rangle$, becomes
\begin{equation}      \label{L2A1}
M |C_{1}^0\rangle=0,
\end{equation}
\begin{equation}                 \label{A1}
|C^0_{1}\rangle + l_{1}^+|\varphi^0_{2}\rangle- l_1|\varphi^0_{1}\rangle =0,
\end{equation}
\begin{equation}\label{S2}
M |\varphi^0_{1}\rangle - |\varphi^0_{2}\rangle = 0,
\end{equation}
\begin{equation}         \label{L2S2}
M |\varphi^0_{2}\rangle=0,
\end{equation}
with residual gauge invariance
\begin{equation}
\delta |\varphi^0_{1}\rangle =l_{1}^+ |\Lambda^0_1\rangle,\;
\delta |\varphi^0_{2}\rangle =l_1 |\Lambda^0_1\rangle,\;
\delta |C^0_{1}\rangle = l_0 |\Lambda^0_1\rangle ,
\end{equation}
where the parameter $|\Lambda^0_1\rangle$ is restricted by the
condition
\begin{equation}
M |\Lambda^0_1\rangle=0.
\end{equation}

Using the equations \p{A1} and \p{S2} one can express
$|\varphi^0_{2}\rangle$ and $|C^0_{1}\rangle$ through
$|\varphi^0_{1}\rangle$
 and
insert them into \p{CL}. The Lagrangian now depends only on the
field $|\varphi^0_{1}\rangle$ and takes the following form
\begin{equation}            \label{LF}
{\cal L}=\langle \varphi^0_{1}| l_0 - l_{1}^+l_{1} - l_{1}^+l_{1}^+ M
- M^+  l_1 l_1 -2 l_0 M^+ M - M^+  l_{1}^+ l_1 M.
|\varphi^0_{1}\rangle
\end{equation}
The field $|\varphi^0_{1}\rangle $, as a consequence of the equations
\p{S2} and \p{L2S2}, is restricted by the condition
\begin{equation} \label{cond}
 M^2 |\varphi^0_{1}\rangle =0,
\end{equation}
making \p{L2A1}  become an identity. After taking the expansion
\begin{equation}
|\varphi^0_{1}\rangle = \varphi_{\mu_1 \mu_2 ...\mu_s}(x)
\alpha^{\mu_1 +} \alpha^{\mu_2 +} ...\alpha^{\mu_s +}|0\rangle
\end{equation}
~we ~find ~that ~the ~Lagrangian \p{LF} ~in ~terms ~of ~the
~fields $ \varphi^0_{1,\mu_1\mu_2...\mu_s}(x) \equiv
\varphi_{\mu_1\mu_2...\mu_s}(x) $ ~coincides ~with ~the ~one
~given  ~by ~Fronsdal \cite{Fronsdal:1978rb}
\begin{eqnarray} \label{actionfr}
{\cal L}&=&  \varphi^{\mu_1 \mu_2 \cdots \mu_s}(x)
\Box
\varphi_{\mu_1 \mu_2 \cdots \mu_s}(x)
 - \frac{s(s-1)}{2}
\varphi_{\mu_1}{}^{\mu_1 \mu_3 \cdots \mu_s}(x)
\Box
\varphi^{\nu_1}{}_{\nu_1 \mu_3 \cdots \mu_s}(x) \\ \nonumber
&&-s \varphi^{\mu_1 \mu_2 \cdots \mu_s}(x) \partial_{\mu_1} \partial_{\nu_1}
\varphi^{\nu_1}{}_{\mu_2 \cdots \mu_s}(x)
+s(s-1) \varphi_{\mu_1}{}^{\mu_1 \mu_3  \cdots \mu_s}(x) \partial_{\nu_1} \partial_{\nu_2}
\varphi^{ \nu_1 \nu_2}{}_{\mu_3 \cdots \mu_s}(x) \\ \nonumber
&&-\frac{s(s-1)(s-2)}{4}
\varphi_{\mu_1}{}^{\mu_1 \mu_3 \mu_4 \cdots \mu_s}(x) \partial_{\mu_3} \partial_{\nu_1}
\varphi^{\nu_1 \nu_2}{}_{\nu_2 \mu_4 \cdots \mu_s}(x)
\end{eqnarray}
 As a
consequence of the condition \p{cond} the field $ \varphi_{\mu_1
\mu_2...\mu_n}(x) $ has a vanishing second trace $
\varphi^{\mu\nu}{}_{\mu\nu\mu_5 \mu_6,...\mu_n}(x) $ and the
Lagrangian is invariant under the gauge transformation
\begin{equation}
\delta\varphi_{\mu_1 \mu_2 ...\mu_n}(x) =\partial_{(
\mu_1}\Lambda_{\mu_2\mu_3...\mu_n ) }(x)
\end{equation}
with constrained parameter $\Lambda^0_{1, \mu_1 ...\mu_{s-1}}
\equiv \Lambda_{ \mu_1... \mu_{s-1}},$
 $\Lambda^{\mu}{}_{\mu\mu_3...\mu_{s-1}}(x)=0$.

To summarize, we have derived the Lagrangian for a single  field
with an arbitrary integer spin $s$ without any off -shell
constraint either on the field or on the parameter of gauge
transformations. This Lagrangian contains however a finite set of
auxiliary fields, and the number of these fields depends on the
value of the spin under  consideration. The Fronsdal Lagrangian
can be obtained from the one we are considering after the
particular choice of the gauge. The unconstrained Lagrangian
description for the
 fermionic higher spin fields, which is analogous to the one described above, has been given in
\cite{Buchbinder:2004gp}.

Let us note that here we use the ``hermitian'' auxiliary representations
as in \cite{Pashnev:1998ti}, \cite{Sagnotti:2003qa}, when the operators $M_{(aux)}$ and $M^+_{(aux)}$
are hermitian conjugate to each other. Alternatively, one can use ``non- hermitian''
auxiliary representations as in \cite{Buchbinder:2001bs}, \cite{Buchbinder:2004gp}--
\cite{Buchbinder:2005cf}, \cite{Burdik:2000ee}.
 The hermiticity of the Lagrangian is maintained by introducing
an extra kernel operator in the definition of the scalar product
in the Fock space.

{\bf Example $s=3$.}

To illustrate the procedure let us consider in detail the simple
example of the irreducible massless higher spin field with spin
$s=3$. Let us first determine the field content. As
explained after equation (\ref{equation1}) one has
\begin{equation}
|\varphi_1 \rangle = (\frac{1}{3!}\varphi^0_{1, \mu_1 \mu_2 \mu_3}(x)\alpha^{\mu_1 +}
\alpha^{\mu_2 +} \alpha^{\mu_3 +} + \varphi_{1, \mu_1}^1 (x) \alpha^{\mu_1 +} d^+)|0\rangle
\end{equation}
\begin{equation}
|\varphi_2 \rangle = \varphi_{2, \mu_1}^0 (x) \alpha^{\mu_1 +}  |0\rangle,
\quad  |\varphi_4 \rangle = -i \varphi_{4}^0 (x)  |0\rangle, \quad
|\varphi_5 \rangle = -i \varphi_{5}^0 (x)  |0\rangle
\end{equation}

\begin{equation}
|C_1 \rangle = (\frac{-i}{2!} C^0_{1, \mu_1 \mu_2}(x)\alpha^{\mu_1 +}
\alpha^{\mu_2 +}  -i C_1^1 (x) d^+)|0\rangle
\end{equation}
\begin{equation}
|C_2 \rangle= - C_2^1 (x) |0\rangle
\end{equation}
the other fields being zero.

The gauge transformation parameters have the form
\begin{equation}
|\Lambda_1 \rangle  = (\frac{i}{2!} \Lambda^0_{1, \mu_1 \mu_2}(x)\alpha^{\mu_1 +}
\alpha^{\mu_2 +}  + i\Lambda_1^1 (x) d^+)|0\rangle, \quad |\Lambda_2 \rangle = \Lambda_{2,\mu_1}^0(x)
\alpha^{\mu_1 +} |0 \rangle
\end{equation}
with all other gauge parameters being zero. The gauge
transformation rules for these fields can be written down from
(\ref{gauge1})
\begin{equation}\label{g3f}
\delta \varphi_{1, \mu \nu \rho}^0 = \partial_{( \mu}\Lambda^0_{1,
\nu \rho)} + g_{(\mu \nu }\Lambda^0_{2 \rho)}, \quad \delta
\varphi^1_{1, \mu}= \partial_\mu \Lambda_1^1 + \sqrt{\frac{{\cal
D}}{2}} \Lambda_{2, \mu}^0,
\end{equation}
\begin{equation}
\delta \varphi^0_{2, \mu}= \partial^\nu \Lambda^0_{1, \mu \nu}  +\Lambda^0_{2, \mu},
\quad \delta \varphi^0_4= \partial^\mu \Lambda^0_{2, \mu}, \quad \delta \varphi^0_5=  \sqrt{\frac{{\cal D}}{2}}
\Lambda_1^1 - \frac{1}{2}\Lambda^{0,\mu}_{1,\mu}.
\end{equation}
\begin{equation} \label{g3l}
\delta C^0_{1, \mu \nu} = \Box \Lambda_{1,\mu \nu}^0, \quad \delta C^1_{1} = \Box \Lambda_{1}^1, \quad
\delta C^0_{2, \mu } = \Box \Lambda_{2,\mu }^0
\end{equation}
From (\ref{equation1}) one can write down the equations of motion
\begin{equation}\label{s3f}
\Box \varphi^0_{1,\mu \nu \rho}=\partial_{(\mu}C^0_{1, \nu \rho)} + g_{(\mu \nu}  C^0_{2, \rho)}
\end{equation}
\begin{equation}
\Box \varphi^1_{1,\mu }=\partial_{\mu}C^1_{1} +  \sqrt{\frac{{\cal D}}{2}} C^0_{2, \mu },
\quad
\Box \varphi^0_{2,\mu }=\partial^{\nu}C^0_{1, \mu \nu} + C^0_{2, \mu }
\end{equation}
\begin{equation}
\Box \varphi^0_{5}=\sqrt{\frac{{\cal D}}{2}} C^1_{1} -\frac{1}{2} C^{0,\mu}_{1, \mu },
\quad
\Box \varphi^0_{4}=\partial^\mu C^0_{2, \mu}
\end{equation}
\begin{equation}
\partial^\rho \varphi^0_{1, \mu \nu \rho}- C^0_{1, \mu \nu}- \partial_{(\mu}\varphi^0_{2, \nu)}
-g_{\mu \nu} \varphi^0_4
=0
\end{equation}
\begin{equation}
\partial^\mu \varphi^1_{1, \mu} -C_1^1-
\sqrt{\frac{{\cal D}}{2}}
 \varphi^0_4
=0
\end{equation}
\begin{equation} \label{s3l}
\varphi^0_{2, \mu}- \partial_\mu \varphi^0_5 - \frac{1}{2}\varphi^{0, \nu}_{1, \mu \nu} +
 \sqrt{\frac{{\cal D}}{2}}\varphi^1_{1, \mu}=0.
\end{equation}
The field equations (\ref{s3f})--(\ref{s3l}) are Lagrangian
equations and can be obtained from the Lagrangian in
(\ref{ireduciblel}). One can show combining the field equations
and gauge transformations (\ref{g3f})-- (\ref{g3l})
 that after complete gauge fixing the only propagating components
are physical components of the tensor $\varphi^0_{1, \mu \nu \rho}$ i.e., physical components of
the spin 3 field.

{\bf Compensator equations}
Another way to describe irreducible higher spin fields is to take triplet equations
 (\ref{flattriplet}) and add manually the extra condition
\begin{equation}
\varphi^\prime- 2D = \partial \alpha,
\end{equation}
where the field  $\alpha$ which has rank $s-3$ is called the
compensator field \cite{Francia:2002aa}, \cite{Sagnotti:2003qa}. In order to maintain
gauge invariance the transformation law for the compensator has to
be
\begin{equation} \label{extra}
\delta \alpha = \Lambda^\prime.
\end{equation}
If one gauges away the compensator via the trace of the parameter
$\Lambda$ one obtains the description of a single irreducible
higher spin field as one can check using (\ref{flattriplet}) and
(\ref{extra}). One can check that these fields only (the triplet
fields and the compensator) are not enough  for the Lagrangian
description of the system. On the other hand one can identify the
field $-2\varphi_5^0$ with the compensator and thus the Lagrangian
derived in this chapter is precisely the Lagrangian for the
compensator field. After elimination of the fields $C$ and $D$ the
compensator equations can be written in the form
\begin{eqnarray}
&& \Box \varphi - \partial \partial \cdot \varphi + \partial^2 \varphi^\prime \
 = \ 3\, \partial^{\; 3} \, \alpha \ , \nonumber \\
&& \varphi^{''} \ = \ 4\,  \partial \cdot \alpha \ + \ \partial \,
\alpha^{'} \ , \label{flatbosecompens}
\end{eqnarray}
and are invariant under the \emph{unconstrained} gauge
transformations
\begin{eqnarray}
&& \delta \; \varphi \ = \ \partial \; \Lambda \ , \\
&& \delta \; \alpha \ = \ \Lambda^{'} \ .
\end{eqnarray}
Let us note that a Lagrangian for the compensator fields can be obtained in terms of a smaller
(minimal)
number auxiliary fields \cite{Francia:2005bu}.
 This Lagrangian (see \cite{Francia:2007qt} for a generalization to the case of an $AdS$ space)
contains higher derivatives, however the number of derivatives can
be  reduced to the ordinary one at the expence of introducing two
more auxiliary fields \cite{Francia:2007ee}.

One can write compensator equations for the fermionic higher spin field as well
and also deform both fermionic and bosonic compensator equations to the AdS space \cite{Sagnotti:2003qa},
thus describing irreducible massless integer or half integer spin modes.

\subsection{Irreducible representations of the AdS group}
The description of irreducible massless representations of the $AdS_{\cal D}$ group
can be performed  in a similar way \cite{Buchbinder:2001bs}.
 All we have to do is to combine the results of subsections 2.2 and 2.3.
Namely, start with the nonlinear algebra (\ref{tripletalg1})
--(\ref{tripletalg2}), but now introduce the ghost variables
$(c_M, b_{M}^+)$ and $(c_{M}^+, b_M)$ for the operators $M$ and
$M^+$ as well. Then construct the nilpotent BRST charge again
formally including the operator $N$ along with its corresponding
ghost-antighost pair $(c_N, b_N)$.

This procedure leads to
the family of ``bare" nilpotent
 BRST charges, which  turn out to depend on
three free parameters $k_1,k_2,k_3,$ i.e, the expressions of terms
which contain higher degrees in the ghost variables turn out not
to be uniquely defined. The possible solution of the problem (not
necessarily the most general one) has the form
\begin{equation}\label{family}
  \tilde Q^1=\tilde Q^1_0+k_1 \tilde Q^1_{k_1}+
   k_2 \tilde Q^1_{k_2}+k_3 \tilde Q^1_{k_3},
\end{equation}
where
\begin{eqnarray}
\tilde Q^1_0 &=&  c_0 (\tilde l_0 + \frac{6}{L^2})
+ c_1l_{1}^+ +
    c_MM^+ + c_{1}^+l_1 + c_{M}^+M - c_N(N -3) \nonumber \\
        &&  +\frac{2}{L^2} c_0c_{1}^+b_1
        -\frac{8}{L^2}
     c_0c_{M}^+b_M
      + \frac{2}{L^2} c_0b_{1}^+c_1 \nonumber \\
    &&-\frac{8}{L^2} c_0b_{M}^+c_M -
    c_{1}^+c_1b_0
    + c_{1}^+c_Nb_1 + c_{1}^+b_{1}^+c_M
    -\frac{4}{L^2} c_{1}^+b_Nc_1 \nonumber \\
    &&- c_{M}^+c_1b_1
    +2c_{M}^+c_Nb_M - c_{M}^+b_Nc_M +
    b_{1}^+c_Nc_1 + 2b_{M}^+c_Nc_M \nonumber \\
&& -\frac{8}{L^2} c_0c_1b_1M^+
    -\frac{4}{L^2} c_0c_{1}^+b_1N
     + \frac{8}{L^2}
     c_0c_{1}^+b_{1}^+M \nonumber \\
&&-\frac{4}{L^2} c_0b_{1}^+c_1N_0
-\frac{12}{L^2}
c_0c_{1}^+b_{1}^+c_1b_1
\\
\tilde Q^1_{k_1} &=&  -2c_0(N - 3)  - 6 c_0c_{M}^+b_M
- 6c_0b_{M}^+c_M
- 3c_{1}^+b_Nc_1
     - 2c_0c_1b_1M^+ \nonumber \\
&&- c_0c_{1}^+b_1N
    -2
     c_0c_{1}^+b_Ml_{1}^+ + 2
     c_0c_{1}^+b_{1}^+M
      - c_0c_{1}^+b_Nl_1  \nonumber \\
     &&-
      c_0b_{1}^+c_1N
    -2
     c_0b_{M}^+c_1l_1 \nonumber \\
    && - c_0b_Nc_1l_{1}^+
 - 6
     c_0c_{1}^+b_{1}^+c_1b_1,
\\
\tilde Q^1_{k_2}&=&-c_0(N -3)
- c_0c_{1}^+b_1
     -2c_0c_{M}^+b_M \nonumber \\
      && - c_0b_{1}^+c_1
     - 2c_0b_{M}^+c_M
-  c_{1}^+b_Nc_1
\\
\tilde Q^1_{k_3}&=&c_0b_{M}^+c_M +c_0c_{M}^+b_M + c_0c_1b_1M^+
+ c_0c_{1}^+b_Ml_{1}^+
-c_0c_{1}^+b_{1}^+M
+c_0b_{M}^+c_1l_1 \nonumber \\
&&  +
    c_0c_{1}^+c_{M}^+b_1b_M +
    2 c_0c_{1}^+b_{1}^+c_1b_1
+c_0c_{1}^+b_{1}^+b_Nc_M -
    c_0c_{1}^+b_{M}^+c_Mb_1 \nonumber \\
     &&-
    c_0c_{M}^+b_{1}^+c_1b_M -
    c_0c_{M}^+b_Nc_1b_1 +
    c_0  b_{1}^+b_{M}^+c_1c_M.
    \end{eqnarray}

All these operators are nilpotent and mutually anti-commuting so that their
sum is  nilpotent as well.
Let us note  that the particular choice of  parameters
$ k_1=0, k_2=0, k_3=8$
leads to the BRST charge constructed in \cite{Buchbinder:2001bs}, which includes
terms up to the fifth order in ghosts.

Further, the
 procedure goes on in complete analogy with the case of the
flat space--time background. Namely, we first build  auxiliary
representations for $M^\pm$ and $N$, then define the new operators
as the sum of old and auxiliary ones
 and finally after the transformation
(\ref{unitary}) we construct the
 Lagrangian (\ref{L}). Though the BRST charge
 is not unique, one can show along the lines of
\cite{Buchbinder:2001bs}
 that all these BRST charges, after making a partial gauge fixing
in the Lagrangian \p{LF}, lead to a unique final form for the
Lagrangian, which contains only one double traceless physical
field $|\varphi \rangle$ ($M^2|\varphi \rangle=0$)
\begin{eqnarray} \label{ALF}
\cal L&=&\langle \varphi|
 \tilde l_0 - l_{1}^+l_1 - 2 M^+  \tilde l_0  M
 +M^+  l_1  l_1
     +l_{1}^+ l_{1}^+  M   - M^+ l_{1}^+ l_1 M \nonumber \\
 &&- \frac{1}{L^2}(6  - 4    N + 10    M^+ M      -
    4  M^+  N  M)
|\varphi \rangle
\end{eqnarray}
or equivalently
\begin{eqnarray} \label{action}
S&=& \int d^Dx \sqrt{-g} (\varphi^{\mu_1 \mu_2 \cdots \mu_s}(x)
(\nabla^2 - \frac{1}{L^2}(s^2 + s{\cal D} - 6s - 2{\cal D} +6))
\varphi_{\mu_1 \mu_2 \cdots \mu_s}(x) \nonumber \\
&& - \frac{s(s-1)}{2}
\varphi_{\mu_1}{}^{\mu_1 \mu_3 \cdots \mu_s}(x)
(\nabla^2 - \frac{1}{L^2}(s^2 +s{\cal D} - 4s - {\cal D} +1))
\varphi^{\nu_1}{}_{\nu_1 \mu_3 \cdots \mu_s}(x)  \nonumber \\
&&-s \varphi^{\mu_1 \mu_2 \cdots \mu_s}(x) \nabla_{\mu_1} \nabla_{\nu_1}
\varphi^{\nu_1}{}_{\mu_2 \cdots \mu_s}(x) \nonumber \\
&&+s(s-1) \varphi_{\mu_1}{}^{\mu_1 \mu_3  \cdots \mu_s}(x) \nabla_{\nu_1} \nabla_{\nu_2}
\varphi^{ \nu_1 \nu_2}{}_{\mu_3 \cdots \mu_s}(x) \nonumber \\
&&-\frac{s(s-1)(s-2)}{4}
\varphi_{\mu_1}{}^{\mu_1 \mu_3 \mu_4 \cdots \mu_s}(x) \nabla_{\mu_3} \nabla_{\nu_1}
\varphi^{\nu_1 \nu_2}{}_{\nu_2 \mu_4 \cdots \mu_s}(x))
\end{eqnarray}
which is invariant under the gauge transformations \be \delta
\varphi_{\mu_1 \mu_2 \cdots  \mu_s}(x) = \nabla_{ ( \mu_s}
\Lambda_{\mu_1 \cdots \mu_{s-1} )}(x) \ee with totally symmetric
and traceless $\Lambda^{\mu_1}{}_{\mu_1 \mu_3 \cdots ...
\mu_{s-1}}(x)=0$ parameter.

\setcounter{equation}0\section{Free Massive Fields }

 \subsection{Reducible representations of the Poincare group}
The description of massive  reducible higher spin modes can be carried out in
a  similar way to that of reducible massless higher spin modes \cite{Burdik:2000kj}.
 Namely, we
need to consider the following set of operators
\begin{equation} \nonumber
l_0 ={p_{\mu}}^2+m^2,
\end{equation}
\begin{equation} \nonumber
l_1 =\alpha^{\mu}p_{\mu}.\;\;\; l_{1}^+=\alpha^{\mu +}p_\mu.
\end{equation}
~However, ~there ~is  ~a ~complication ~since ~due ~to ~the ~nonzero ~value ~of
~the ~mass ~parameter\footnote{One can make the mass parameter $m$ to
depend on spin thus describing a ``Regge trajectory''
\cite{Pashnev:1997rm}}
 the operators $l_{1}, l_{1}^+$ and $l_0$ no longer
form a closed algebra and therefore BRST construction is more
complicated.

Let us therefore consider the massless case (where, as we know, this problem does not exist)
 in one dimension
higher and define the usual set of operators $L_0, L_1, L_{1}^+$
which satisfy the algebra (\ref{al1}) i.e., consider in ${\cal
D}+1$ dimensions the following set of operators
\begin{equation} \nonumber
L_0=p_{\mu}^2+p_D^2= l_0 + p_D^2,  \;\;\;\;\mu=0,1,...,D-1,
\end{equation}
\begin{equation} \nonumber
L_1= \alpha^\mu p_\mu+\alpha^D p_D = l_1 + \alpha^Dp_D,
\end{equation}
\begin{equation} \nonumber
L_{1}^+= \alpha^{\mu +} p_\mu+\alpha^{D +} p_D = l_{1}^+ +
\alpha^{D +}p_D.
\end{equation}
The BRST charge for this system is given by (\ref{BRST})
\begin{equation} \label{BRSTm}
Q= c_0 L_0 + c_1 L_{1}^++c_{1}^+ L_1 -c_{1}^+c_1 b_0.
\end{equation}
In order to describe the massive fields we fix the following $x_D$
dependence of the Fock space vector $| \Phi \rangle $ in (\ref{L})
\begin{equation}               \label{sub}
|{\Phi}\rangle = U|\Phi^{\prime}\rangle =
 e^{ ix_D m}|\Phi^{\prime}\rangle.
\end{equation}
The result of the substitution of (\ref{sub}) in the expression
(\ref{L}) is
\begin{equation} \label{L12}
{\cal L}=\int d c_0 \langle {\Phi}^{\prime} |\tilde{Q}|
{\Phi}^{\prime} \rangle.
\end{equation}
The new BRST - charge $\tilde{Q}$  is nilpotent due to unitarity
of the transformation $\tilde{Q}=U^{-1} Q U$. Our choice of $x_D$-
dependence in the exponent in (\ref{sub}) leads to the presence in
the BRST charge of the correct operator $l_0+m^2$ for the massive
case
\begin{equation}
\tilde Q = c_0 (l_0 +m^2) + c_1(l_{1}^++ m \alpha^{D+}) +
c_{1}^+(l_1 + m \alpha^D)- c_{1}^+c_1 b_0.
\end{equation}

 Let us note that the unitarity transformation (\ref{unitary})
has the same form as the transformation we are considering now.
  It is exactly in the heart of the approach and makes it possible
to construct the nilpotent BRST charge in the presence of
operators which do not form a closed algebra (second class
constraints).

The expansion of the field $|\Phi \rangle$ and of the parameter of
gauge transformations $|\Lambda \rangle$ in terms of ghost variables
is again (\ref{trexp}) and (\ref{lambdat}) but now one deals with a
massive triplet. After the substitution of equation (\ref{sub}) into
the Lagrangian (\ref{L12}) and integration over the ghost variables
one gets
\begin{eqnarray}\label{MM1}
{\cal L}& =&  \langle \varphi | l_0 + m^2   | \varphi \rangle -
\langle D | l_0 + m^2  | D \rangle + \langle C|| C \rangle
 - \langle \varphi | l_{1}^+ + m \alpha^{D +} | C \rangle \nonumber
 \\
 &&+ \langle
D | l_1 + m \alpha^D|C \rangle - \langle C | l_1 + m \alpha^D |
\varphi \rangle + \langle C| l_{1}^+ + m \alpha^{D +} | D \rangle.
\end{eqnarray}
The  equations of motion for the massive triplet are:
\begin{equation}\label{trf1m}
(l_0 +m^2)|\varphi \rangle = (l_{1}^++ m \alpha^{D +})|C \rangle
\end{equation}
\begin{equation}\label{trf2m}
(l_0 +m^2) |D \rangle = (l_1 + m \alpha^D)|C \rangle
\end{equation}
\begin{equation}\label{trf3m}
|C \rangle = (l_{1}^+ + m \alpha^{D +}) |D \rangle - (l_{1} +m
\alpha^D)|\varphi \rangle
\end{equation}
while the gauge transformation rule (\ref{GT}) gives
\begin{equation}\label{GTt1m}
\delta |\varphi \rangle = (l_{1}^++m \alpha^{D +})|\lambda \rangle,
\quad \delta |D \rangle = (l_{1} + m \alpha^D)|\lambda \rangle,
\quad \delta |C \rangle = (l_{0} +m^2)|\lambda \rangle.
\end{equation}
Then, using the gauge transformations, one can
show \cite{Hussain:1988uk}, \cite{Pashnev:1997rm},
 that
the fields $|C \rangle$ and $|D \rangle$ as well as $\alpha^{D+}$
dependence in $|\varphi \rangle$ can be gauged away. Finally, one
obtains conditions
\begin{equation}
(l_0  +m^2)|\varphi\rangle = l_1|\varphi\rangle = 0
\end{equation}
 as the result of the equations of motion.
In the simplest example of a spin one massive triplet which describes
a massive vector field one has the expansion
\begin{equation}
|\varphi \rangle = A_\mu(x) \alpha^{\mu +}|0 \rangle + iA_D(x)
\alpha^{D +}|0 \rangle, \quad |C \rangle = -i C(x)|0 \rangle
\end{equation}
and
\begin{equation}
|\Lambda \rangle = i \lambda |0 \rangle.
\end{equation}
Therefore, for gauge transformations we get
\begin{equation}
\delta A_\mu = \partial_ \mu \lambda, \quad \delta A_D = m \lambda,
\quad \delta C = (\Box - m^2 ) \lambda.
\end{equation}
After the elimination of field $C$ via its equations of motion one
obtains the Lagrangian which contains the physical field $A_\mu$
and the Stueckelberg field $A_D$. After gauging away the latter
field one obtains the usual  Lagrangian for a massive spin $1$
field ( see also \cite{Bianchi:2005ze} for a discussion in the
context of AdS/CFT correspondence).

The  Lagrangian  (\ref{MM1})  describes a chain of
massive states with mass equal to $m$.
 Due to the absence of the zero trace  constraint, each state $|
\varphi \rangle$ describes a chain of massive states with spins
$s.\; s-2,\; s-4\; ...1/0$ as  was in the case of massless
triplet. This is to be expected, since we are describing the
Kaluza-Klein reduction on a circle $S^1$ of a massless triplet,
which as we know describes a chain of massless states, in ${\cal
D}+ 1$ dimensions.

\subsection{Irreducible representations of the Poincare group}
A Lagrangian for irreducible massive representations  of the
Poincare group was given in \cite{Singh:1974qz}.
 In this
description the Lagrangian contains a massive spin $s$ physical
field and auxiliary fields with spins $s-1,s-2,..,1,0$ all of them having zero traces.
Un ``unconstrained'' BRST formulation
for massive irreducible higher spin fields is given in \cite{Buchbinder:2005ua}.
 Apart from
these descriptions there is an alternative one \cite{Pashnev:1989gm}
which contains the physical field with spin $s$ and only three
auxiliary fields with spin $s-1,s-2,s-3$. This approach is based
again on the method of dimensional reduction.

Since for this computation ``the mostly minus'' signature is
slightly more convenient we shall use this signature in this
subsection. The operators are again $l_0 = p^\mu p_\mu$, $l_{ 1}=
\alpha^\mu p_\mu$, $M = \frac{1}{2} \alpha^\mu\alpha_{\mu }$, and
their hermitian conjugates but $N= - \alpha^{\mu + } \alpha_{\mu}
+ \frac{\cal D}{2}$, $[\alpha_\mu, \alpha_\nu^+]= -g_{\mu \nu}$,
with $g_{\mu \nu}= (1,-1,...-1.)$.

Let us start with Fronsdal
Lagrangian in ${\cal D}+1$ dimensions
\begin{eqnarray} \label{LFD}
{\cal L} &=& \langle \varphi_{{\cal D}+1}|
  { {L}_0} + { L}_{1}^+{L}_1 - 2 {\overline M}^+  \tilde { L}_0  {\overline
 M} \\ \nonumber
&& +{\overline M}^+  { L}_1  { L}_1
     +{ L}_{1}^+ { L}_{1}^+ {\overline M}   + {\overline M}^+ { L}_{1}^+ { L}_1 {\overline M}
|\varphi_{{\cal D}+1} \rangle.
\end{eqnarray}
where
\begin{equation}
{\overline M}= M - \frac{1}{2} \alpha_D \alpha_{ D}
\end{equation} The basic field satisfies once more the condition
\begin{equation}\label{DTD}
{\overline M}^2|\varphi \rangle = 0,
\end{equation}
 the gauge transformation rule is
\begin{equation} \label{GTD}
\delta |\varphi \rangle = { L}_{1}^+|\Lambda \rangle,
\end{equation}
 and the parameter of the
gauge transformations is traceless
\begin{equation}\label{TD}
{\overline M} |\Lambda \rangle =0.
\end{equation}
  The next step is to perform the dimensional
reduction procedure in a  similar way  to the previous subsection.
Namely we take $|\varphi_{{\cal D}+1} \rangle=e^{ix_{\cal D}m}
|\varphi \rangle$. One can solve the  condition (\ref{DTD})
explicitly. The solution is given in
  terms of four completely unrestricted vectors
  \begin{eqnarray} \label{SOLD}
|\varphi \rangle &=& \sum_{k=0}^\infty
[{(\alpha^+_D)}^{2k}(\frac{2^k}{(2k)!}M^k |\varphi_0\rangle +
\frac{ 2^{k-1}}{(2k-1)!}M^{k-1} |\varphi_2\rangle)
 \\ \nonumber
 &&+
 [{(\alpha^+_D)}^{2k+1}(\frac{2^k}{(2k+1)!}M^k |\varphi_1\rangle +
\frac{k 2^k}{(2k+1)!}M^{k-1} |\varphi_3\rangle)].
  \end{eqnarray}
In a similar manner the equation (\ref{TD}) can be solved to give
\begin{equation}
|\Lambda \rangle =\sum_{k=0}^\infty
[{(\alpha^+_D)}^{2k}\frac{{(2M)}^k }{(2k)!} |\Lambda_0\rangle
+{(\alpha^+_D)}^{2k+1} \frac{{(2M)}^{k} }{(2k+1)!}
|\Lambda_1\rangle].
\end{equation}
Putting (\ref{SOLD}) back into the Lagrangian (\ref{LFD}) and
performing the normal ordering with respect to the oscillators
$\alpha_D^+$, thus integrating them out, one arrives at the
Lagrangian
\begin{eqnarray} \label{LMD}
{\cal L}&=& \sum_{k=0}^\infty [( \langle \varphi_0| {(M^+)}^{k}2^k +
2k \langle \varphi_2| {(2M^+)}^{k-1})(\frac{{(2M)}^k}{(2k)!}|T_0
\rangle \\ \nonumber
&&+ \frac{{(2M)}^{k-1}}{(2k-1)!}(\frac{1}{2}-N)|T_2 \rangle) \\
\nonumber &&+ (\langle \varphi_1| {(2M^+)}^{k} + 2k \langle
\varphi_3| {(2M^+)}^{k-1})(\frac{{(2M)}^k}{(2k+1)!}|T_1 \rangle \\
\nonumber &&+ \frac{2k{(2M)}^{k-1}}{(2k+1)!} (\frac{1}{2} + N)|T_3
\rangle ) ]
\end{eqnarray}
where
\begin{eqnarray}
|T_0 \rangle &=&[l_0 - m^2 + l_{1}^+l_{1} + M^+ l_{1} l_1 + 2m^2 M^+
M ]|\varphi_0 \rangle \\ \nonumber &&- m[l_{1}^++ 2M^+ M]|\varphi_1
\rangle  +[2l_0 M^+ - l_{1}^+l_{1}^+- M^+ l_{1}^+l_{1}]|\varphi_2
\rangle + m M^+ l_{1}^+ |\varphi_3 \rangle
\end{eqnarray}
\begin{eqnarray}
|T_1 \rangle &=&-m[ l_1 + 2l_{1}^+M + 4 M^+ l_{1}M ]|\varphi_0
\rangle
\\ \nonumber &&+[l_0 + l_{1}^+l_1 + M^+ l_1 l_1 + 2m^2 M^+M ]|\varphi_1
\rangle \\ \nonumber &&+m[-3 M^+ l_1 + 2 M^+ l_{1}^+M]|\varphi_2
\rangle + [(2l_0 - m^2)M^+ - l_{1}^+l_{1}^+- l_{1}^+M^+ l_1] |\varphi_3
\rangle
\end{eqnarray}
\begin{equation}
|T_2 \rangle =[l_1 l_1 + 2 m^2 M]|\varphi_0 \rangle - 2m l_{1}
|\varphi_1 \rangle  +[2l_0  - l_{1}^+l_{1}]|\varphi_2 \rangle \\
 + m  l_{1}^+ |\varphi_3 \rangle
\end{equation}
\begin{eqnarray}
|T_3\rangle &=&4 ml_1 M|\varphi_0 \rangle - [l_1 l_1 +2 m^2
M]|\varphi_1 \rangle  -m[2l_{1}^+M-3l_1]|\varphi_2 \rangle \\
\nonumber &&+[l_{1}^+l_1 + m^2 - 2l_0]  |\varphi_3 \rangle.
\end{eqnarray}
The equations of motion which can be obtained from the Lagrangian
(\ref{LMD}) are
\begin{equation}
|T_i \rangle =0, \quad i=0,1,2,3.
\end{equation}
In addition to this there are two Jacobi identities
\begin{equation}
l_1 |T_0 \rangle - M^+ l_1 |T_2 \rangle - m M^+ |T_3 \rangle  -
m|T_1 \rangle =0
\end{equation}
\begin{equation}
2m M |T_0 \rangle + l_1 |T_1 \rangle - m(1 - 2 M^+ M)|T_2 \rangle +
M^+ l_1 |T_3 \rangle=0.
\end{equation}
The gauge transformation rules are
\begin{equation}
\delta |\varphi_0 \rangle = l_{1}^+|\lambda_0 \rangle, \quad \delta
|\varphi_1 \rangle=l_{1}^+|\lambda_1 \rangle - m |\lambda_0 \rangle,
\end{equation}
\begin{equation}
\delta |\varphi_2 \rangle=-l_{1}|\lambda_0 \rangle + m |\lambda_1
\rangle, \quad \delta |\varphi_3 \rangle=-l_{1}|\lambda_1 \rangle
+2m M |\lambda_0 \rangle.
\end{equation}
One can see that the system decouples into quartets. For each
physical field with mass $m$ and spin $s$ contained in the vector
$|\varphi_0 \rangle$, there are three auxiliary fields contained
in vectors $|\varphi_1 \rangle$, $|\varphi_2 \rangle$, $|\varphi_3
\rangle$ which have ranks $s-1, s-2,$ and $s-3$. After gauge
fixing one can prove that in each quartet there is only one
physical polarization with mass $m$ and spin $s$.

\subsection{Irreducible representations of the AdS group}
As we noted before the Lagrangian description of a higher spin
field with spin $s$ and nonzero mass contains a set of auxiliary
fields with spins $s-1,s-2,...,1,0$ \cite{Singh:1974qz}. The
generalization of this result to the (A)dS case
 was given in
\cite{Zinoviev:2001dt} while the unconstrained Lagrangian
formulation for massive higher spin fields on AdS has been carried
out in \cite{Buchbinder:2006ge} using the BRST formalism.

\setcounter{equation}0\section{Interactions}
\subsection{General method}
In this section we discuss the general construction
\cite{Buchbinder:2006eq} of the cubic vertex for massless higher
spin fields on flat and AdS spaces which is based on a
generalization of the BRST method. This approach is analogous in
some aspects to the cubic vertex construction in string field
theory, however, in our case there is no analog of the overlap
conditions on the three-string interaction vertex that would
strongly restrict its form. In the case of interacting massless
higher spin  fields the only guiding principle is gauge invariance
which manifests itself as the requirement of BRST invariance of
the vertex (see also \cite{Bekaert:2005jf}).

There is one crucial point regarding interacting higher spin
fields. It appears that a length parameter is necessary for the
construction of the interaction vertex, so that the latter has the
right dimensions. For higher spin fields in flat space there is no
obvious candidate
 for this length parameter. One possibility would be
to consider higher spin gauge fields emerging in the tensionless
limit of string theory, in which case the role of the above
mentioned parameter is played by the inverse of the string tension
$\alpha^\prime$. On the other hand, for higher spin fields in
curved space-times such a dimensionful parameter is naturally
given by the inverse curvature. In particular, in the case of
higher spin gauge fields on AdS space-times this parameter is
naturally associated with the AdS radius $L$. Note that the zero
radius limit of such a construction is the large-curvature limit.

After these remarks  we will proceed along the  lines of
\cite{Bengtsson:1987jt}, \cite{Cappiello:1988cd}. We wish to construct the
most general cubic vertex;
for that we use three copies of the
triplet  defined in (\ref{trexp}) as $|\Phi_i\rangle$, $i=1,2,3$.
If we studied the quartic vertex we would use four copies of the
Higher Spin  functional $|\Phi \rangle$  etc
\cite{Bengtsson:2006pw}. The tensor fields in $|\Phi_i\rangle$
are all at the same space-time point. Then, the $|\Phi_i\rangle$
interacting among each other are expanded in terms of the set of
oscillators $\alpha^{i+}_\mu, c^{i+}$ and $b^{i+}$
\begin{equation}
[\alpha_\mu^i, \alpha_\nu^{j,+} ] = \delta^{ij} g_{\mu \nu},
\quad \{ c^{i,+}, b^j \} = \{ c^i, b^{j,+} \} = \{ c_0^i , b_0^j \} = \delta^{ij}\,,
\end{equation}
in complete analogy to the free field case.
 The BRST charge of our construction
consists of three copies of the free BRST change $\tilde Q = Q_1+Q_2+Q_3$.
The full interacting Lagrangian can be
written as \cite{Neveu:1986mv} -- \cite{Gross:1986ia}
\begin{equation} \label {LIBRST}
{L} \ = \ \sum_i \int d c_0^i \langle \Phi_i |\, Q_i \, |\Phi_i
\rangle \ + g( \int dc_0^1 dc_0^2  dc_0^3 \langle \Phi_1| \langle
\Phi_2|\langle \Phi_3||V \rangle + h.c)\,, \end{equation} where
$|V\rangle$ is the cubic vertex and $g$ is a dimensionless
coupling constant\footnote{ Each term in the Lagrangian
(\ref{LIBRST}) should have length dimension $- {\cal D}$. This
requirement holds true for each space-time vertex contained in
(\ref{LIBRST}) after multiplication by an appropriate power of the
length scale of the theory, as discussed before.}.

It is straightforward to show that the Lagrangian (\ref{LIBRST}) is
 invariant up to terms of order $
g^2$ under the nonabelian gauge transformations
\begin{equation}\label{BRSTIGT1}
\delta | \Phi_1 \rangle  =  Q_1 | \Lambda_1 \rangle  - g \int
dc_0^2 dc_0^3[(  \langle \Phi_2|\langle \Lambda_3| +\langle
\Phi_3|\langle \Lambda_2|) |V \rangle] + O(g^2)\,,
\end{equation}
\begin{equation}\label{BRSTIGT2}
\delta | \Phi_2 \rangle  =  Q_2 | \Lambda_2 \rangle  - g \int
dc_0^3 dc_0^1[(  \langle \Phi_3|\langle \Lambda_1| +\langle
\Phi_1|\langle \Lambda_3|) |V \rangle] + O(g^2)\,,
\end{equation}
\begin{equation}\label{BRSTIGT3}
\delta | \Phi_3 \rangle  =  Q_3 | \Lambda_3 \rangle  - g \int
dc_0^1 dc_0^2[(  \langle \Phi_1|\langle \Lambda_2| +\langle
\Phi_2|\langle \Lambda_1|) |V \rangle] + O(g^2)\,,
\end{equation}
provided that the vertex $V$ satisfies the BRST invariance condition
\begin{equation}\label{VBRST}
\sum_i Q_i |V \rangle=0\,.
\end{equation}
Indeed the invariance for the terms of zeroth
 order in g is guaranteed by the nilpotence of the BRST charges
 $Q_i$ and the invariance for  the terms of first order in g is
guaranteed by the BRST invariance of the vertex. In a similar way
the closure of the algebra of gauge transformations for the terms
linear in g is guaranteed by the BRST invariance condition of the
vertex. The gauge transformations
(\ref{BRSTIGT1})--(\ref{BRSTIGT3}) are nonlinear deformations of
the previously considered abelian gauge transformations. We assume
here that the tensor fields obtained after the expansion of the
$|\Phi_i\rangle$ functionals in terms of the oscillators
$\alpha^{i+}_\mu$ are different from each other. One can also
consider cases when two or all three higher spin functionals
contain the same tensor fields, as we show at the end of this
chapter.

In order to ensure zero ghost number for the Lagrangian, the cubic
vertex must have ghost number $3$. We make the following ansatz
for the cubic vertex
\begin{equation}\label{Vertex1}
| V \rangle= V |- \rangle_{123} \,
\end{equation}
where the vacuum $|-\rangle $, with ghost number $3$, is defined
as the product of
the individual Fock space ghost vacua
\begin{equation}\label{Defghost}
|-\rangle_{123}= c^1_0 c^2_0 c^3_0\ |0 \rangle_{1} \otimes |0
\rangle_{2} \otimes |0 \rangle_{3}\,.
\end{equation}
The function $V$ has  ghost number $0$ and it is a function of the
rest of the creation operators as well as of the operators
$p_\mu^i$. In  Open String Field Theory the r.h.s. of (\ref{Defghost}) is
multiplied by $\delta^{\cal D}(\sum_i p_i)$ which imposes momentum
conservation on the three string vertex. In our case the analogous
constraint is to discard total derivative terms of the Lagrangian
which is certainly true for flat and AdS space-times. So in what
follows we will impose "momentum" conservation in the sense
described above.

The condition of BRST invariance (\ref{VBRST})  does not
 completely fix the cubic vertex. There is an enormous
freedom due to Field Redefinitions (FR) just like in any field
theory Langrangian. It is clear in the free theory case that
any FR of the form
\begin{equation}\label{FREx}
\delta \Phi_i= F(\Phi_i)\,,
\end{equation}
gives a gauge equivalent set of equations of motion for the fields
$\Phi_i$.
Lagrangians obtained from the free one after the field redefinition
(\ref{FREx}) yield additional ``fake interactions`` and should be discarded.
For the interacting case at hand we see, from
(\ref{VBRST}), that the modified gauge variation (\ref{BRSTIGT1}) --(\ref{BRSTIGT3})
can only determine the cubic vertex up to $\tilde{Q}$-exact cohomology
terms:
\begin{equation}\label{VFR}
\delta |V \rangle= \tilde{Q} |W \rangle\,,
\end{equation}
where   $|W \rangle$  is a state with total ghost
charge $2$. We will see in what follows that this FR
freedom can lead to major simplifications for the functional
form of the vertex.

Next, we expand  the vertex operator $|V\rangle$ and the
function $|W \rangle$ in terms of ghost variables or equivalently
in terms of the following two ghost  quantities with ghost number
zero
\begin{equation}\label{Defab}
\gamma^{ij,+}=c^{i,+} b^{j,+}, \ \ \ \ \beta^{ij,+}=c^{i,+} b^j_0\,.
\end{equation}
These are $3\times 3$ matrices of fields with no symmetry
properties. For the cubic vertex we have the expansion
\begin{eqnarray}\label{VExp}
&|V\rangle= \Bigl\{X^1 + X^2_{ij} \gamma^{ij,+}+ X^3_{ij} \beta^{ij,+} +
X^4_{(ij);(kl)}\gamma^{ij,+}\gamma^{kl,+}+
X^5_{ij;kl}\gamma^{ij,+}\beta^{kl,+}+ \nonumber \\
&+X^6_{(ij);(kl)}\beta^{ij,+}\beta^{kl,+}
+X^7_{(ij);(kl);(mn)}\gamma^{ij,+}\gamma^{kl,+}\gamma^{mn,+}+
X^8_{(ij);(kl);mn}\gamma^{ij,+}\gamma^{kl,+}\beta^{mn,+}+ \nonumber \\
&+X^9_{ij;(kl);(mn)}\gamma^{ij,+}\beta^{kl,+}\beta^{mn,+}+
X^{10}_{(ij);(kl);(mn)}\beta^{ij,+}\beta^{kl,+}\beta^{mn,+}\Bigl\}
|-\rangle_{123}\,,
\end{eqnarray}
since  the function  $V$  in (\ref{Vertex1}) has ghost number
zero.
  In our notation we put in parentheses pairs of
indices which are symmetric under mutual exchange. For example,
$X^4_{(ij);(kl)}$ is symmetric under $(ij) \leftrightarrow  (kl)$.
The coefficient $X^4_{(ij);(kl)}$ is also antisymmetric under $ i
\to k$  since $ \{c^{i,+}, c^{k,+} \} =0$
but we have not indicated these symmetries in order
to avoid clustering notation.

In a similar manner we have the following expansion:
\begin{eqnarray}\label{WExp}
&|W\rangle_{123}= \Bigl\{W_i^1b^{i,+} + W_i^2 b^i_{0}+ W^3_{i;jk}b^{i,+}
\gamma^{jk,+}+ W^4_{i;jk}b^{i,+} \beta^{jk,+}+ W^5_{i;jk}b^i_{0}
\beta^{jk,+}+
  \nonumber \\
&W^6_{i;(jk);(lm)}b^{i,+}\gamma^{jk,+}\gamma^{lm,+}+W^7_{i;jk;lm}b^{i,+}\gamma^{jk,+}\beta^{lm,+}+
W^8_{i;(jk);(lm)}b^{i,+}\beta^{jk,+}\beta^{lm,+}+ \nonumber \\
&W^9_{i;(jk);(lm)}b^i_{0}\beta^{jk,+}\beta^{lm,+}
+W^{10}_{i;(jk);(lm);pn}b^{i,+}\gamma^{jk,+}\gamma^{lm,+}\beta^{pn,+}
+
\nonumber \\
&W^{11}_{i;jk;(lm);(pn)}b^{i,+}\gamma^{jk,+}\beta^{lm,+}\beta^{pn,+} +
W^{12}_{i;(jk);(lm);(pn)}b^{i,+}\beta^{jk,+}\beta^{lm,+}\beta^{pn,+}\Bigl\}|-\rangle_{123}\,.
\end{eqnarray}
for the FR functional $W$.

We consider higher spin fields in flat space-time first. Each
component of the vertex in (\ref{VExp}) has an oscillator
expansion in terms of matter oscillators $\alpha^{i,+}_{\mu}$ and
derivatives $p_\mu^i$, where the latter act to the left. We will
restrict our study to the case of totally symmetric massless
higher spin fields  and therefore we have only to consider three
different sets of oscillators and momenta. The generalization to
the case of the fields that belong to the reducible mixed symmetry
representations of the Poincare group will be given in the last
subsection.

 The interaction vertex glues together three Fock
spaces and for this reason it is convenient to define, in complete
analogy to the free case, the following generators
\begin{eqnarray}\label{VGen}
&l^{ij}=\alpha^{\mu i} p_{\mu}^j, \ \ l^{ij,+}=\alpha^{ \mu , i+}
p_{\mu}^j, \ \ l_0^{ij}=p^{\mu i} p_{\mu}^j \,,\nonumber \\
&M^{ij}= \frac{1}{2}\alpha^{\mu i}\alpha_{\mu}^{j}, \ \ M^{ij,+}=
\frac{1}{2}\alpha^{  \mu ,i +}\alpha_{\mu}^{j +} \,,\nonumber \\
&N^{ij}= \alpha^{ \mu ,i +} \alpha_{\mu}^j +
\delta^{ij}\frac{{\cal D}}{2}\,.
\end{eqnarray}
We see that generators (\ref{VGen}) are indexed by integers $i, j
=1,2,3$. The three values for i and j originate from the fact that
we consider a three field interaction. In the general case of an
n-field interaction, we should take the same generators with $i,j
= 1,2 .., n$. Using the generators above one can build all
possible interaction terms between symmetric higher spin fields.
Therefore, our ansatz for the vertex is that of the most general
polynomial made out from the operators $l^{ij}_0$, $l^{ ij,+}$ and
$M^{ij,+}$. This corresponds to the usual derivative expansion for
the vertex, since the operators $l_{0}^{ij}$ have dimensions
[Length]$^{-2}$ and the operators  $l^{ij,+}$ have dimension
[Length]$^{-1}$. To make sense of such an expansion one needs to
introduce a physical length parameter. In flat space-times it is
not clear where  such a length scale may come from,
nevertheless the hope is that it
 would be connected to the length scale of a fundamental theory such as string or M-theory.

The commutator algebra of the operators in (\ref{VGen}) is:
\begin{eqnarray}\label{VGenA}
&[l^{ij}, l^{kl,+}]= \delta^{ik} l_0^{jl},
\ \ [N^{ij},l^{kl}]=-\delta^{ik}l^{jl}\,, \nonumber \\
&[M^{ij,+},l^{kl}]= -\frac{1}{2} ( \delta^{jk}
l^{il,+}+\delta^{ik}l^{jl,+}), \ \
[N^{ij},M^{kl}]=-(\delta^{ik}M^{jl}+\delta^{il}M^{kj})\,, \nonumber
\\
&[M^{ij},M^{kl,+}]=- \frac{1}{4}
(\delta^{jk}N^{il}-\delta^{jl}N^{ik}-\delta^{ik}N^{jl}-\delta^{il}N^{jk})\,.
\end{eqnarray}
Let us consider the constraints imposed by momentum conservation on
the vertex. Clearly, not all generators in (\ref{VGen}) are
linearly independent once we consider the operatorial equation
$\sum_i p^\mu_i=0$, which means that we omit total derivatives. A convenient set of linearly
independent generators is the following:
\begin{eqnarray}\label{LIVGen}
&l_0^{ij}=(l_0^{11},l_0^{22},l_0^{33})=(l_0^{1},l_0^{2},l_0^{3})
\nonumber \\
\nonumber \\
&l^{ij,+}= (l^{1,+},I^{1,+},l^{2,+},I^{2,+},l^{3,+},I^{3,+}), \nonumber
\\
&l^{i,+}=l^{ii,+}, \ \ I^{1,+}=\alpha^{ \mu, 1
+}(p_{\mu}^2-p_{\mu}^3) ,  \nonumber
\\
&I^{2,+}=\alpha^{ \mu, 2 +}(p_{\mu}^3-p_{\mu}^1) \ \
I^{3,+}=\alpha^{ \mu, 3 +}(p_{\mu}^1-p_{\mu}^2)
\nonumber \\
&M^{ij,+}= (M^{11,+},M^{22,+},M^{33,+},M^{12,+},M^{13,+},M^{23,+}).
\end{eqnarray}

Based on  the above analysis we can write the most general form of
the expansion coefficients $X^l_{(\dots)}$:
\begin{eqnarray}\label{XEXP}
&X^{l}_{(\dots)}=
X^{l}_{n_1,n_2,n_3;m_1,k_1,m_2,k_2,m_3,k_3;p_1,p_2,p_3,r_{12},r_{13},r_{23}(\dots)}
\nonumber \\
&(l_0^1)^{n_1} \dots (l^{+,1})^{m_1} (I^{+,1})^{k_1} \dots
(M^{+,11})^{p_1}\dots (M^{+,12})^{r_{12}} \dots
\end{eqnarray}
Using the explicit form of the BRST charges:
\begin{equation}\label{QBRST}
Q^i=c_0^i l_0^i+ c^i l^{i, +}+ c^{i,+}
l^i-c^{i,+}c^ib_0^i\,,\,\,\,\,\,\mbox{(no sum)}
\end{equation}
and equations (\ref{VBRST}), (\ref{XEXP}) we arrive to the
following set of equations:
\begin{eqnarray}\label{VEQN}
&c^{i,+}[l^i X^1-l^{s,+} X^2_{is}-l_0^sX^3_{is}]=0  \\
&c^{i,+} \gamma^{jk,+}[l^iX^2_{jk}-2l^{s,+}X^4_{(is);(jk)} -l_0^s X^5_{jk;is}]=0
\nonumber \\
&c^{i,+} \beta^{jk,+}[-\delta_{jk}X^2_{ij}+ l^i
X^3_{jk}-l^{s,+}X^5_{is;jk}-2l_0^sX^6_{(is);(jk)}]=0 \nonumber
\\
&c^{i,+}\gamma^{jk,+} \gamma^{lm,+}[l^iX^4_{(jk);(lm)}- 3l^{s,+}X^7_{(is);(jk);(lm)}
-l_0^sX^8_{(jk);(lm);is}]=0 \nonumber \\
&c^{i,+} \gamma^{jk,+} \beta^{lm,+}[-2\delta_{lm}X^4_{(il);(jk)}+
l^iX^5_{jk;lm}-2l^{+s}X^8_{(is);(jk);lm}-2l_0^sX^9_{jk;(is);(lm)}]=0
\nonumber \\
&c^{i,+} \beta^{jk,+} \beta^{lm,+}[-\delta_{jk} X^5_{ji;lm}+ l^i
X^6_{(jk);(lm)}-l^{s,+}X^9_{is;(jk);(lm)}-3l_0^sX^{10}_{(is);(jk);(lm)}]=0 \nonumber\,.
\end{eqnarray}
To simplify the analysis of these equations we define the
operator:
\begin{equation}\label{DefN}
\tilde N=\alpha^{\mu, i+} \alpha^i_{\mu}+b^{i,+} c^{i}+
c^{i,+}b^i\,.
\end{equation}
This operator commutes with the BRST charges $Q_i$ and its
eigenvalues count the degree of the $X^l_{(...)}$s in the
$\alpha_\mu^{i,+}$ oscillator expansion. Namely, as  can be seen
from equation (\ref{VExp}), if the degree of the coefficient $X^1$
in oscillators $\alpha_\mu^{i ,+}$ is $K$, then the rest of the
coefficients have the following degrees in the oscillators
$\alpha_\mu^{i,+}$
$$ X^1(K), \quad
X^2(K-2), \quad X^3(K-1),  \quad X^4(K-4), \quad X^5(K-3),$$ $$
X^6(K-2), \quad  X^7(K-6), \quad  X^8(K-5), \quad X^9(K-4), \quad
X^{10}(K-3).
$$
For example, the first equation has degree $K-1$, since $l^{ij}$
reduces the value of $K$ by one, $l^{ij,+}$ increases it by one
and $l_0^{ij}$ leaves it unchanged.

There is yet another number which can be used in a manner similar
to $K$. Namely, if a term in  the expansion of $V$ has powers of
operators $l_0^{ij}, l^{ij,+}, M^{ij,+}, \gamma^{ij,+}$ and
$\beta^{ij,+}$ equal to $s_1, s_2, s_3, s_4$ and $s_5$
respectively, then the total number $s= s_1+ s_2 +s_3 +s_4 +s_5$
is unchanged under the action of the BRST charge.

The above observations   can be used to classify
  equations (\ref{QBRST}) according to their degree $K$ and the number $s$. This means
  that the vertex can be expanded in a sum of contribution with fixed degrees $K$ and $s$ as
  \begin{equation}
  |V\rangle =\sum_{K,s}|V(K,s)\rangle\,.
  \end{equation}
Therefore,  the  equation (\ref{VBRST}) can be split into an
infinite set of equations
\begin{equation}\label{VBRST1}
\sum_i Q_i V(K,s) =0
\end{equation}
for each value of $K$ and $s$.

To construct the vertex on AdS we use the same procedure as in the
flat case, in particular we solve the same equation (\ref{VBRST}).
In this case, however, care is needed when trying to extend the
algebra (\ref{VGenA}) to a nontrivial background.

For the construction of the interaction vertex on AdS it is
convenient to slightly modify the definition of the operator
(\ref{covcurve}) \cite{Buchbinder:2006ge} as
\begin{equation}
\label{pop}
p_\mu \ = \  -\; i \, \left(
\nabla_\mu + \omega_{\mu}^{ab} \, \alpha_{\; a}^+\,
  \alpha_{ \; b} \right) \ ,
\end{equation}
where $\nabla_\mu$ is AdS covariant derivative. The reason behind
this modification is the following: in the free case, one is
working with only one Fock space and all indexes of the tensor
fields are contracted with the corresponding oscillators.
 Therefore, in the free case the
last term in (\ref{pop}) is enough for $p^\mu$ to act covariantly on Fock space states.
However, in the interacting case, where we have three different
Fock spaces, expressions of the form $\varphi^1_\mu(x) \alpha^{\mu
3 +}$ have a free index $\mu$ with respect to the first and third
Fock spaces. Therefore, $p_\nu^1$ should act as a covariant
derivative $\nabla_{\nu}^1$ on $\varphi^1_\mu(x)$ instead of a
partial one. With this modification the operators of the type
$l_0$ read simply
\begin{equation}
l^{ij}_0 = p^{\mu, i} p^j_\mu,
\end{equation}
while the definition of the operators $l^{ij}, l^{ij,+}$ is given
in (\ref{VGen}) . In order to compute the algebra it is useful to
recall how various operators defined previously  act on physical
states. For example operator $l_0^{12}= p_{\mu}^1 p_{\mu}^2 $,
where $p_{\mu}$ is the operator (\ref{pop}), acts as follows
\begin{eqnarray} \nonumber
l_{0}^{12} |\Phi_1\rangle \otimes |\Phi_2\rangle &=&
\frac{i}{(s_1)!}\alpha^{\mu_1, 1 +} \ldots \alpha^{\mu_s,1 +}
\nabla^\mu \, \varphi^1_{\mu_1\mu_2...\mu_{s_{1}}}(x) |0\rangle_1
\otimes \\ \nonumber &&\frac{i}{(s_2)!}\alpha^{\nu_1,2 +} \ldots
\alpha^{\nu_s, 2 +} \nabla_\mu \,
\varphi^2_{\nu_1\nu_2...\nu_{s_{2}}}(x) |0\rangle_2.
\end{eqnarray}
The operators $p_{\mu}^i$ act only on $i$ -th Fock space and
therefore
  \begin{equation} \label{COMU1} [p^i_\mu,p^j_\nu]
= \delta^{ij} \, (-[\nabla^i_\mu,\nabla^i_\nu]+  \frac{1}{L^2}\;
(\alpha_{\; \mu}^{i,+} \, \alpha_{\; \nu}^i \, -\, \alpha_{\;
\nu}^{i,+} \, \alpha_{\; \mu}^i) \, ) = \delta^{ij} D_{\mu \nu}^i\
.
\end{equation}

The other operators are defined in an analogous way. For example the
operator $l^{12} = \alpha^{\mu,1 } p_{\mu}^2$ acts as
\begin{eqnarray} \nonumber
l^{12} |\Phi_1\rangle \otimes |\Phi_2\rangle &=&
\frac{1}{(s_1-1)!}\alpha^{\mu_2,1 +} \ldots \alpha^{\mu_s,1 +}  \,
\varphi^{1 \mu}{}_{\mu_2...\mu_{s_{1}}}(x) |0\rangle_1 \otimes \\
\nonumber &&-\frac{i}{(s_2)!}\alpha^{\nu_1,2 +} \ldots
\alpha^{\nu_s,2 +} \nabla_\mu \,
\varphi^2_{\nu_1\nu_2...\nu_{s_{2}}}(x) |0\rangle_2,
\end{eqnarray}
the operator $l^{12+} = \alpha^{\mu,1 + } p_{\mu}^2$ acts as
\begin{eqnarray} \nonumber
l^{12+} |\Phi_1\rangle \otimes |\Phi_2\rangle &=& \frac{1}{(s_1)!}
\alpha^{\mu, 1 +} \alpha^{\mu_1,1 +} \ldots \alpha^{\mu_s,1 +}  \,
\varphi^{1}_{\mu_1...\mu_{s_{1}}}(x) |0\rangle_1 \otimes \\ \nonumber
&&-\frac{i}{(s_2)!}\alpha^{\nu_1,2 +} \ldots \alpha^{\nu_s,2 +}
\nabla_\mu \, \varphi^2_{\nu_1\nu_2...\nu_{s_{2}}}(x) |0\rangle_2,
\end{eqnarray}
and the operator  $M^{12}= \frac{1}{2}\alpha^{\mu,1} \alpha_{\mu}^2 $ acts as
\begin{eqnarray} \nonumber
M^{12} |\Phi_1\rangle \otimes |\Phi_2\rangle &=& \frac{1}{2}
\frac{1}{(s_1-1)!}\alpha^{\mu_2,1 +} \ldots \alpha^{\mu_s,1 +}  \,
\varphi^{1 \mu}{}_{\mu_2...\mu_{s_{1}}}(x) |0\rangle_1 \otimes \\
\nonumber &&\frac{1}{(s_2-2)!}\alpha^{\nu_2,2 +} \ldots
\alpha^{\nu_s +}  \, \varphi^2_{\mu \nu_2...\nu_{s_{2}}}(x)
|0\rangle_2.
\end{eqnarray}

At this point  it is instructive to present an explicit example of
a computation. Let us compute the commutator between $l^{11}$ and
$l^{12+}$ acting on $|\Phi_1\rangle \otimes |\Phi_2\rangle$, where,
for clarity, we take $|\Phi_1 \rangle$ to be a vector and $|\Phi_2 \rangle$
to be a scalar.
\begin{eqnarray} \label{example}
&&[\alpha^{\mu, 1}p_{\mu}^1, \alpha^{\nu,1 +} p_{\nu}^2]
\,\varphi^1_\rho \alpha^{\rho,1 +}|0\rangle_1 \otimes
\varphi^2|0\rangle_2 = \nonumber \\
&&=-i\Bigl(\alpha^{\mu,
1}[p_\mu^1,\alpha^{\nu,1+}]\phi_\rho^1\alpha^{\rho,1+}+
[\alpha^{\mu,1},\alpha^{\nu,1+}]p_\mu^1\phi_\rho^1\alpha^{\rho,1+}\Bigl)
(\nabla_\nu\phi^2)|0\rangle_1\otimes |0\rangle_2\nonumber \\
&&=-\alpha^{\rho,1+}(\nabla_\nu \varphi^1_\rho)(\nabla_\nu
\varphi^2)|0\rangle_1 \otimes |0\rangle_2
\end{eqnarray}
In obtaining the above result it was crucial that $p^i_\mu$, as
defined in (\ref{pop}), commutes with $\alpha^{ \nu, j + }$.

Proceeding this way one obtains the algebra of operators
\begin{equation}\label{hatpdef}
l_0^{ij}=p^{\mu, i} p^j_\mu \qquad l^{ij}= \alpha^{\mu, i} p^j_\mu
\qquad l^{ij,+}= \alpha^{ \mu i, +} p^j_\mu
\end{equation}
 on AdS
  for the interacting case
\begin{equation}\label{AdSGenA}
[l^{ij}, l^{mn,+}] = \delta^{im}l_{0}^{jn} -\delta^{jn}
\alpha^{\mu m, +} D^j_{\mu \nu} \alpha^{\nu i}
\end{equation}
\begin{equation}\label{AdSGenA2}
[l^{mn}, l^{kl}] =  \delta^{nl} \alpha^{\mu m} D^j_{\mu \nu}\alpha^{\nu k}
\end{equation}
\begin{equation} \label{AdSGenA4}
[l^{ij}_0, l^{mn}] = \delta^{jn} \alpha^{\nu m} D^j_{\mu \nu}
p^{\mu, i} + \delta^{in} \alpha^{\nu m}  D^i_{\mu \nu} p^{\mu,
j}-{(1-{\cal D})\over L^2} \delta^{ij} \delta^{jn} l^{mi}
\end{equation}
\begin{eqnarray}\label{AdSGenA3}
[l_0^{ij},l_0^{kl}]&=& \delta^{jk}{p}^{\mu, i} D^j_{\mu \nu}p^{\nu
,l}+ \delta^{ik}{p}^j_\mu D^i_{\mu \nu}{p}^l_\nu
-\delta^{jl}p^{\mu, k} D^j_{\mu
\nu}p^{\nu, i} \nonumber \\
&&-\delta^{il}p^{\mu, k} D^i_{\mu \nu}p^{\nu j} +\frac{(1-{\cal
D})}{L^2}\delta^{ik}\delta^{ij}l_0^{il} -\frac{(1-{\cal
D})}{L^2}\delta^{jl}\delta^{ij}l_0^{ki}
\end{eqnarray}
supplemented by the part of the algebra (\ref{VGenA}) which
involves commutators of $M^{ij}$, $M^{ij,+}$ and $N^{ij}$. We will
call the algebra (\ref{AdSGenA}) -- (\ref{AdSGenA3}) the symmetry
 algebra of interacting higher spin theory in AdS space-time.

The commutation relations above differ from the corresponding flat
space -- time ones (\ref{VGenA}), in that they involve extra terms
which when acting on states give $O(1/L^2)$ contributions. These
terms are sub-leading in the $L\rightarrow \infty$ limit, hence
the algebra (\ref{AdSGenA}) contracts to the flat space--time
algebra (\ref{VGenA}) in the small curvature limit. This implies
that free higher spin gauge fields in flat space-time can be
viewed as the zero curvature limit of free higher spin gauge
fields on AdS. However, interacting higher spin gauge fields on
AdS do not have a smooth $L\rightarrow\infty$ limit since
interaction vertices contain positive powers of $L$. Nevertheless,
as we shall see below, the functional form of the cubic vertex of
higher spin gauge fields on AdS differs from the cubic vertex in
flat space-time by terms which are sub-leading as
$L\rightarrow\infty$.

The action of the algebra (\ref{AdSGenA}-\ref{AdSGenA3}) on Fock
space states is more complicated than the free case one and
appears in \cite{Buchbinder:2006eq}. We find it again useful to
demonstrate how
 calculations are done
 in the interacting case on AdS with an example:
\begin{eqnarray}\label{example2}
&&[l^{12}, l^{22,+}]\ l^{12,+}\varphi^1_\rho \alpha^{\rho,1
+}|0\rangle_1 \otimes \varphi^2|0\rangle_2=\alpha^{2,+}_\mu
D^2_{\mu \nu} \alpha^1_\nu  \ l^{12,+} \varphi^1_\rho \alpha^{\rho
1, +}|0\rangle_1 \otimes \varphi^2|0\rangle_2=
\nonumber \\
&&-\frac{1}{L^2}(l^{22,+} (N^{11}-1+\frac{{\cal D}}{2})-2M^{12,+}
l^{12}) \ \varphi^1_\rho \alpha^{\rho,1 +}|0\rangle_1 \otimes
\varphi^2|0\rangle_2= \nonumber \\
&&\frac{i}{L^2} \alpha^{2,+}_{\mu} \alpha^{1,+}_\nu({\cal D}
\varphi^\nu_1(\nabla^\mu \varphi_2) - g^{\mu \nu}\varphi_1^\rho
(\nabla_\rho \varphi_2))|0\rangle_1 \otimes |0\rangle_2.
\end{eqnarray}
There is only one $p^2_\lambda$ from the second Fock space
involved in the example above. In the first equality we used
(\ref{AdSGenA}). In the second equality we acted with $D^2_{\mu
\nu}$ on the $p^2_\lambda$ of the $l^{12,+}$ operator using
(\ref{COMU1}) and (\ref{comut}). This was the only "tensor"
operator in the 2nd Fock space, since $\varphi^2$ is a scalar.
Consequently, we commuted operators $\alpha^i_\sigma$ and
$p^2_\sigma$ past each other to bring the result to the second
line of (\ref{example2}). Finally, we used the action of the
operator (\ref{COMU1}) on scalar Fock space states to
complete the calculations since no other "vector" operator, in the
the second Fock space, was left for $l^{22+}$ or $l^{12}$ to act
upon.

From the manipulations above we conclude the following: the
algebra of constraints being obviously more complicated than in
the case of flat space-time shares its main property-- namely it
preserves the polynomial form of (\ref{VExp}), (\ref{WExp}),
(\ref{XEXP}). Therefore, we can proceed in an analogous manner as
in the flat case.

The next step is to choose an expansion for the cubic vertex in
terms of the AdS generators (\ref{hatpdef}) and (\ref{VGen}).
 In the AdS case the
creation generators of (\ref{LIVGen}) do not commute among each other, unlike the flat
case, as one can see from e.g. (\ref{AdSGenA4}). Nevertheless,
 we can choose a ${\it standard}$ ordering as in (\ref{XEXP}). All other possible
orderings can be brought into the
${\it standard}$ form ( i.e.,
use an analogue of the Weyl ordering in quantum mechanics),
using the algebra
(\ref{AdSGenA}-\ref{AdSGenA3}) and the manipulations described in the previous
subsection, modulo $\frac{1}{L^2}$ terms.
The action of
$D^i_{\mu \nu}$ on "tensors" produces terms proportional to $\frac{1}{L^2}$ as on can
easily verify from (\ref{COMU1})
and (\ref{comut})
that affect lower dimension terms in the $L^2$ expansion of $X^n$.
These latter terms can again
be brought into the ${\it standard}$ form following the same procedure and finally can be
absorbed into the definition of the matrix elements with lower dimension than the one we
started from.

In addition, although naively we do not have momentum conservation
in AdS space-time, we can still make use of the equation $\sum_i
p^\mu_i=0$, since it leads to total derivative terms in
the Lagrangian.

To conclude, one can construct the same linearly independent set
of generators as in (\ref{LIVGen}). The expansion of the
coefficients is exactly the same as in (\ref{XEXP}) with all
generators the AdS equivalent of the flat ones. In order to write
down the BRST invariance condition in a simpler form let us write
 the
 BRST charge on AdS in a compact form
\begin{equation}\label{QAdS}
Q =
c_0\hat{l}_0 +c l^{+}+c^+ l -\frac{8}{L^2} c_0(\gamma^+ M + \gamma M^+) - c^+cb_0
\end{equation}
with
\begin{eqnarray}\label{l0AdS}
l_0 \rightarrow \hat{l}_0&=& {p}^\mu {p}_\mu +
\frac{1}{L^2}({(\alpha^{\mu +} \alpha_\mu)}^2 + {\cal
D}\alpha^{\mu +} \alpha_\mu -6 \alpha^{\mu +} \alpha_\mu -2{\cal
D} +6
-4 M^+ M +\nonumber \\
&& c^+b (4 \alpha^{\mu +} \alpha_\mu +2{\cal D} -6 ) + b^+c (4 \alpha^{\mu +} \alpha_\mu +2{\cal D} -6) +
12 c^+b b^+c).
\end{eqnarray}
Using the explicit
form of (\ref{QAdS}) it is straightforward to write down the
equations resulting from (\ref{VBRST}). They are the same as in
flat case with the substitution $l_0 \to \hat{l}_0$ as in
(\ref{l0AdS})
 along with some modifications which appear because of the explicit $\frac{1}{L^2}$
dependence of the BRST charge. The final result is:
\begin{eqnarray}\label{VAdSEQN}
&&c^{i,+}[l^i X^1-l^{s,+} X^2_{is}-\hat{l}_0^sX^3_{is}
+\frac{16}{L^2}M^{s,+}X^5_{is;ss}]=0  \nonumber \\
&&c^{i,+} \gamma^{ jk,+}[l^iX^2_{jk}-2l^{s,+}X^4_{(is);(jk)}
-\hat{l}_0^s X^5_{jk;is}
 \nonumber \\
&&+\frac{8}{L^2}(\delta_{jk} M_j X^3_{jk} -6 M^{s,
+}X^8_{(ss);(jk);is})]=0\\
 &&c^{i,+}
\beta^{jk,+}[-\delta_{jk}X^2_{ij}+ l^i
X^3_{jk}-l^{s,+}X^5_{is;jk}-2\hat{l}_0^sX^6_{(is);(jk)}
-\frac{32}{L^2}M^{s, +}X^9_{ss;(jk);(is)}]=0 \nonumber
\end{eqnarray}
\begin{eqnarray}
&&c^{i,+} \gamma^{jk,+}\gamma^{lm,+}[l^iX^4_{(jk);(lm)}- 3l^{s,+}X^7_{(is);(jk);(lm)}
-\hat{l}_0^sX^8_{(jk);(lm);is} \nonumber \\
&&-\frac{8}{L^2}\delta_{jk} M^j
X^5_{lm;ij}]=0 \nonumber \\
&&c^{i,+} \gamma^ {jk,+}\beta^{ lm,+}[-2\delta_{lm}X^4_{(il);(jk)}+
l^iX^5_{jk;lm}-2l^{s,+}X^8_{(is);(jk);lm}-2\hat{l}_0^sX^9_{jk;(is);(lm)} \nonumber \\
&&+\frac{16}{L^2} \delta_{jk} M^j X^6_{(lm);(ij)}]=0
\nonumber \\
&&c^{i,+} \beta^{jk,+}\beta^{lm,+}[-\delta_{jk} X^5_{ji;lm}+ l^i
X^6_{(jk);(lm)}-l^{s,+}X^9_{is;(jk);(lm)}-3\hat{l}_0^sX^{10}_{(is);(jk);(lm)}]=0. \nonumber
\end{eqnarray}
 Combinations involving the operator $\hat{l}_0^s$
should be understood as follows: for example the term in the first equation
$c^{i+}\hat{l}_0^sX^3_{is}$ is  a result of an  action of the operator
$c^{i +}_0 \hat{l}_0^i$ at $X^3_{mn} \beta^{+mn}$ and using the expression
(\ref{l0AdS})
\begin{eqnarray}\label{l0AdS11}
c^{i,+}_0 \hat{l}_0^i X^3_{mn} \beta^{mn, +}&=&-c^{i,+}
 ({p}^{\mu, s} {p}^s_\mu + \frac{1}{L^2}({(\alpha^{\mu, s +} \alpha^s_\mu)}^2 + {\cal
D}\alpha^{\mu, s +} \alpha^s_\mu -6 \alpha^{\mu, s +} \alpha^s_\mu
-  \\ \nonumber &&{(2{\cal D} -6)}^s -4 M^{s,+} M^s)) X^3_{is} -
 \frac{1}{L^2}c^{i,+}(4 \alpha^{\mu, i +} \alpha^i_\mu + {(2{\cal D} -6)}^i) X^3_{ii}.
 \nonumber
\end{eqnarray}
The equations in (\ref{VAdSEQN}) are  more difficult to analyze
compared to the flat case despite their apparent similarity.
  The main reason is obvious from the algebra
(\ref{AdSGenA})--(\ref{AdSGenA3}) which has nontrivial commutators
containing $D^i_{\mu \nu}$. This causes more of a technical
difficulty rather than a conceptual one. It would be interesting
to find a solution in a closed compact form (if such a solution
exists of course) but at the present moment we are content to have
a well defined iteration procedure and a system of equations which
can be straightforwardly solved via this procedure as we shall
demonstrate with a couple of examples below.

\subsection{Some explicit examples}
{\bf Spin-$1$ with two scalars}
Let us work out in detail the most trivial example of a vector field interacting with two scalars
i.e., the case of  scalar electrodynamics.
Let us put the scalars in the first and the second Fock spaces respectively, and the vector field in
the third Fock space.
 Since the oscillators
$\alpha_\mu^{i,+}, c^{i,+}$ and $b^{i,+}$ occur only in the third
Fock space we omit the index $i$ for them in what follows. The fields we are using are
\begin{equation} \label{phi12}
|\Phi_1\rangle = \phi_1(x)|0\rangle, \quad |\Phi_2\rangle =
\phi_2(x)|0\rangle,
\end{equation}
\begin{equation}
|\Phi_3\rangle = ( A_{\mu}(x) \alpha^{\mu +}
 - i C
b^+) |0 \rangle\,,
\end{equation}
\begin{equation}
|\Lambda \rangle = i\lambda  b^+ |0 \rangle\,.
\end{equation}
Then, according to the discussion after  equation (\ref{DefN}),
in order to saturate the last term in (\ref{LIBRST})
we need the expansion of the vertex at $K=1$ and ghost number zero.
Obviously the unique possibility is
\begin{equation}
V= a_i p_\mu^i \alpha^{\mu +} + d_i c^+ b_0^i, \quad i=1,2,3
\end{equation}
where $a_i$ and $d_i$ are constants to be determined.
However, one can show that some of these constants are redundant.
Let us consider the cohomology of the BRST charge
\begin{equation}
Q= Q_1+ Q_2 + Q_3 = c_0^1 l^1_0 + c^2_0l_0^3 + c_0^3 l_0^3 +
c^+ l^{3}+ c l^{+,3}- c^+cb^3_0.
\end{equation}
Recalling that the vertex is determined modulo $Q W$ and taking (the unique option)
\begin{equation}
W=b^+ w, \quad QW = c^+ b_0^3
\end{equation}
one `gauges away'' the parameter $d_3$. Furthermore, since we have the
 momentum conservation law $p_\mu^1 + p_\mu^2 + p_\mu^3=0$, we
can express $p_\mu^3$ in terms of the other two. This means that
the parameter $a_3$ is redundant as well. So we have four
parameters $a_1, a_2, d_1, d_2$. The condition for BRST invariance
of the vertex gives:
\begin{equation}
c^+(-d_1p^1_\mu p^1_\mu - d_2p^2_\mu p^2_\mu + a_1 p^1_\mu p^3_\mu + a_2 p^2_\mu p^3_\mu )=0.
\end{equation}
Applying momentum conservation to the first two terms one arrives
at the equations
\begin{equation}
d_1+d_2=0, \quad d_1+a_1=0, \quad a_2-d_1=0,
\end{equation}
i.e., we can choose
\begin{equation}
a_1= ig, \quad a_2=-ig, \quad d_1=-ig, \quad d_2=ig.
\end{equation}
Let us write down the interaction vertex
\begin{eqnarray} \nonumber
V&=& \int dc_0^1 dc_0^2 dc_0^3 \langle 0|\phi_1(x)
 \langle 0|\phi_2(x)  \langle 0|(A_\mu \alpha^\mu \\ \nonumber
&&+ i Cb c_0^3)(a_i p^i_\mu \alpha^{\mu, +} +b_i c^+ b_0^j)c_0^1
c_0^2 c_0^3 |0\rangle  |0\rangle  |0\rangle + h.c.\\ \nonumber &&=
\int dc_0^1 dc_0^2 dc_0^3 \langle 0|\phi_1  \langle 0|\phi_2
\langle 0|
A_\mu (a_i )p^i_\mu  c_0^1 c_0^2 c_0^3 |0\rangle  |0\rangle  |0\rangle + h.c. \\
&& = - \phi_1 \phi_2 A_\mu  a_i p^i_\mu + h.c.
= - g (\partial_\mu \phi_1)\phi_2 A^\mu + g(\partial_\mu \phi_2)\phi_1 A^\mu + h.c.
\end{eqnarray}
which is the standard vertex for scalar electrodynamics. For the
gauge transformations we get
\begin{eqnarray}
\delta \phi_1  &=& - \int  dc_0^2 dc_0^3  \langle 0|\phi_2(x)  \langle 0|\lambda b (-i)
d_1 c^+ b_0^1  c_0^1 c_0^2 c_0^3 |0\rangle  |0\rangle  |0\rangle \\ \nonumber
&&=\int  dc_0^2 dc_0^3 \langle 0|\phi_2(x)  \langle 0|\lambda i d_1 c_0^2 c_0^3
|0\rangle  |0\rangle  |0\rangle \\ \nonumber
 &&=
-i d_1 \phi_2(x) \lambda = -g\phi_2(x) \lambda
\end{eqnarray}
\begin{equation}
\delta \phi_2= g\phi_1(x) \lambda, \quad \delta A_\mu = \partial_\mu \lambda.
\end{equation}
The analysis for the case of  $AdS_{\cal D}$ is absolutely the
same and gives the same result.

{\bf Spin-$2$ triplet with two scalars}

We
assign the field with spin two to the third Fock space, so we have
\cite{Fotopoulos:2007yq}
\begin{equation}
|\Phi_3\rangle = (\frac{1}{2!} h_{\mu \nu}(x) \alpha^{\mu +}
\alpha^{\nu +} + D(x)c^+ b^+ - i C_\mu(x) \alpha^{\mu +} c_0^3
b^+) |0 \rangle\,,
\end{equation}
\begin{equation}
|\Lambda \rangle = i\lambda_\mu (x) \alpha^{\mu +} b^+ |0 \rangle\,.
\end{equation}
In this case we need the expansion of the vertex at $K=2$.  Following the same procedure as in the previous case
we get
the Lagrangian
\begin{equation}\label{le}
L= L_{free} + L_{int}\,,
\end{equation}
\begin{eqnarray}\label{lf200}
L_{free}&=& (\partial_\mu \phi_1) (\partial^\mu \phi_1)
+(\partial_\mu \phi_2) (\partial^\mu \phi_2) +m^2(\phi_1^2+\phi_2^2)+ (\partial_\rho h_{\mu
\nu}) (\partial^\rho h^{\mu \nu})\nonumber\\
&&-4 (\partial_\mu h^{\mu \nu})
C_\nu - 4 (\partial_\mu C^\mu) D -2(\partial_\mu D)
(\partial^\mu D) + 2C_\mu C^\mu\,,
\end{eqnarray}
\begin{eqnarray}\label{li200}
L_{int}&=& C_{2,0} \ (  h^{\mu \nu}(\partial_{\mu}
\partial_{\nu} \phi_1)\phi_2 + h^{\mu
\nu}(\partial_{\mu}
\partial_{\nu} \phi_2)\phi_1
-  2h^{\mu \nu}(\partial_{\mu}  \phi_1)(\partial_{\nu} \phi_2)) \nonumber \\
 && - C_{2,1} \ \phi_1 \phi_2 (h^\mu_\mu -2D)\,,
\end{eqnarray}
and the relevant gauge transformations
\begin{equation}
\label{gt21}
\delta \phi_1 = C_{2,0} \ (2 \lambda^\mu \partial_\mu \phi_2 +
\phi_2
\partial_\mu \lambda^\mu)\,,
\end{equation}
\begin{equation}\label{gt22}
\delta \phi_2 = C_{2,0} \ (2 \lambda^\mu \partial_\mu \phi_1 +
\phi_1
\partial_\mu \lambda^\mu)\,,
\end{equation}
\begin{equation}\label{g200}
\delta h_{\mu \nu} =  \partial_\mu \lambda_\nu +\partial_\nu
\lambda_\mu, \quad \delta C_\mu  = \Box \lambda_\mu, \quad \delta D
=\partial_\mu \lambda^\mu\,,
\end{equation}
where $C_{2,0}$ and $C_{2,1}$ are arbitrary real constants.
Note that we have added a mass-term for the scalars in the
Lagrangian. Curiously enough the Lagrangian describing the
interaction of two massless scalars with a spin two triplet is
still gauge invariant after the addition of the mass terms for the
scalar. This opens the interesting possibility  to start with the
Lagrangian for the free massive scalars and gauge its symmetries.
In this way one recovers the Lagrangian given above after gauging
the symmetries generated by a parameter $\lambda_\mu$. A similar
result holds for the case of two scalars interacting with a spin
$3$ gauge field. In this case one gauges the symmetries
 of the free Lagrangian generated by a parameter $\lambda_{\mu \nu}$
\cite{Fotopoulos:2007yq} (see also \cite{Eastwood:2002su},
\cite{Bekaert:2007mi}).

According to our general construction  we have
 obtained the cubic vertex which involves two different scalars and the
 triplet with higher spin 2.
To obtain the interaction of a single scalar with the spin-2 field
we need to set  $\phi_1=\phi_2$. Note that setting i.e.
$\phi_2=0$ is meaningless since in our formalism that would mean
to consider two Fock spaces, hence no cubic interaction vertex.
It should also be noted that for $\phi_1=\phi_2$ (\ref{li200})
is equivalent to the linearized interaction of a scalar field with
gravity.
 The generalization for the coupling of a spin-2 triplet with an arbitrary
 number of scalar fields $n$ goes in an analogous manner.

In $AdS_{{\cal D}}$ we replace ordinary derivatives  with
covariant ones. There will be no other changes for the gauge
transformation rules (\ref{g200}) (i.e., for all fields
$\delta_{AdS}= \delta$)
 except for
\begin{equation}
\delta_{AdS}C_\mu = \delta C_\mu + \frac{1-{\cal D}}{L^2}\lambda_\mu\,.
\end{equation}
The free Lagrangian is modified to include the standard AdS ``mass-terms" of order $1/L^2$
\begin{equation}
\Delta L_{free}= -\frac{1}{L^2} (2h_\mu^\mu h_\nu^\nu - 16 h_\mu^\mu
D + 2h_{\mu \nu} h^{\mu \nu} + (4 {\cal D} +12) D^2 + (2{\cal D}-6)\
( \phi_1^2+\phi_2^2))\,.
\end{equation}
The interaction part also changes and gets an additional piece
\begin{equation}
\label{lintAdS200}
\Delta L_{int.} = C_{2,0}\  \frac{{\cal D}-1}{L^2} D \phi_1 \phi_2\,.
\end{equation}
This is an additional interaction of the $D$ scalar with a ``spin-0" current.

{\bf Spin-$3$ triplet with two scalars}

The spin-3 triplet is described by the field \cite{Fotopoulos:2007yq}
\begin{equation}
|\Phi_3\rangle = (\frac{1}{3!} h_{\mu \nu \rho}(x) \alpha^{\mu +}
 \alpha^{\nu +}\alpha^{\rho +} + D_\mu(x) \alpha^{\mu +} c^+ b^+ -
 \frac{i}{2} C_{\mu \nu}(x) \alpha^{\mu +}\alpha^{\nu +} c_0^3 b^+)
|0 \rangle,
\end{equation}
\begin{equation}
|\Lambda \rangle = \frac{i}{2}\lambda_{\mu \nu} (x) \alpha^{\mu +}
\alpha^{\nu +} b^+ |0 \rangle\,.
\end{equation}
Again, solving the BRST invariance condition for the vertex at $K=3$ we get
the relevant gauge transformations
\begin{equation}
\label{gt31}
\delta \phi_1 = 3i \ C_{3,0}\ (4 \lambda^{\mu \nu} \partial_\mu
\partial_\nu \phi_2 + \phi_2
\partial_\mu \partial_\nu\lambda^{\mu \nu} +
4 (\partial_\mu \phi_2)(\partial_\nu \lambda^{\mu \nu} )) + i \
C_{3,1}\ \phi_2 \lambda_\mu^\mu\,,
\end{equation}
\begin{equation}\label{gt32}
\delta \phi_2 = -3i \ C_{3,0}\ (4 \lambda^{\mu \nu}
\partial_\mu
\partial_\nu \phi_1 + \phi_1
\partial_\mu \partial_\nu\lambda^{\mu \nu} +
4 (\partial_\mu \phi_1)(\partial_\nu \lambda^{\mu \nu} )) - i\
C_{3,1}\  \phi_1 \lambda_\mu^\mu\,,
\end{equation}
\begin{equation}\label{g300}
\delta h_{\mu \nu \rho} =  \partial_\mu \lambda_{\nu \rho}
+\partial_\nu \lambda_{\mu \rho} +\partial_\rho \lambda_{\mu \nu} ,
\quad \delta C_{\mu \nu}  = \Box \lambda_{\mu \nu}, \quad \delta
D_\mu =\partial_\nu \lambda^\nu_\mu\,.
\end{equation}
and the free and interacting parts of the Lagrangian
\begin{eqnarray}
L_{free}&=& (\partial_\mu \phi_1) (\partial^\mu \phi_1)+
(\partial_\mu \phi_2) (\partial^\mu \phi_2) +m^2(\phi_1^2+\phi_2^2)+  (\partial_\tau h_{\mu
\nu \rho})(\partial^\tau h^{\mu \nu \rho})
\\ \nonumber
&&- 6 (\partial_\rho h^{\mu \nu \rho}) C_{\mu \rho} -12(\partial_\mu
C^{\mu \nu})D_\nu - 6 (\partial_\mu D_\nu) (\partial^\mu D^\nu) +
3C_\mu C^\mu\,,
\end{eqnarray}
\begin{eqnarray} \label{lint300}\nonumber
L_{int.}&=&i\  C_{3,0}\ ( h^{\mu \nu \rho} \phi_1
\partial_\mu
\partial_\nu \partial_\rho \phi_2 -
h^{\mu \nu \rho} \phi_2 \partial_\mu
\partial_\nu \partial_\rho \phi_1 -3
h^{\mu \nu \rho} (\partial_\mu \partial_\nu \phi_2)(\partial_\rho
\phi_1) \\ \nonumber && +  3 h^{\mu \nu \rho} (\partial_\mu
\partial_\nu
\phi_1)(\partial_\rho \phi_2)) \\
&&   +i \ C_{3,1}\ (h^{\mu \nu}_\nu -2D^\mu) (\phi_1
\partial_\mu \phi_2 -\phi_2
\partial_\mu \phi_1  ) + h.c.,
\end{eqnarray}
where $C_{3,0}$ and $C_{3,1}$ are arbitrary pure imaginary constants.
Note that in this case, had we set $\phi_1=\phi_2$ the interaction
would have vanished. Unlike the previous example for the case of an
interacting triplet with higher spin $3$  with two scalars  one
cannot put the scalars $\phi_1$ and $\phi_2$ to be equal to each
other so one needs a complex scalar in analogy with scalar
electrodynamics. There is one more difference with respect to the
previous example, namely when doing the deformation to the
$AdS_{\cal D}$ case , apart from changing ordinary derivatives to
covariant ones, both the Lagrangian and gauge transformation rules
for scalars get deformed
\begin{eqnarray}
&&\Delta L_{free}=- \frac{1}{L^2} (6h_{\mu}^{\mu \rho} h_{\nu
\rho}^\nu - 48 h_\mu^{\mu \nu} D_\nu -({\cal D} -3) h_{\mu \nu
\rho} h^{\mu \nu \rho} + \nonumber \\
&&+18( {\cal D} +3) D^\mu D_\mu + (2{\cal D}-6)\ (
\phi_1^2+\phi_2^2) )
\end{eqnarray}
\begin{equation}\label{irrAdSV2}
\Delta L_{int} = i \ C_{3,0}\ \frac{6{\cal D}}{L^2}\ D^\mu \ (
\phi_1 \nabla_\mu \phi_2 - \phi_2 \nabla_\mu \phi_1)+ h.c.
\end{equation}
\begin{equation}\label{gtAdS}
\delta_{AdS} \phi_1 = \delta_0 \phi_1 - i \ C_{3,0}\
\frac{6}{L^2}\lambda^\mu_\mu \phi_2, \quad \delta_{AdS} \phi_2 =
\delta \phi_2 +i \ C_{3,0}\ \frac{6}{L^2}\lambda^\mu_\mu \phi_1,
\end{equation}
\begin{equation}
\delta_{AdS}C_{\mu \nu} = \delta C_{\mu \nu} + \frac{2(1-{\cal
D})}{L^2}\lambda_{\mu \nu}+ \frac{2}{L^2}g_{\mu \nu}
\lambda^\rho_\rho.
\end{equation}

\subsection{An exact vertex}

In this subsection we will give a solution to the cubic vertex
which is exact to all orders in the constant $g$
\cite{Fotopoulos:2007nm}. We begin first with the simple case of a
vertex for totally symmetric fields. This means we consider only
one set of oscillators as in (\ref{B4}). The  form of the vertex
can be deduced  from the high energy limit of the corresponding
vertex of OSFT. In bosonic OSFT the cubic vertex has the form
\begin{eqnarray}\label{SFT}
&&| V_3 \rangle = \int \ dp_1\ dp_2\ dp_3\ (2\pi)^d \
\delta^d(p_1+p_2+p_3) \\
&&\times exp \ \left( {1\over 2} \sum_{i,j=1}^3 \
\sum_{n,m=0}^\infty \ \alpha^{+, i}_{n, \mu} \ N^{ij}_{nm}
\alpha^{+, j}_{m, \nu} \ \eta^{\mu \nu} +  \sum_{i,j=1}^3 \
\sum_{n \geq 1, m \geq 0}\ c^{+,i}_{n} X^{ij}_{nm} b_{m}^{+,j}
\right) \ |-\rangle_{123}, \nonumber
\end{eqnarray}
where the solution is given in terms of the Neumann coefficients
and all string modes contribute. The oscillators
$\alpha^i_{0,\mu}$ are proportional to the momenta $p^{i}_\mu$.
The vertex is  invariant under the  action of the BRST charge
(\ref{B1}). In addition, the action (\ref{LIBRST}) with the vertex
(\ref{SFT}) is invariant under the gauge transformations
(\ref{BRSTIGT1})  to all orders in $g$.

Furthermore, since the BRST charge
 can be truncated to contain any finite number of oscillator
variables \cite{Sagnotti:2003qa}, it is possible to look for the
BRST invariant vertex that describes the interaction among only
totally symmetric tensor fields of arbitrary rank, without the
inclusion of modes with mixed symmetries.
 One possibility is to start from the SFT vertex
(\ref{SFT}) and keep in the exponential
 only terms proportional to at least one momentum $p^{r}_\mu$,
 therefore dropping all trace operators $(\alpha^r_\mu \eta^{\mu \nu}
\alpha^{s}_\nu)$,
  as one does when obtaining the BRST charge (\ref{hebrst})
  from (\ref{B1}) since
they are leading in the $\alpha ' \to \infty $ limit. However, since
these terms  are exponentiated and  the term $\alpha^{+, r}_{n, \mu}
N^{rs}_{n0} p^{s}_\mu$ is of the same order as $\alpha^{+, r}_{n,
\mu} N^{rs}_{n0} p^{s}_\mu \ (\alpha^{+, r}_{n, \mu} \ N^{rs}_{nm}
\alpha^{+, s}_{m, \nu})^p, \ m,n \geq 1$, a priori one can keep them
both . The same is true  regarding the ghost part where, although the
term $c^{+,r}_{n} b_{0}^{s}$ is leading compared to the term
$c^{+,r}_{n} X^{rs}_{nm} b_{m}^{s}, \ n,m \geq 1$, one can not
neglect the latter one in the exponential. Let us stress that all
these terms will be essential to maintain the off shell closure of
the algebra of gauge transformations and complete gauge invariance
of the action.

Based on the discussion above one can make the following ansatz
for the vertex which describes interactions between massless
totally symmetric fields with an arbitrary spin
\begin{equation}\label{KOmod}
|V \rangle=  V^{1} \times V^{mod}|-\rangle_{123}
\end{equation}
where the vertex  contains two parts: a part considered in
\cite{Koh:1986vg}
\begin{equation}\label{KOansatz}
V^{1}= exp\ (\ Y_{ij} l^{+,ij} + Z_{ij} \beta^{+,ij}\ )\,,
\end{equation}
and the part which ensures the closure of the nonabelian algebra
\begin{equation}\label{Vmod}
V^{mod}=exp\ (\ S_{ij} \gamma^{+,ij} + P_{ij} M^{+, ij}\ ),
\end{equation}
where $P_{ij}=P_{ji}$.
 Putting this ansatz into the BRST invariance condition
and
using momentum conservation $p_\mu^1+ p_\mu^2 + p_\mu^3=0$ one can
obtain a solution for $Y^{rs}$ and $Z^{rs}$
\begin{equation}\label{KOsolution}
Z_{i,i+1}+Z_{i,i+2}=0
\end{equation}
$$
Y_{i,i+1}= Y_{ii}-Z_{ii} -1/2(Z_{i,i+1}-Z_{i,i+2})
$$
$$
Y_{i,i+2}= Y_{ii}-Z_{ii} +1/2(Z_{i,i+1}-Z_{i,i+2}).
$$
\begin{eqnarray}\label{KOmsol}
&& S_{ij}= P_{ij}=0 \qquad i\neq j \\
&& P_{ii} - S_{ii}=0 \qquad i=1,2,3. \nonumber
\end{eqnarray}
In what follows we will assume cyclic symmetry in the three
Fock spaces which implies along with (\ref{KOsolution})
\begin{equation}\label{cyclic}
Z_{12}=Z_{23}=Z_{31}=Z_a, \quad
Z_{21}=Z_{13}=Z_{32}=Z_b=-Z_a
\end{equation}
$$ Y_{12}=Y_{23}=Y_{31}=Y_a, \quad
Y_{21}=Y_{13}=Y_{32}=Y_b
$$
$$
Y_{ii}=Y, \quad \ Z_{ii}= Z, \quad P_{ii}=S_{ii}=S=P
$$
Having determined the form of the vertex from (\ref{KOsolution}) and
(\ref{KOmsol}) we will proceed in computing the commutator of two
gauge transformations with gauge parameters $|\Xi \rangle$ and
$|\Lambda \rangle$. In general, closure of the algebra to order
$O(g)$ implies
\begin{eqnarray}\label{comGT2}
&&[\delta_\Lambda, \ \delta_\Xi] |\Phi_1\rangle =
\delta_{\tilde{\Lambda}} |\Phi_1\rangle= Q_1
|\tilde{\Lambda}_1\rangle -g[( \langle \Phi_2|\langle
\tilde{\Lambda}_3| +\langle \Phi_3|\langle \tilde{\Lambda}_2|) |V
\rangle] + \ O(g^2) \nonumber \\
\end{eqnarray}
where
\begin{equation}\label{GP}
|\tilde{\Lambda}_1 \rangle= g ( \langle \Lambda_2 | \langle
\Xi_3|+\langle \Lambda_3 | \langle \Xi_2|)|V\rangle +\ O(g^2).
\end{equation}
It should be emphasized that unlike the case of  free triplets
where the total Lagrangian splits into an infinite sum of
individual ones \cite{Sagnotti:2003qa}, for the case of
interacting triplets
 the fields
$|\Phi_i\rangle$ of (\ref{trexp}) need to be composed of an
infinite tower of higher spin triplet fields, at least when the
vertex is defined via (\ref{KOmod}) or (\ref{KOansatz}). In other
words:
\begin{equation}\label{PhiKO}
|\Phi_i\rangle \to \sum_{s=0}^{\infty} \ |\Phi_i^{(s)}\rangle.
\end{equation}
Let us give the full nonabelian gauge transformations based
on (\ref{BRSTIGT1})
\begin{eqnarray}\label{GTKOm}
&&\delta (|\varphi_1 \rangle + c_1^+ \ b_1^+\ |d_1\rangle + c^0_1\
b_1^+\ |c_1\rangle)= (l^+_{11}  + l_{11}\ c_1^+ \ b_1^+ + c^0_1\
b_1^+ \ l_0^{1})\ |\lambda_1\rangle  + \nonumber \\
&& + \ g \ e^{(S \ c^+_1  b^+_1)} \ \{- Z_{a}(  \langle
\varphi_2|\ \langle \lambda_3|\ |A\rangle - S\ \left( \langle
d_2|\ \langle
\lambda_3|\ |A\rangle \right)) +  \\
&&+\left(Z\ Z_{a}-Z^2_{b} \right)\ \langle c_2|\ \langle
\lambda_3|\ |A\rangle + \ (2 \leftrightarrow 3, \ Z_a \to Z_b) \}
\nonumber
\end{eqnarray}
where for convenience we have defined the matter part of the
vertex
\begin{equation}\label{A}
|A \rangle= exp\ (\ Y_{ij} l^{+,ij} +P \sum_{i=1}^3 M^{+,ii} )\
|0\rangle.
\end{equation}
From this transformation rule it is very simple to prove the
exactness of the vertex. Namely, the crucial point is the presence
of the term $e^{S c_1^+ b_1^+}$. Indeed, taking the variation with
respect to say $|\Phi_1 \rangle$ in the interaction term and
considering the term of order $g^2$ one obtains a term $\langle
0_1|e^{-Sc_1b_1} $. This term should be saturated by the term
$e^{S c_1^+ b_1^+}|0_1 \rangle$ from the vertex.
Therefore, the whole expression
vanishes if
\begin{equation}\label{condition}
{|S|}^2=1
\end{equation}

In a similar manner one can prove the closure of the algebra at
order $g^2$. The commutator of two gauge transformations is
\begin{eqnarray}\label{comGT}
&&[\delta_\Lambda,  \delta_\Xi] |\Phi_1\rangle = Q_1
|\tilde{\Lambda_1}\rangle \\
&&+ g^2  [ \langle V| \left( |\Phi_1\rangle|\Lambda_3\rangle
+|\Lambda_1\rangle|\Phi_3\rangle \ \right) \langle \Xi_3|
|V\rangle+ \langle V| \left( |\Phi_1\rangle|\Lambda_2\rangle +
|\Lambda_1\rangle|\Phi_2\rangle
\right) \langle \Xi_2| |V\rangle  \nonumber \\
&&-\langle V| \left( |\Phi_1\rangle|\Xi_3\rangle +
|\Xi_1\rangle|\Phi_3\rangle \right) \langle \Lambda_3| |V\rangle-
\langle V| \left( |\Phi_1\rangle|\Xi_2\rangle +
|\Xi_1\rangle|\Phi_2\rangle \right) \langle  \Lambda_2| |V\rangle
 \nonumber
\end{eqnarray}
where we have suppressed the integrations over the ghost fields of
(\ref{BRSTIGT1}).
One can show that the condition (\ref{condition}) leads to
\begin{equation}\label{comGT8}
[\delta_\Lambda,  \delta_\Xi] |\Phi_1\rangle =0,
\end{equation}
or rather to
\begin{equation}\label{comGT9}
\delta_\Lambda \delta_\Xi | \Phi_1 \rangle=0.
\end{equation}
In other words we can consider the vertex (\ref{KOmod}) as a field
dependent deformation of the  BRST charge in (\ref{hebrst}), which
can be written schematically
\begin{equation}\label{newbrst}
Q'= Q + gV(\Phi)
\end{equation}
with the nilpotency property
\begin{equation}\label{nilbrst}
Q'^2= Q^2 + 2g Q V(\Phi) + g^2 V(\Phi)^2=0.
\end{equation}
 Proceeding further in analogy with String Field Theory one
can make both the string functional and gauge transformation
parameters to be matrix valued (i.e., introduce Chan --Paton
factors). The resulting theory will still satisfy (\ref{comGT9}).

The case of arbitrary mixed symmetry fields is completely analogous
to the construction  for totally symmetric fields. As in
(\ref{KOmod}) we make the ansatz
\begin{eqnarray}\label{KOmmod}
&&V= exp\ (\sum_{n=1}^\infty \ Y^{(n)}_{ij} l_{ij}^{+,(n)} +
 \ Z^{(n)}_{ij} \beta_{ij}^{+,(n)})\times  \\
&&exp\ (\sum_{n,m=1}^\infty\ S^{(nm)}_{ij} \gamma^{+,(nm)}_{ij} +
P^{(nm)}_{ij} M^{+,(nm)}_{ij}) \nonumber
\end{eqnarray}
where in this case we are summing over $n,m$  as well. We put the
oscillator level indices in parentheses in order to distinguish them
from the Fock space ones. The oscillator algebra takes the form
\begin{equation}\label{oscialg}
[\alpha_\mu^{(m),i}, \alpha_\nu^{+,(n),j} ] =
\delta^{mn}\delta^{ij} g_{\mu \nu}, \quad \{ c^{+,(m),i},
b^{(n),j} \} = \{ c^{(m),i}, b^{+,(n),j} \} =
\delta^{mn}\delta^{ij}.
\end{equation}
BRST invariance with respect to (\ref{hebrst}) implies
\begin{eqnarray}\label{BRSTM1}
&&\sum_{i=1}^3 \ \sum_{r=0}^\infty \ c^{+,(r),i} ( Y^{(r)}_{is}
l_0^{is}-Z^{(r)}_{is} l_0^{ss})|-\rangle_{123} =0
\end{eqnarray}
\begin{eqnarray}\label{BRSTM2}
&&\sum_{i=1}^3\ \sum_{r=0}^\infty \{  \ c^{+,(r),i} \
\Bigl({1\over 2}
(P_{il}^{(rs)}\ l^{+,(s),li} + P_{li}^{(sr)}\ l^{+,(s),li}) +  S^{(rs)}_{ik} \ l^{+,(s),kk} \Bigr)  - \nonumber \\
&& b^i_0 \ c^{+,(p),i}\ c^{+,(r),m} \ p\  S^{(rp)}_{mi}
\}|-\rangle_{123} =0
\end{eqnarray}
where the summation over repeated indexes is assumed. Solving
(\ref{BRSTM1}) we get
\begin{equation}\label{KOMsolution}
Z^{(r)}_{i,i+1}+Z^{(r)}_{i,i+2}=0
\end{equation}
\begin{equation}
Y^{(r)}_{i,i+1}= Y^{(r)}_{ii}-Z^{(r)}_{ii} -{1 \over
2}(Z^{(r)}_{i,i+1}-Z^{(r)}_{i,i+2})
\end{equation}
\begin{equation}
Y^{(r)}_{i,i+2}= Y^{(r)}_{ii}-Z^{(r)}_{ii} +{1\over
2}(Z^{(r)}_{i,i+1}-Z^{(r)}_{i,i+2}).
\end{equation}
Equation (\ref{BRSTM2}) gives
\begin{eqnarray}\label{KOMmsol}
&& S^{(ps)}_{ij}= P^{(ps)}_{ij}=0 \qquad i\neq j \ or \ p \neq s \\
&&  P^{(ss)}_{ii} - S^{(ss)}_{ii}=0 \qquad i=1,2,3. \nonumber
\end{eqnarray}
 We can choose once more a
cyclic solution in the three Fock spaces as in (\ref{cyclic}) and
in this way get  an obvious generalization of (\ref{KOMsolution}).
Finally as in the case of only one oscillator, the complete
invariance of the vertex requires.
\begin{equation}\label{clMGT}
|S^{(r)}|\ ^2=1,
\end{equation}
for all $r$ being integer numbers.

Let us conclude with several remarks

\begin{itemize}
\item Dropping the cyclicity  constraint does not seem to alter the
conclusions. In this case we will have $|S_{ii}|^2=1, \ i=1,2,3$.

\item Despite the algebra being trivial it seems the vertex cannot
be obtained from the free Lagrangian via some field redefinition.
In other words the vertex (\ref{KOmod}) is not an exact cohomology
state of the BRST charge (\ref{hebrst}): $|V\rangle \neq
Q|W\rangle$ for any $|W \rangle$. One can show that only terms
diagonal in the Fock spaces $i,j$ can be removed from the exponent
of (\ref{KOmod}) via a specific field redefinition scheme
\cite{Buchbinder:2006eq}.

\item The infinite tower of triplets is essential for the closure.
The nonabelian part of the gauge transformation of each component
of $|\varphi\rangle$ is cancelled against the same rank tensor
component of $|d \rangle$. However, the two tensors belong to
different triplets.

\end{itemize}

\setcounter{equation}0\section{Fermions}

In this section we briefly describe the generalization of the
constructions given in the previous chapters to the case of the
fermionic fields.

In order to describe massless reducible representations of the
Poincare group with an arbitrary half-integer spin let us start
with
 the open sector of the type-I superstring,
(closed superstrings could be treated in a similar way). Let us
first perform the $\alpha^\prime \rightarrow \infty$ limit in the
BRST charge for the open superstring
\begin{eqnarray} \nonumber
Q&=& \sum_{-\infty}^{+\infty}
      \, \left[ L_{-n} \, C_n \ + \ G_{-r}\, \Gamma_r
\ - \ \frac{1}{2}\;
(m-n):C_{-m}\, C_{-n}\, B_{m+n}: \right. \\
&+& \left. \left(\frac{3 n}{2} \ + \ m \right) :C_{-n}\, {\cal
B}_{-m}\, \Gamma_{m+n}: \ - \ \Gamma_{-n}\, \Gamma_{-m} \, {\cal
B}_{m+n} \right] \ - \ a \, C_0 \ ,
\end{eqnarray}
where $a$ is the intercept and the super-Virasoro generators
\begin{eqnarray} \label{VS}
&& L_k \ = \ \frac{1}{2}\, \sum_{l= -\infty}^{+\infty}\,
\alpha_{k-l}\, \alpha_l \ + \ \frac{1}{4}\, \sum_r \, (2r-l)\,
\psi_{l-r}\, \psi_r\ ,\nonumber
\\
\label{VF} && G_r \ = \ \sum_{l=-\infty}^{+\infty} \alpha_l \,
\psi_{r-l}\ ,
\end{eqnarray}
obey the super-Virasoro algebra
\begin{eqnarray}
&&[L_k, L_l] \ = \ (k-l)\, L_{k+l} \ +  \ \frac{{\cal D}}{8}\, \,
(k^3-k) \ ,
\nonumber \\
&&[L_k, G_r] \ = \ \left( \frac{k}{2} \, -\, r \right)\, G_{k+r} \
,
\nonumber \\
&&  \{ G_r, G_s\} \ =\ 2\, L_{r+s} \ +\ \frac{{\cal D}}{2}\,
\left(r^2 \, - \, \frac{1}{4}\right) \, \delta_{rs} \ .
\end{eqnarray}
Here $(k,l)$ are integers for both the Neveu-Schwarz (NS) and
Ramond (R) sectors, while $(r,s)$ are integers for the R sector
and half-odd integers for the NS sector, ${\cal D}$ denotes once
more the space-time dimension (${\cal D}=10$ for the tensile
string) and
 $\alpha_0^\mu = \sqrt{2 \alpha^\prime} \, p^\mu $.
The fermionic oscillators $\psi^\mu_r$ and the ghosts $\Gamma_r$
and antighosts ${\cal B}_r$ satisfy
\begin{equation}
\{ \psi^\mu_r, \psi^\nu_s \} \ = \ \delta_{r+s,0}\, \eta^{\mu \nu} \ ,
\qquad
[\Gamma_r , {\cal B}_s] \ = \ i \, \delta_{r+s,0}\ ,
\end{equation}
and the intercept is $ a=0$ in the R
sector and  $ a=\frac{1}{2}$ in the NS sector.

Rescaling the ghost variables as \be \gamma_{-r} \ = \
\sqrt{2\alpha^\prime} \ \Gamma_{-r}\ , \qquad \beta_{r} \ =\
\frac{1}{\sqrt{2\alpha^\prime}}\ {\cal B}_r \ee and then taking
the $\alpha^\prime \rightarrow \infty$ limit, one obtains the BRST
charge for the NS sector \be \label{BRSTNS} Q_{NS} \ = \ c_0 \,
\ell_0 \ + \ \tilde Q_{NS} \ - \ M_{NS} \, b_0 \ , \ee with
\begin{eqnarray}
&& \label{TQ} \tilde Q_{NS}\ = \ \sum_{k \neq 0} \; \left [\;
c_{-k} \, \ell_k   +  \gamma_{-r}\, g_r \right ]\ , \nonumber \\
\label{TN} && M_{NS}\ = \ \frac{1}{2} \, \sum_{-\infty}^{+\infty}
\; \left [ \, k \, c_{-k}\, c_k  + \gamma_{-r} \, \gamma_r \,
\right ] \ ,
\end{eqnarray}
and
\be \label{GB}
 g_r \ = \ p \cdot \psi_{r} \ . \ee
In a similar fashion, the limiting BRST charge for the R sector
reads
 \be Q_R \ = \ c_0 \, \ell_0  +  \gamma_0 \, g_0  +  \tilde Q_R  -  M_R
b_0  -  \frac{1}{2}\, \gamma_0^2 \, b_0 \ , \ee where $\tilde
Q_R$ and $M_R$ are again given by \p{TN},
 the only difference being that their sums are over
half-odd integer modes for fermionic Virasoro generators and
bosonic (anti)ghosts. Both BRST charges are again identically
nilpotent, independently of the space-time dimension ${\cal D}$.

For the type I superstring, the string field is invariant under
the action of the BRST invariant GSO projection operators for the
NS sector
 \be \label{GSONS}
 P_{NS} \ = \
\frac{1}{2} \, \left[ 1 \ - \ (-1)^{\psi_p^\dagger \; \psi_p \,
+\, i \gamma_p^\dagger \; \beta_p \, - \, i\; \gamma_p\;
\beta_p^\dagger}\right] \ee and for the R sector
 \be \label{GSOR}
 P_R \ = \ \frac{1}{2}\,\left[ 1\ +\
\gamma_{11}\, (-1)^{\psi^\dagger_r \; \psi_r \, + \, i\;
\gamma^\dagger_r \; \beta_r \, - \, i \; \gamma_r \;
\beta^\dagger_r \, + \, i \; \gamma_0 \; \beta_0}\right] \ , \ee
where $\gamma_{11}$ is the ten-dimensional chirality matrix.
Expanding the NS string field and gauge parameter in terms of the
fermionic ghost zero mode as
\begin{eqnarray}
&& |\Phi^{NS} \rangle \ = \ |\Phi_1^{NS}\rangle \ + \ c_0|\Phi_2^{NS}
\rangle \ , \nonumber \\
&& |\Lambda^{NS} \rangle \ = \ |\Lambda_1^{NS} \rangle
\ + \ c_0 |\Lambda_2^{NS} \rangle  \ , \end{eqnarray}
and making use of the BRST charge
\p{BRSTNS}, one obtains the field equations
\begin{eqnarray} \label{EMNS1}
&& \ell_0 |\Phi_1^{NS} \rangle \ - \ \tilde Q_{NS}|\Phi_2^{NS} \rangle \ = \ 0
\ , \nonumber  \\
&& \tilde Q_{NS} |\Phi_1^{NS} \rangle \ - \ M_{NS} |\Phi_2^{NS}
\rangle \ =\ 0 \ , \label{EMNS2} \end{eqnarray} along with the
gauge transformations
\begin{eqnarray}
&& \delta |\Phi_1^{NS} \rangle \ = \ \tilde Q_{NS}
|\Lambda_1^{NS} \rangle \ - \ M_{NS} |\Lambda_2^{NS} \rangle \ , \nonumber \\
\label{NSG3}
&& \delta |\Phi_2^{NS} \rangle \ =\
\ell_0 |\Lambda_1^{NS} \rangle \ - \ \tilde Q_{NS}|\Lambda_2^{NS} \rangle \ .
\end{eqnarray}
The R sector is more complicated, due to the presence of the
bosonic ghost zero mode $\gamma_0$. However, one can work with the
truncated string field
\begin{equation} \label{truncated}
|\Phi^R \rangle \ = \ |\Phi_1^R\rangle \ + \ \gamma_0 \,  |\Phi_2^R \rangle
\ + \ 2 \, c_0 \, g_0\, | \Phi^R_2\rangle \ ,
\end{equation}
while still preserving the relevant portion of the gauge symmetry
and, of course, not affecting the physical spectrum \cite{KNNW1}.
The resulting, consistently truncated, field equations
\begin{eqnarray} \label{EMR1}
&& g_0 \, |\Phi_1^R \rangle \ + \ \tilde Q_{R}|\Phi_2^R \rangle \ =\ 0 \ ,
\nonumber \\
\label{EMR2} && \tilde Q_{R} \, |\Phi_1^R \rangle \ - \ 2 \, M_R
\, g_0 \, |\Phi_2^R \rangle \ = \ 0 \ ,
\end{eqnarray}
are then invariant under the gauge transformations
\begin{eqnarray} \label{GTR1}
&& \delta\, |\Phi_1^R\rangle \ = \ \tilde Q_R |\Lambda_1^R \rangle
\ +\ 2 \,M_R \, g_0 \, |\Lambda_2^R \rangle \ , \label{GT1}
\nonumber \\
\label{GTR2}
&& \delta|\Phi_2^R\rangle \ = \ g_0 \, |\Lambda_1^R \rangle \ - \ \tilde
Q_R \, |\Lambda_2^R \rangle \ . \label{GT2}
\end{eqnarray}

{\bf Symmetric spinor-tensors}

If, as for the bosonic string, one considers fields $|\Phi^{R,1}
\rangle$ and  $|\Phi^{R,2} \rangle$ depending only on the bosonic
oscillator $\alpha^{\mu +}$ and on the fermionic ghost variables
$c_{-1}$ and $b_{-1}$, the expansions
\begin{eqnarray}\label{Rtrip}
&& |\Phi_1^R \rangle \ = \ \frac{1}{n!}\ \psi_{\mu_1\mu_2\, ...\,
\mu_n}(x)\, \alpha^{\mu_1+} \, \alpha^{\mu +} \, ...\,
\alpha^{\mu_n +}\, |0\rangle \nonumber \\ \nonumber && \qquad
\qquad +\ \frac{1}{(n-2)!}\ \lambda_{\mu_1\mu_2\, ...\,
\mu_{n-2}}(x)\, \alpha^{\mu_1+} \, \alpha^{ \mu_2+} \, ..\,
\alpha^{ \mu_{n-2}+} \, |0\rangle \ ,
\\
&&|\Phi_2^R\rangle \ = \ - \ \frac{1}{\sqrt{2}\; (n-1)!}\
\chi_{\mu_1\mu_2\, ...\, \mu_{n-1}}(x) \, \alpha^{\mu_1+} \,
\alpha^{ \mu_2+} \, ...\, \alpha^{\mu_{n-1}+}\, |0\rangle
\end{eqnarray}
define spinor-tensor fields $\psi$, $\chi$ and $\lambda$ totally
symmetric in their tensor indices and of spin $(n+1/2)$, $(n-1/2)$
and $(n-3/2)$, respectively. Substituting these expressions in the
field equations \p{EMR2} then yields precisely the fermionic
triplet equations of \cite{Francia:2002pt}:
\begin{eqnarray}
&& \dsll \psi \ = \ \partial \chi \ , \nonumber \\
&& \partial \cdot \psi \ - \ \partial \lambda
\ = \ \dsll \chi \ , \nonumber \\
&& \dsll \lambda \ =\
\partial \cdot \chi \  \label{fermitriplet}
\end{eqnarray}
where $\dsll= \gamma^\mu \partial_\mu$. The BRST gauge invariance
involves an unconstrained parameter,
\begin{equation}
|\Lambda^\prime_1 \rangle \ =\ \frac{1}{(n-1)!}\
\epsilon_{\mu_1\mu_2...\mu_{n-1}}(x) \, \alpha^{\mu_1+} \,
\alpha^{\mu_2+} ... \alpha^{\mu_{n-1}+}\, |0\rangle \ ,
\end{equation}
and determines the gauge transformations
\begin{eqnarray}
&& \delta \psi \ = \ \partial \, \epsilon \ , \nonumber \\
&& \delta \Lambda \ = \ \partial \cdot \epsilon \ ,
\nonumber \\
&& \delta  \chi \ = \ \dsll \epsilon \ ,
\end{eqnarray}
in agreement with \cite{Francia:2002pt}.

Let us note, however, that the totally symmetric bosonic triplets
  do not arise directly in the NS sector of the
open superstring, since all states containing only bosonic
$\alpha^\mu$ oscillators and fermionic $b,c$ ghosts are eliminated
by the GSO projection operator \p{GSONS}. However, they can emerge
from tensors with mixed symmetry, or even directly if the GSO
projection is modified to correspond to type-0 strings
\cite{type0} (see \cite{Angelantonj:2002ct} for a review).
 One can also consider generalized triplets for spinor-tensors in complete analogy to the case
of the bosonic string \cite{Sagnotti:2003qa}.

{\bf Space--time Supersymmetry}

Generalized triplets of mixed symmetry are actually
the superpartners of symmetric fermionic triplets in the type-I
superstring. Below we briefly outline  how the supersymmetry
for the triplets can be established.

We consider the case where the fields in the Ramond sector consist
of totally symmetric fields, while the field in the NS sector,
apart from the oscillators $\alpha^\mu$, $c$ and $b$, contain at
most one creation operator $\psi_{1/2}^{+,\mu}$ along with the
$\beta$ and $\gamma$ ghosts. This case corresponds to the $N=1$
SUSY.
 Fixing the number of oscillator
$\psi^{+,\mu}$ and $\alpha^{+\mu}$, will determine the content of
the generalized triplet. Namely, one can show that the bosonic
ghosts and their conjugate momenta $\gamma^+_{1/2}$ and
$\beta^+_{1/2}$ can appear only once in the expansion of
$|\Phi^{NS} \rangle$. Then one can show that the total Lagrangian
\begin{eqnarray} \nonumber
{\cal L}_{tot.}&=&  \langle \Phi^{NS}_1|l_0|\Phi^{NS}_1\rangle -
 \langle \Phi^{NS}_2|\tilde Q_{NS}|\Phi^{NS}_1\rangle -
 \langle \Phi^{NS}_1|\tilde Q_{NS}|\Phi^{NS}_2\rangle +
 \langle \Phi^{NS}_2|M_{NS}|\Phi^{NS}_2\rangle \\
&&+
 \langle \Phi^{R}_1|g_0|\Phi^{R}_1\rangle +
 \langle \Phi^{R}_2|\tilde Q_{R}|\Phi^{R}_1\rangle +
 \langle \Phi^{R}_1|\tilde Q_{R}|\Phi^{R}_2\rangle -
2 \langle \Phi^{NS}_2|M_R g_0|\Phi^{R}_2\rangle
\end{eqnarray}

is invariant under the supersymmetry transformations
\begin{equation}
\delta |\Phi^{NS}_1 \rangle = u^+ |\Phi^R_1 \rangle -
\gamma_{1/2}^+ u^+ |\Phi_2^R \rangle, \quad \delta |\Phi^{NS}_2
\rangle = 2 u^+ g_0 |\Phi^R_2 \rangle
\end{equation}
\begin{equation}
\delta |\Phi^{R}_1 \rangle = -2 g_0 u |\Phi^{NS}_1 \rangle +
\gamma_{1/2} u |\Phi_2^{NS} \rangle, \quad \delta |\Phi^{R}_2
\rangle =  u  |\Phi^{NS}_2 \rangle
\end{equation}
with
\begin{equation}\label{w}
u = \langle 0^{NS}|\ exp\ (\psi^\mu_0 \psi_{\mu 1/2} +
\frac{i}{2}\gamma_{1/2} \beta_{1/2} ) |0^R\rangle.
\end{equation}
In this manner the generalized triplet in the NS sector with
physical field (top spin) which has one $\psi^\mu_{-1/2}$
oscillator and $n$ bosonic $a^{\mu +}$ oscillators is superpartner
of the R sector triplet in (\ref{Rtrip}) with $n$ bosonic
oscillators for the physical field (top spin). Therefore the
expansion of the exponential in (\ref{w}) will have a finite
number of terms in its Taylor expansion.

One can possibly try to consider an arbitrary oscillator content
in both sectors. For this reason one has to construct the operator
which is similar to that of \cite{KNNW1} (which is an analog of
the fermion emission vertex operator of \cite{Friedan:1985ge})
which transforms states of the NS sector to states of the R sector and
vice versa. This operator should have the property
\begin{equation} \label{susy}
Q_R U = U Q_{NS}.
\end{equation}
However this problem along with the problem of finding irreducible
supermultiplets for the generalized triplet is still open.

{\bf Compensator  equations}

Similarly to  the case of bosonic fields, one can write the compensator equations
for the fermionic fields as well. Namely, introducing the
fermionic Fang-Fronsdal operator \cite{fangfronsd}
\begin{equation}
{\cal S} \ = \ i \, \left( \dsll \psi \ - \ \partial \psisl
\right)
\end{equation}
and using the short-hand notations of section 1 one can write the
compensator equations in the form
\begin{eqnarray}
&& {\cal S} \ = \ - \ 2 \, i \, \partial^2 \, \xi \ , \nonumber \\
&& \psisl^{\ '} \ = \ 2 \, \partial \cdot \xi \ + \ \partial \,
\xi^{\ '} \ + \ \dsll \xisl \ , \label{compfermiflat}
\end{eqnarray}
where the  field $\xi$ is the compensator.
These equations are then invariant under the gauge transformations
\begin{eqnarray}
&& \delta \psi \ = \ \partial \, \epsilon \ , \nonumber \\
&& \delta \xi \ = \ \esl \ ,
\end{eqnarray}
involving an unconstrained gauge parameter, and are consistent,
since the first implies the second via the Bianchi identity
\be
{\cal S} \ - \ \frac{1}{2} \, \partial \ {\cal S}{\; '} \ - \
\frac{1}{2} \dsll  \ssl \
 = \ i \ \partial^{\; 2} \psisl\;' . \   \label{bianchifermi}
\end{equation}
The corresponding Lagrangian description of compensator equations
are derived in \cite{Buchbinder:2004gp}, while the AdS deformation
of both fermionic and bosonic compensator equations can be found
in \cite{Sagnotti:2003qa}.

\vspace{1cm}

\noindent {\bf Acknowledgements.} We are most grateful to late A.
Pashnev, who supervised the PhD thesis of one of the authors
(M.T.) for introducing him to the subject, for a long and  most
valuable collaboration, for his discussions and explanations. We
would  like to  specially  acknowledge I.L. Buchbinder and A.
Sagnotti for their collaboration on many topics presented in this
review.
 We would like also to thank N.Irges and A.Petkou
with whom we obtained the results on the interaction vertexes
and  X. Bekaert for
the collaboration on the gauge invariant description of massive higher
spin fields. It
is a pleasure to acknowledge I. Bandos,  P. Benincasa,
F. Cachazo, D. Francia, C. Iazeolla, P. Pasti, K.L. Panigrahi,
D.Sorokin, M. Tonin, P.Sundell, M. Vasiliev and P.West for
most valuable discussions. The work of A. F. is partially
supported by the European Community's Human Potential Programme
under contract MRTN-CT-2004-005104 and by the Italian MIUR under
contracts PRIN-2005023102 and PRIN-2005024045. The work of M.T.
has been supported by the Austrian Research Funds project
P18679-N16 ``Non Perturbative effects in String
Compactifications''.


\renewcommand{\thesection}{A}

\setcounter{equation}{0}

\renewcommand{\theequation}{A.\arabic{equation}}

\section{Some Formulas in Ambient Space}\label{Ap}

For simplicity we shall put the radius of the AdS space equal to
$1$.
 One can check some useful relations for an ambient space
 \be
\theta^{AB} \theta^{BC} = \theta^{AC},  \quad \nabla^C\theta^{AB} =
\theta^{CA}y^B + \theta^{CB} y^A,  \quad \nabla^A\theta^{AB} = {\cal
D}y^B, \ee
 \begin{equation}
 [\nabla_A,\nabla_B] =    -y_A \nabla_B +y_B \nabla_A,
\ee
\begin{equation}
 [\nabla^2,y^A] = 2 \nabla_A +{\cal D} y^A,  \quad  [\nabla^2,\nabla^A] =
 (2 -{\cal D} )\nabla^A + 2 y^A \nabla^2.
\ee

\begin{equation}
 y^A \nabla_A =0,  \quad  y^A  \theta_A^B =0,  \quad  \nabla^A y_A = {\cal
D}. \ee The induced metric, its inverse and Christofell connection
look as follows: \be g_{\mu \nu} = (\partial_\mu y^A) (\partial_\nu
y^A),  \quad g^{\mu \nu} = (\nabla^A x^\mu)(\nabla^A x^\nu), \quad
\Gamma^\lambda_{\mu \nu}=
 \frac{\partial x^\lambda}{\partial y^A} \frac{\partial^2 y^A}{\partial
x^\mu \partial x^\nu}. \ee
 We have also
\be \label{thetadelta} \theta^A_B = \frac{\partial y^A}{\partial
x^\mu} \frac{\partial x^\mu}{\partial y^B}, \quad \delta^\mu_\nu =
\frac{\partial y^A}{\partial x^\nu} \frac{\partial x^\mu}{\partial
y^A}, \ee as well as \be \theta^{AB} = g^{\mu\nu} ({\partial_\mu
y^A}) ({\partial_\nu y^B}), \ee which follow from the
differentiation rules
 \be \theta^{AB}\frac{\partial }{\partial
y^B}
 = \eta^{AC} \frac{\partial x^\mu}{\partial y^C} \frac{\partial }{\partial
x^\mu},
 \quad
\frac{\partial }{\partial x^\mu}
 =  \frac{\partial y^A}{\partial x^\mu} \frac{\partial }{\partial
 y^A} \label{derivatives}.
\ee
Finally, it is straightforward to derive the following relations
\be \theta^{AB}\frac{\partial x^\mu}{\partial y^B} =
g^{\mu\nu}\frac{\partial y^A}{\partial x^\nu}, \quad
 \nabla^2  x^\mu = - \Gamma^\mu_{\nu \rho}g^{\nu \rho}.
 \end{equation}
\begin{equation}
\nabla^A \nabla_A \Phi^{C_1 C_2 ...C_s} = \nabla^\mu\nabla_\mu
\Phi^{C_1 C_2...C_s}, \quad
 {\nabla}_\mu \frac{\partial y^A}{ \partial
x^\nu}= g_{\mu \nu}y^A. \label{imptr}\ee

\end{document}